\documentclass[10pt,letterpaper]{article}

\usepackage[T1]{fontenc}
\usepackage{amsmath}
\usepackage{amssymb}
\usepackage[tt=false, type1=true]{libertine}
\usepackage[varqu]{zi4}

\usepackage[libertine]{newtxmath}

\usepackage{setspace}
\makeatletter
\renewcommand\normalsize{  \@setfontsize\normalsize{9}{11}  \abovedisplayskip 8pt \@plus 4pt \@minus 4pt
  \abovedisplayshortskip \z@ \@plus 3pt
  \belowdisplayshortskip 5pt \@plus 3pt \@minus 3pt
  \belowdisplayskip \abovedisplayskip
  \let\@listi\@listI}
\normalsize
\renewcommand\small{\@setfontsize\small{8}{10}}
\renewcommand\footnotesize{\@setfontsize\footnotesize{7}{8}}
\renewcommand\scriptsize{\@setfontsize\scriptsize{6}{7}}
\renewcommand\tiny{\@setfontsize\tiny{5}{6}}
\renewcommand\large{\@setfontsize\large{10}{12}}
\renewcommand\Large{\@setfontsize\Large{10.95}{13}}
\renewcommand\LARGE{\@setfontsize\LARGE{12}{14}}
\renewcommand\huge{\@setfontsize\huge{14.4}{17}}
\renewcommand\Huge{\@setfontsize\Huge{17.28}{20}}
\makeatother

\usepackage{textcase}
\makeatletter
\renewcommand\section{\@startsection{section}{1}{\z@}  {-.75\baselineskip \@plus -2\p@ \@minus -.2\p@}  {.25\baselineskip}  {\sffamily\bfseries\MakeTextUppercase}}
\renewcommand\subsection{\@startsection{subsection}{2}{\z@}  {-.75\baselineskip \@plus -2\p@ \@minus -.2\p@}  {.25\baselineskip}  {\sffamily\bfseries}}
\renewcommand\subsubsection{\@startsection{subsubsection}{3}{\z@}  {-.5\baselineskip \@plus -2\p@ \@minus -.2\p@}  {-3.5\p@}  {\sffamily\itshape}}
\renewcommand\paragraph{\@startsection{paragraph}{4}{\parindent}  {-.5\baselineskip \@plus -2\p@ \@minus -.2\p@}  {-3.5\p@}  {\itshape}}
\makeatother

\def\ifarxiv{}

\usepackage{graphicx}
\usepackage{booktabs}
\usepackage{xcolor}
\usepackage{framed}
\usepackage{tikz}
\usetikzlibrary{shapes.geometric, arrows, positioning, arrows.meta, calc}
\usepackage{pgfplots}
\pgfplotsset{compat=1.18}
\usepgfplotslibrary{groupplots}
\usepackage{algorithm}
\usepackage{algpseudocode}

\usepackage{geometry}
\usepackage{microtype}
\usepackage[bookmarksnumbered,unicode]{hyperref}
\hypersetup{hidelinks, pdflang={en}, pdfdisplaydoctitle}

\newcommand{\apptablesize}{\small}

\makeatletter
\renewenvironment{abstract}{  \small
  \@afterindentfalse
  \@afterheading
}{  \par\medskip
}
\makeatother

\usepackage{caption}
\makeatletter
\def\fps@figure{tbp}
\def\fps@table{tbp}
\makeatother

\makeatletter
\renewcommand\@maketitle{  \newpage
  \null
  \vskip 1em  {\raggedright\LARGE\sffamily\bfseries \@title \par}  \vskip 1.5em  {\raggedright\Large\sffamily\lineskip .5em
   \@author\par}  \vskip 1.5em}
\makeatother

\title{Ab Initio Accuracy Beyond the Cutoff: Distributed Message-Passing Equivariant Attention at Tens-of-Millions-of-Atom Scale}

\newcommand{\artaffil}{IQuest Research, UBio Team and HPC Team, China}
\newcommand{\corrnote}{\textsuperscript{*}Corresponding authors:
  \texttt{arthursjiang@iquestlab.com}, \texttt{zhaojp@gmail.com},
  \texttt{jialrs.z@iquestlab.com}. The authors above are listed in random
  order.}
\author{  \parbox{0.94\textwidth}{\centering
    Allex Kan, Ji Zhang, Hao Yin, Hao Li, Owen Fu, Xunxing Chen,
    Lin Huang, Zion Wang, Lucas Li, Jiajun Cheng,
    Arthur Jiang\textsuperscript{*}, Junping Zhao\textsuperscript{*},
    and Jia Zhang\textsuperscript{*}\\[10pt]
    \normalsize\itshape \artaffil}}
\date{}

\begin{document}

\maketitle

\begingroup
\let\thefootnote\relax
\footnotetext{\corrnote}
\endgroup

\begin{abstract}
Molecular dynamics predicts how molecules move and interact by computing the force on every atom and integrating its motion. Density functional theory supplies those forces from first principles, but its $O(N^3)$ cost stops at thousands of atoms; empirical force fields reach millions and miss bond breaking and charge transfer. Message-passing equivariant Transformers regress the \textit{ab initio} surface at linear cost, and their stacked layers carry information past any single cutoff, compounding four layers into a receptive field of up to $23$\,\AA{}. That is where the accuracy frontier now sits. Getting there at scale is a communication problem: every layer exchanges boundary features across devices, the conservative force repeats that exchange in the backward pass, and the neighbor graph is re-derived from moving atoms, so tensor shapes keep changing under a runtime that would rather capture them once. We present \textsc{Tessera}, a distributed runtime that holds the workload's shape fixed between rebuilds. It sorts atoms along a space-filling curve and cuts the sorted order into equal-count shards: every rank holds the same number of atoms, every tensor has one shape, and each rank's peer set grows sublinearly, roughly as $K^{0.5}$ rather than $K$ over the device counts we measure. A conservative neighbor skin keeps that topology valid across many coordinate updates, so the forward and backward passes of an entire rebuild interval replay as one captured CUDA Graph. Fused kernels, a three-level neighbor search, and an overlapped All-to-All complete it. On 512 GPUs \textsc{Tessera} sustains $30.7$M atoms at $89.7\%$ weak-scaling efficiency ($77$--$90\%$ across three per-GPU loads), with per-step time rising $1.11$--$1.29\times$ as machine and problem grow $16\times$. Graph reuse cuts full-step time $1.87\times$, and neighbor-search cost grows $1.7\times$ over an $8\times$ device increase where a Morton-order baseline grows $3.0\times$. On a live $10.24$M-atom salt solution, energy drift stays at $0.42\,\mu$eV/atom/ps over $10$\,ps of NVE, and structure matches a serial reference. A $10.06$M-atom Norovirus capsid then holds a $100$\,ps NVT trajectory at $\approx\!8$\,ps/day on $256$ A100 GPUs, with no broken bonds, no anomalously close pairs, and only three compiled variants across $200{,}000$ steps---no graph break, no fallback to eager execution.
\end{abstract}


\section{Introduction}

\input{figures/main_fig_tikz}

Some computations are hard to distribute for a reason that arithmetic alone does not explain: the data they need is not where they compute. Predicting the force on every atom is one of them, and it is where molecular simulation begins.

Atoms are not independent. An atom's force is set by its whole environment: electrostatic interactions decay slowly, and the effects that decide chemical behaviour---polarization, charge transfer, the making and breaking of bonds---are neither pairwise nor confined to a few neighbours. The energy is therefore a function of the entire configuration, and it admits no division into independent per-atom problems that keeps those effects.

For any machine that computes it, two consequences follow. The work is global: each atom's evaluation gathers information from around it, so pulling the system apart across $K$ machines does not pull the dependency apart with it. Data has to cross whatever boundary the machine imposes, and the further the physics reaches, the more of it has to move. The pattern is also not fixed. Positions change every step, so which atoms depend on which changes with them, and a plan built for one configuration cannot be assumed valid for the next.

The systems where this matters most are the solvated, charged ones: salt solutions, electrolyte interfaces, ion channels, the binding of a ligand to a protein. There, polarization and charge transfer reach past any single cutoff, and a model that stops at one cutoff cannot represent them. Those systems also have to be large, because the interface, the pore and the binding site are properties of a whole assembly rather than of a few hundred atoms.

Empirical force fields have long been the practical answer. Their force law is cheap: fixed partial charges, bonded springs and torsions, a fitted van der Waals term. That cheapness carries the method far---a single GPU holds millions of atoms, and a distributed run reaches a hundred million~\cite{abraham2015gromacs, thompson2022lammps}---which is why classical MD became the default rather than a compromise. What the force law cannot do is change. Charges do not polarize, bonds do not break, and charge transfer never happens, so the model is wrong wherever the chemistry is interesting. Nor is the size of that error known in advance for a new system, which matters because the quantities that decide a biological outcome---a conformational free-energy difference, an ion-selectivity ratio---are often smaller than the error itself~\cite{huang2026ubiomolfm}.

\textit{Ab initio} methods, density functional theory among them~\cite{kohn1965self}, supply the missing physics at a cost that grows as $O(N^3)$: in practice they stop at hundreds to a few thousand atoms, too small to hold a solvated interface and too slow to follow one for long.

Machine-learning force fields are the route that has worked~\cite{behler2007generalized, schutt2017schnet, unke2021machine, batzner2022nequip, batatia2022mace}. They regress the \textit{ab initio} energy surface and evaluate it at a cost linear in atom count, and the strictly local ones have taken that to a hundred million atoms: Deep Potential MD's Gordon-Bell-winning Summit run~\cite{jia2020pushing, zhang2018deep} confines each atom's energy to its own cutoff ball, $6$ and $8$\,\AA{} for water and copper, and Allegro uses cutoffs as short as $4$\,\AA{} in some benchmarks, forgoing message passing by design~\cite{musaelian2023learning}. The choice buys scale at the cost of the property the physics needs: information never travels the molecular graph, so a learned feature never propagates past a single cutoff~\cite{grisafi2019incorporating, yue2021short, ko2021fourth}.

Message-passing equivariant Transformers restore the propagation~\cite{liao2022equiformer, tholke2022torchmdnet, huang2026e2formerlsr}. Stacked finite-radius attention compounds radii, so a few layers reach $20$\,\AA{} and beyond, $23$\,\AA{} in the stack used here, and this class now sets the accuracy frontier in molecular simulation. It is also the class the argument above bites hardest on: the further a model reaches into the system, the more of it each evaluation needs, and the more communication the machine must carry. To our knowledge, no published system has taken the class past tens of GPUs; the closest, SevenNet~\cite{park2024sevennet}, reports $32$.

This paper asks a single question. \textit{How can we propagate graph attention across a live molecular graph at tens-of-millions-of-atom scale without letting communication and variable topologies dominate the step?} Three difficulties stand in the way, one per layer of the stack.

\textbf{Acceleration-hostile equivariant operators.} Accuracy hinges on $E(3)$-equivariant Transformers, which keep forces rotating with the system. Their components---Clebsch--Gordan tensor products, radial-basis encodings, equivariant gating---match neither the dense kernels FlashAttention~\cite{dao2022flashattention} targets nor the BF16 precision it assumes. Single-GPU efficiency therefore has to be rebuilt from custom kernels.

\textbf{A decomposition that carries its own cost.} Ten million atoms exceed any single node, so the graph must be cut, yet cutting it does not make the per-layer exchange small. The graph is also the physical configuration itself. Nobody partitions it offline, once, the way a training pipeline partitions a dataset. Topology, neighbor lists, and plan are therefore re-derived from live coordinates mid-trajectory.

\textbf{Recapture on the critical path.} Every re-derivation changes edge counts, which is what forbids replay. Conservative forces make the recapture expensive, because the full forward--backward execution \emph{is} the step. There is no smaller unit to fall back on.

Adjacent systems each address a subset. LLM training scales to thousands of GPUs~\cite{narayanan2021megatron}, but over dense, statically shaped collectives whose routing is fixed for the run. Classical domain decomposition~\cite{plimpton1995fast, thompson2022lammps} assumes shape-flexible updates that a CUDA Graph cannot tolerate. Distributed GNN training~\cite{zheng2020distdgl, jia2020roc} partitions a \emph{fixed} graph once, where ours is re-derived from moving coordinates every step. The equivariant models that reach biomolecular size do so by forgoing propagation~\cite{musaelian2023scaling}. Universal MLFFs such as UMA are released for single-GPU inference~\cite{wood2025uma}, with no distributed runtime reported.

No single optimization suffices. The three difficulties sit at different layers of the stack, and a fix at one pays off only if the others cooperate. We therefore treat the model and the distributed runtime as one object of design, an \textbf{algorithm--system co-design} resting on a common foundation: how atoms are placed on devices. Standard rules force a choice there. Equal-volume cells bound each rank to 26 neighbouring cells, 27 with its own, but counts drift with density, and every tensor shape with them. Equal atom shares give exact counts but ignore geometry, so they destroy locality and leave per-edge work unbalanced. One construction avoids the choice. Quantize atomic coordinates onto a space-filling curve~\cite{springel2005gadget}, sort, and cut the sorted list into equal-count contiguous segments, one per rank. The cut fixes per-rank counts, and with them every node-indexed tensor shape; edge-indexed tensors and the exchange plan are pinned separately by the rebuild epoch. The curve then confines each segment's neighbors to a set of adjacent subdomains that grows far more slowly than the machine. At $25$k atoms per GPU the heaviest rank requests candidates from $6$, $11$, $14$, and $19$ source ranks at $64$, $128$, $256$, and $512$ GPUs, an exponent near $0.5$ rather than $1$ (Section~\ref{sec:morton_partition}). The confinement is tight for Hilbert, where Z-order shards fragment.

Two further leverage points come from the physics. At $300$\,K, hydrogen moves about $0.1$\,\AA{} between one rebuild and the next, well inside the $0.25$\,\AA{} displacement trigger that the $0.5$\,\AA{} skin allows, and about $1$\,\AA{} over the $50$\,fs between re-partitions, small against the $\sim\!67$\,\AA{} subdomain. A molecular graph's \emph{topology} therefore outlives its \emph{geometry} on two clocks: nine steps in ten reuse the edge set and the communication plan, and the partition is re-cut only every hundred steps. Neither reuse stands alone. A rebuild turns over the edge shapes, so every rank must rebuild on the \emph{same} step. An \texttt{all\allowbreak\_reduce(MAX)} over local triggers agrees that step, which keeps the exchange plans mutually consistent and the captured graph valid. Partition, neighbor search, and graph capture hold or fail together.

Propagation is not a design preference here: the distributed cost follows from the model class, not from one implementation choice (Section~\ref{sec:bg-gnn}).

We call this runtime \textsc{Tessera}, after the mosaic tile.\footnote{A tessera is a small block, cut to the same size as its neighbours and laid beside them so that many make a whole; it has fixed faces and a definite boundary; and in Rome it was also a token a messenger carried, telling the next person what to do. The design has one of each: the equal-count shards are the blocks, the fixed tensor shapes are their faces, and the epoch every rank agrees on before a rebuild is the token.} Figure~\ref{fig:main_architecture} traces one step through it, from the equal-count partition to the decentralized integrator. The four contributions below build it from the bottom up, kernels to trajectory.

\textbf{High-performance equivariant kernels.} A unified MCH layout, dual CSR/CSC indexing, and deep operator fusion in custom Triton kernels, together with \texttt{torch.\allowbreak compile}, reduce step latency $2.6\times$ and peak memory $4.4\times$ on a $10$k-atom single-GPU ablation; adding TF32 and recomputation brings these to $3.2\times$ and $9.2\times$, at the accuracy cost Section~\ref{sec:experiments} characterizes.

\textbf{Live-graph decomposition with topology reuse.} An equal-count space-filling-curve cut fixes every rank's atom count and keeps its shard local, supplying static shapes and a peer set that grows as roughly $K^{0.5}$ rather than $K$. A conservative AABB/chunking/FAISS~\cite{johnson2019billion-scale} hierarchy then rebuilds the graph from live coordinates at $\sum_i Q_iC_i$ rather than $Q_{\mathrm{loc}}C_{\mathrm{cand}}$ cost. Separating reusable topology from per-step geometry cuts neighbor cost by an order of magnitude at 512 GPUs, and its growth over an $8\times$ device increase falls to $1.7\times$, against $3.0\times$ for a Morton-order all-gather baseline.

\textbf{Sparse, overlapped, decentralized communication.} Plan reuse, sender-side SKV projection, and a pre-registered buffer pool strip non-payload overhead. Paired \texttt{pre}/\texttt{post} autograd operators keep the exchange hidden in both passes, worth $84.7\%$ against $72.9\%$ efficiency at 512 GPUs. An on-GPU decentralized engine replaces the rank-0 scatter/reduce loop with a constant-payload barrier.

\textbf{Demonstration and validation at scale.} \textsc{Tessera} sustains the full force-evaluation step (forward, exchange, autograd backward) at $30.7$M atoms on 512 GPUs at $89.7\%$ \emph{weak-scaling} efficiency. Per-step time rises $1.11$--$1.29\times$ across a $16\times$ growth in machine and problem. On a live $10.24$M-atom trajectory it holds energy drift to $0.42\,\mu$eV/atom/ps and matches the serial reference's structure. A matched pair against a $10.06$M-atom Norovirus capsid (protein-dense against water) then locates the cost of an atom-count cut: $13$--$33\%$ more per step across trajectory phase, $14.5\%$ in the reported window, in edges per atom.

To our knowledge, no prior system has run a message-passing equivariant Transformer at tens of millions of atoms on hundreds of GPUs. The forces are taken as the analytic gradient, over a neighbor graph re-derived from live coordinates, with information propagating along that graph rather than stopping at a single cutoff. The exchange each GPU performs is set by local geometry and grows only slowly with device count, so the exchange cost extends in principle toward billion-atom systems, though the $K$- and $N$-dependent terms of Section~\ref{sec:discussion} still grow.

\textbf{Roadmap.} Section~\ref{sec:background} supplies the background the rest assumes. It covers equivariant message passing and the receptive-field arithmetic, the conservative-force path and why it doubles the exchange, why static graph execution is fragile, and the three requirements a partition must satisfy. Section~\ref{sec:engine} develops the co-design against them. Section~\ref{sec:experiments} evaluates it, Section~\ref{sec:related} positions it, and Section~\ref{sec:discussion} states what the measurements do and do not establish. The derivations the argument turns on---the overlap bound, the local-neighbor fraction, the payload and plan-reuse arithmetic, and the cache exactness invariant---are developed in place, in the sections that use them.

\section{Background and Problem Formulation}
\label{sec:background}

This section supplies the two bodies of background the paper assumes---equivariant message passing and static graph execution---and states, as requirements, what a partition must deliver for the second to survive the first. The receptive field and the force path are derived here because the design choices of Section~\ref{sec:engine} are only legible against them.

\subsection{Equivariant Message Passing}
\label{sec:bg-gnn}

A molecular configuration is a set of $N$ atoms with positions $\mathbf{r}_i\in\mathbb{R}^3$ and species $z_i$, under periodic boundary conditions. An equivariant message-passing potential represents each atom not by a flat vector but by a collection of \emph{irreducible representations} of the rotation group: for each degree $l$ up to a maximum $l_{\max}$, a block of $2l+1$ components that mix among themselves under a rotation as the spherical harmonics $Y_l^m$ do. A block carries $C$ channels, so the feature at atom $i$ after layer $s$ is a family $\mathbf{h}_i^{s,(l)}\in\mathbb{R}^{(2l+1)\times C}$ for $l=0,\dots,l_{\max}$. The deployed model uses $l_{\max}=2$ and $C=256$ (Appendix~\ref{sec:app-config}), so one atom's feature holds $(l_{\max}+1)^2C=9\times256=2{,}304$ scalars---the count that reappears as the equivariant Value width in Section~\ref{sec:reduce_volume}.

One layer performs three operations. First, \emph{edge construction}: an edge $(i,j)$ exists when the minimum-image displacement $\mathbf{r}_{ij}=\mathbf{r}_j-\mathbf{r}_i$ satisfies $|\mathbf{r}_{ij}|\le r_c$, described by a radial basis encoding of the distance (256 Gaussian-smear functions here) and the spherical harmonics of its direction. Second, \emph{Clebsch--Gordan tensor products}~\cite{thomas2018tensorfield, geiger2022e3nn}: features of degrees $l_1$ and $l_2$ combine into a feature of degree $L_{\mathrm{out}}$ by
\begin{equation}
\bigl(\mathbf{x}^{l_1}\otimes\mathbf{y}^{l_2}\bigr)^{L_{\mathrm{out}}}_{M}
=\sum_{m_1=-l_1}^{l_1}\sum_{m_2=-l_2}^{l_2}
G^{L M}_{l_1 m_1 l_2 m_2}\,
x^{l_1}_{m_1}\,y^{l_2}_{m_2},
\qquad L_{\mathrm{out}}=|l_1-l_2|,\dots,l_1+l_2,
\end{equation}
with $G^{LM}_{l_1m_1l_2m_2}$ the Clebsch--Gordan coefficients. The product is what makes the representation equivariant: rotating both inputs rotates the output by the same rotation, so a network built only from such products, scalar contractions, and gating maps a rotated configuration to a rotated prediction. It is also not a dense matrix multiply: the index space is a direct sum over degree triples, and the number of surviving paths grows with $l_{\max}$, which is why the kernels of Section~\ref{sec:single_gpu} are custom.

Third, \emph{attention}~\cite{schutt2021equivariant, liao2022equiformer}: for each edge the layer forms a scalar score $s_{ij}$ as an invariant contraction of a query feature on $i$ with a key feature on $j$; with $H=128$ heads of scalar width $d_k=8$ the score is a per-head scalar, and a softmax over $i$'s neighbors turns the scores into weights $\alpha_{ij}$. The message is an equivariant Value---the degree-$l$ block of atom $j$, possibly tensor-produced with the edge's spherical harmonics---aggregated as $\sum_j\alpha_{ij}\mathbf{V}_j$ and combined with the residual stream. The score and the message are per-edge objects, so the attention cost scales with the number of edges rather than the number of atoms, while the per-atom projections and gating still scale with $N$. That is what Section~\ref{sec:density-heterogeneity} confronts when an atom-count cut meets a protein-dense region.

\textbf{Receptive field.} Stacking layers compounds their cutoffs. Let layer $s$ have cutoff $r^{(s)}$. After one layer, $\mathbf{h}_i^{1}$ depends on atoms within $r^{(1)}$ of $i$. If $\mathbf{h}_j^{s}$ depends on atoms within $\sum_{s'\le s}r^{(s')}$ of $j$, then $\mathbf{h}_i^{s+1}$ depends on every atom within $r^{(s+1)}$ of some such $j$, hence on atoms within $\sum_{s'\le s+1}r^{(s')}$ of $i$: the bound is additive by induction, because each layer spends its cutoff to reach one more hop. Our stack has three short-range layers at $5$\,\AA{} and one long-range layer at $8$\,\AA{}, so the per-atom receptive field is bounded by
\begin{equation}
3\times5\,\text{\AA}+1\times8\,\text{\AA}=23\,\text{\AA},
\end{equation}
the radius we quote throughout as the nominal receptive field. A strictly local potential has a receptive field equal to its single cutoff---$4$--$8$\,\AA{} in the systems Section~\ref{sec:related} compares---so its per-atom computation touches only its own cutoff ball. Here the energy of one atom can depend on atoms up to $23$\,\AA{} away, through intermediate atoms whose features were themselves computed from remote neighbors.

\subsection{Conservative Forces}
\label{sec:bg-forces}

The potential is trained to reproduce an \emph{ab initio} energy surface, and the dynamics must conserve energy. A network that emitted forces as a second output head would not do: the force field would not be the gradient of the energy it reports, and a trajectory integrated with it would drift for reasons unrelated to the physics. We therefore take the force on atom $i$ as the analytic gradient
\begin{equation}
\mathbf{F}_i=-\frac{\partial E}{\partial\mathbf{r}_i},
\qquad
E=\sum_{i}E_i\bigl(\{\mathbf{r}_j\}_{j\in\mathcal{R}(i)}\bigr),
\end{equation}
where $\mathcal{R}(i)$ is the receptive field of Section~\ref{sec:bg-gnn}---the $\mathcal{R}$ distinguishing it from the one-hop neighbour set $\mathcal{N}(i)$ of a single layer, used in Section~\ref{sec:single_gpu}. Two consequences follow.

First, a step costs a forward \emph{and} a backward pass. The forward pass evaluates $E$; the forces come from reverse-mode differentiation through the same graph, and because every atom's energy depends on its neighbors' features, the gradient with respect to one coordinate collects terms from every edge on which that coordinate appears---as a direct displacement and through the features of intermediate atoms. There is no cheaper local formula: the whole propagated graph participates. The backward pass is a first-class half of the step, comparable in cost to the forward and, as Section~\ref{sec:overlap} shows, separately schedulable.

Second, the backward pass must exchange the \emph{same} boundary features again, in reverse. In the forward pass a remote atom $j$'s Key, source embedding, and Value travel from $j$'s owner to $i$'s owner. Differentiating with respect to $\mathbf{r}_i$ and $\mathbf{r}_j$ requires the upstream gradient of the score, $g_{ij}=\partial E/\partial s_{ij}$, to travel back along the same edge, and requires the Value and edge geometry on $j$'s owner to accumulate $j$'s contribution. A forward-only exchange leaves the backward exchange behind the forward compute, a serial tail no stream juggling repairs---which is why Section~\ref{sec:overlap} splits the interface into a \texttt{pre} operator that communicates in the forward pass and a \texttt{post} operator that communicates in the backward.

Both consequences constrain correctness. Because the force is the exact gradient of the energy actually evaluated, the backward pass must use the neighbor set the forward pass used, with the same active pairs; an approximate query that drops an in-cutoff pair changes the force, not merely the speed. The cache of Section~\ref{sec:multilevel_search} is built to satisfy this: it reuses topology only under a displacement bound that guarantees the reused superset still contains every pair that can be active, and it re-applies the exact cutoff test on every hit.

\subsection{Why Static Graph Execution Is Fragile}
\label{sec:bg-static}

The per-step work is thousands of small, irregular kernels---gathers, scatters, gated multiplications, short tensor-product contractions---rather than a few large matrix multiplies. At that granularity the CPU-side cost of dispatching a kernel is comparable to its GPU-side cost, so the runtime is worth capturing. Two mechanisms do the capturing, and they fail differently.

\textbf{\texttt{torch.compile} and guards.} The compiler traces the model and specializes it~\cite{ansel2024pytorch2}, emitting \emph{guards}: predicates on shapes, strides, and dtypes that must hold for the compiled code to be valid. Every call checks them, and a guard that fails forces a re-trace and often a recompilation; the work is not wrong, but it is paid on the step where the shape moved. Compiling for dynamic shapes---symbolic extents rather than constants---removes the re-trace for extents that vary within the compiled range.

\textbf{CUDA Graph capture and replay}~\cite{nvidia2023cuda}. A captured graph is a different object. Capture records the sequence of kernel launches, their arguments, and their grid and block geometry into a DAG, together with the device addresses the kernels read and write; replay submits that DAG with a single launch, bypassing the dispatcher, the allocator, and the Python layer. That is where the saving comes from, and it is why the mechanism is rigid: replay consults no per-launch shape, because no per-launch host code is left. The addresses are fixed, so the allocator must hand out stable buffers; the launch geometry is fixed, so every tensor extent that determines it must be fixed.

The two compose in one direction only. Compiling for dynamic shapes removes the re-trace, but not the capture: the captured graph still replays the allocations and launch parameters recorded at capture time, so a step whose tensors change extent cannot be replayed and must be captured again. Section~\ref{sec:related} states this once; here is the consequence. The neighbor graph is re-derived from live coordinates, and its edge count moves whenever atoms cross the cutoff; if the node count also moved, no captured graph would survive more than a step, and the runtime would pay capture---not merely dispatch---every step. A design that wants replay must pin shapes by construction.

\subsection{Problem Formulation}
\label{sec:bg-problem}

Let $K$ be the number of ranks (GPUs), $N$ the number of atoms, and $\mathbf{r}=(\mathbf{r}_1,\dots,\mathbf{r}_N)$ the positions in a periodic cell. The stack carries two cutoffs---$5$\,\AA{} for its three short-range layers and $8$\,\AA{} for its long-range one---and at step $t$ the live edge set is
\begin{equation}
\mathcal{E}_t=\bigl\{(i,j): |\mathbf{r}_{ij}|\le r_c^{(s)}\ \text{for the layer } s \text{ that owns the edge}\bigr\},
\end{equation}
with $r_c$ standing for the cutoff of whichever layer is being built. The cache of Section~\ref{sec:multilevel_search} does not store $\mathcal{E}_t$: it stores a rebuild-time superset built at $r_c+\delta$, and a hit re-applies the test in~(4) to recover the active pairs. A partition is a map $\pi_t:\{1,\dots,N\}\to\{1,\dots,K\}$ assigning each atom to a rank; write $P_p$ for the set of ranks from which rank $p$ requests candidates.

A step advances the positions, rebuilds $\mathcal{E}_t$ as needed, and evaluates the force by the forward--backward path of Section~\ref{sec:bg-forces}. The runtime can replay that step only if the partition delivers three properties together.

\textbf{(R1) Exact per-rank atom count.} Every rank owns $\lfloor N/K\rfloor$ or $\lceil N/K\rceil$ atoms, so the shape of every node-indexed tensor is fixed. The cut is a deterministic function of the coordinates at each re-cut, so all ranks derive the same ownership map without a consensus round, and the map is frozen between re-cuts.

\textbf{(R2) A peer set that grows sublinearly in $K$.} The number of ranks a rank must talk to should be set mainly by local density and cutoff geometry, and grow only sublinearly with machine size. Neither classical rule delivers that alone. An equal-volume grid bounds its peers at $27$---the $26$ face-, edge-, and corner-adjacent cells plus the cell itself---but only by letting counts float. An equal-count cut with no geometry keeps the counts exact but degenerates toward $\approx K$ peers. The design requires both properties at once, and Section~\ref{sec:morton_partition} shows that one construction supplies them. There the measured peer degree is $6$, $11$, $14$, and $19$ source ranks at $64$, $128$, $256$, and $512$ GPUs ($25$k atoms/GPU, Hilbert with rank-box): $3.2\times$ over an $8\times$ increase in devices, an exponent near $0.5$ rather than $1$. This is sublinear growth, not a constant. The peer count still rises with the machine, and Section~\ref{sec:discussion} lists it among the terms that do.

\textbf{(R3) Tensor shapes constant within a cache epoch.} Between rebuild events every tensor in both passes keeps a fixed shape, so one capture serves the whole interval. Node tensors are pinned by (R1); edge-indexed tensors follow the cached candidate superset of Section~\ref{sec:multilevel_search}, and both they and the communication plan change only on a rebuild, which is therefore an epoch boundary and must be unanimous: if one rank rebuilds and another does not, the exchange plans disagree and the captured graph is invalid.

These are requirements, not results; Section~\ref{sec:engine} shows that one placement mechanism supplies all three.

\textbf{Notation.} $N$ atoms, $K$ ranks; $\mathbf{r}_i,\mathbf{v}_i,\mathbf{F}_i$ position, velocity, and force of atom $i$; $m$ atomic mass; $\Delta t$ timestep; $r_c$ cutoff, with $5$\,\AA{} and $8$\,\AA{} the two values in our stack; $\delta$ displacement skin ($0.5$\,\AA{}); $I$ rebuild interval ($10$ steps); $\mathcal{E}_t$ the active edge set at step $t$ and $\Pi$ the communication plan, with the rebuild-time candidate superset and the plan cached between rebuilds; $\sigma$ the sort permutation produced by a re-partition (Algorithm~\ref{alg:md-step}); $Q_{\mathrm{loc}}$ and $C_{\mathrm{cand}}$ the local query and candidate counts of the neighbor search (Section~\ref{sec:multilevel_search}), with $Q_i$ and $C_i$ the per-chunk values; $C$ alone is the channel count of Section~\ref{sec:bg-gnn}; $H$ attention heads and $d_k$ scalar head width; $l_{\max}$ maximum degree; $R_r$ the number of Transformer blocks sharing cutoff $r$; $T_c$, $T_\ell$, $T_r$ the exchange, local-branch, and remote-branch times of the overlap analysis; $\rho$ number density.

\section{Algorithm-System Co-Design}
\label{sec:engine}

We develop the co-design in the order its dependencies run. Section~\ref{sec:single_gpu} restores single-GPU efficiency for operators the mainstream stack cannot serve. Everything after it rests on how atoms are placed on devices: the equal-count space-filling-curve cut of Section~\ref{sec:morton_partition} supplies invariant tensor shapes and a peer set that grows sublinearly with device count, and on those two guarantees rest the multi-level neighbor search (Section~\ref{sec:multilevel_search}), the communication path that shrinks, fuses, and overlaps the remaining exchange (Section~\ref{sec:reduce_volume}), and the decentralized integrator that removes the per-step $O(N)$ rank-$0$ gather/scatter path, leaving that loop's collectives at constant payload (Section~\ref{sec:md_engine}). Two threads recur: a graph's \emph{topology} outlives its \emph{geometry}, letting most steps reuse both the graph and the partition; and a rebuild must be unanimous to preserve the shape invariant of Section~\ref{sec:bg-problem}, so these mechanisms hold or fail together rather than composing independently.

\subsection{Single-GPU Optimization}
\label{sec:single_gpu}

E2Former~V2's sparse equivariant attention, Clebsch--Gordan tensor products, and radial-basis encodings neither reduce to the dense matmuls FlashAttention-style kernels target nor tolerate the reduced-precision fast paths those kernels assume, so both throughput and HBM budget must be rebuilt from custom kernels upward.

\textbf{Triton-based Kernel Optimization.}
Custom Triton~\cite{tillet2019triton} kernels, registered as opaque \texttt{torch.\allowbreak compile}~\cite{ansel2024pytorch2} operators so hand-written and compiler fusion compose cleanly, are optimized along four axes; two are standard practice---hardware-aware tile-size autotuning and deep operator fusion---and are deferred to Appendix~\ref{sec:app-kernels}. The other two follow from the two-pass execution of Section~\ref{sec:bg-forces}.

\emph{Dual sparse indexing.} The forward pass and the $d\mathbf{k}$ backward pass traverse the \emph{same} sparse structure in transposed orders. Forward reads query-grouped, accumulating over each query's neighbors,
\begin{equation}
\mathbf{o}_i=\sum_{j\in\mathcal{N}(i)}\alpha_{ij}\mathbf{V}_j,
\end{equation}
while the key gradient $d\mathbf{k}_j=\sum_i g_{ij}\mathbf{q}_i$---with $g_{ij}=\partial E/\partial s_{ij}$ carrying the upstream gradient through the softmax Jacobian---is a sparse matrix--vector product that reads neighbor-grouped, accumulating over each neighbor's queries. The two traversals want opposite index orders: a row-major view is coalesced for one and strided for the other. A single row-major format therefore leaves the backward pass on strided loads, and since the backward pass runs every step, the penalty would be paid every step. We keep a hybrid \textbf{CSR/CSC} representation instead---CSR for the forward, CSC for $d\mathbf{k}$---so each pass reads its natural order. Both are built once per neighbor-list rebuild and cached; because the cache lives as long as the graph does (Section~\ref{sec:multilevel_search}), that construction is charged to rebuild steps only, and a cache hit pays nothing for it.

\emph{A layout that survives both passes.} Equivariant attention indexes atoms, channels, and heads in different orders at different stages: attention iterates over edges and heads, the tensor product contracts over channels and degree, and the residual stream is atom-major. A naive implementation pays repeated \texttt{permute} traffic between these orders, and the traffic lands in exactly the small irregular kernels the GPU is worst at. We adopt a unified \textbf{MCH} ($M\times C\times H$: rows $\times$ channels $\times$ heads) layout end-to-end and block our kernels over it: a C-blocked kernel walks a 3D grid $(M,\lceil C/BC\rceil,\lceil H/BH\rceil)$, loads contiguous $[BC,BH]$ tiles, and accumulates $\text{out}=\sum\alpha\cdot V$ and its gradient entirely in registers. Because each tile is reused across the rows of a CSR row, both passes are held to a single HBM round-trip per tensor; the same layout that removes the transposes also removes the intermediate materialization that would otherwise dominate the peak allocation.

\textbf{Compiler, Precision, and Memory Budget.}
On top, \texttt{torch.\allowbreak compile} fuses pointwise ops and cuts launch overhead, valuable since runtime is dominated by many small gather, scatter, and gating kernels rather than large matmuls. Dense projections and tensor products additionally run in TF32, trading latency for non-trivial energy/force drift. Two further mechanisms buy HBM headroom instead: a compiler-driven activation-recomputation budget lets \texttt{torch.\allowbreak compile} rather than the author place checkpoints, roughly halving the activation footprint at moderate runtime cost; a two-tier allocator policy then keeps fragmentation from eroding that headroom over long runs (Appendix~\ref{sec:app-kernels}). The TF32 trade is not free, which is why the trajectory validation of Section~\ref{sec:md-validation}, where drift \emph{is} the measurement, runs with it off.

\subsection{Spatial Partitioning via Space-Filling Curves}
\label{sec:morton_partition}

A partition over hundreds of GPUs must meet the three goals Section~\ref{sec:bg-problem} stated as requirements: (i) \emph{exact atom-count balance}, (ii) \emph{local, sparse communication}, and (iii) \emph{static tensor shapes} the per-rank computation can be captured under. Neither classical choice reaches all three. An equal-count cut ignoring geometry balances perfectly but lets a rank's neighbors live anywhere, degenerating toward all-to-all ($\approx K$ peers); an equal-volume grid bounds peers at $27$---the $26$ adjacent cells plus its own---but lets counts follow local density and pins no shape without padding to fixed capacity. The count drift is structural, not a nuisance: under uniform density a cell's population is $\mathrm{Binomial}(N,1/K)\approx\mathrm{Poisson}(N/K)$, so the extreme-value spread across $K$ cells is $n_{\max}-n_{\min}\approx2\sqrt{2\lambda\log K}$ with $\lambda=N/K$---at $N=1.024\times10^7$, $K=512$ this is $\approx\!19{,}500$--$20{,}500$ atoms/rank---an estimate that assumes uniform density and therefore degrades further on an inhomogeneous system.

One mechanism reaches all three, with no capacity to guess: quantize the 3D coordinates onto an integer lattice, map them onto a 1D \emph{space-filling curve} (SFC) that keeps spatially-close atoms close in the linear order, radix-sort by SFC code, and cut the sorted list into \emph{equal-count} contiguous segments, one per rank. The cut gives exact balance: $\lfloor N/K\rfloor$ or $\lceil N/K\rceil$ atoms per rank, independent of density. Because work is per-edge rather than per-atom, exact counts balance the load only up to the local edge-density variation Section~\ref{sec:md-validation} measures. The curve then confines each segment's neighbors to a small set of adjacent subdomains. That set grows sublinearly with the machine rather than staying fixed: at $25$k atoms/GPU the heaviest rank requests from $6$, $11$, $14$, and $19$ source ranks at $64$, $128$, $256$, and $512$ GPUs (Section~\ref{sec:neighbor-scaling}; counted per Appendix~\ref{sec:app-neighbor}), an $8\times$ increase in devices costing $3.2\times$ in peers, which is an exponent near $0.5$ rather than $1$. This is the sparse peer set Section~\ref{sec:reduce_volume} relies on, and it is small enough to keep the exchange off the critical path at our scale. Goal (i) is the \emph{means} to goal (iii): only because counts are fixed exactly can shapes be pinned.

\textbf{Static shapes make the step capturable.}
Section~\ref{sec:bg-static} explains why capture fails when shapes move; here is how the cut removes the cause. Once per-device counts move, \texttt{torch.\allowbreak compile} re-triggers guard failures and any captured CUDA Graph is discarded, leaving CPU-side scheduling latency to idle the Streaming Multiprocessors (SMs). The equal-count cut removes that cause for the \emph{node} tensors, and is also \emph{deterministic}---a pure function of the quantized coordinates with fixed offsets, so identical coordinates always yield the identical ownership map. Two timescales then follow. Per-rank atom counts---and with them every node-indexed tensor---change only when ownership is re-cut; edge-indexed tensors and the communication plan follow the neighbor graph, rebuilt on the schedule of Section~\ref{sec:multilevel_search} and constant between rebuilds. Within such a \emph{cache epoch} every tensor in both passes has a fixed shape, so we trace the computation once and capture it as a static CUDA Graph, \emph{replayed} on each intervening step rather than re-dispatched kernel by kernel. Section~\ref{sec:md-validation} reports three compiled variants, no graph break, and no fallback over a $200{,}000$-step run.

\definecolor{cblue}{HTML}{2F6DB5}
\definecolor{cteal}{HTML}{2E9E96}
\definecolor{corange}{HTML}{E08A3C}
\definecolor{cslate}{HTML}{5B6573}
\definecolor{cink}{HTML}{2B2F36}
\definecolor{cgreen}{HTML}{6DA34D}

\providecommand{\sfckeybox}[1]{  \tikz[baseline=-0.55ex]{\draw[rounded corners=0.6pt, fill=#1!22, draw=#1!55,
    line width=0.4pt] (0,-0.075) rectangle (0.20,0.115);}}
\providecommand{\sfckeyaabb}{  \tikz[baseline=-0.55ex]{\draw[cteal!85!black, line width=0.7pt,
    dash pattern=on 1.9pt off 1.3pt] (0,-0.075) rectangle (0.20,0.115);}}
\providecommand{\sfckeycross}{  \tikz[baseline=-0.55ex]{\draw[cink!60, line width=0.45pt]
    (-0.045,-0.025) -- (0.045,0.065)  (-0.045,0.065) -- (0.045,-0.025);}}
\providecommand{\sfckeyjump}{  \tikz[baseline=-0.55ex]{\draw[corange!85!black, line width=0.8pt,
    -{Stealth[length=3pt,width=2.8pt]}] (0,0.02) -- (0.22,0.02);}}

\input{figures/figlayout}

\begin{figure}[t]
\centering
\resizebox{\figSfcScale\linewidth}{!}{\begin{tikzpicture}[
    font=\sffamily,
    x=1cm, y=1cm,
    every node/.style={inner sep=1.2pt},
    ttl/.style={font=\sffamily\bfseries\footnotesize, text=cink},
    sub/.style={font=\sffamily\fontsize{6.2}{7.0}\selectfont, text=cslate, align=left},
    tiny/.style={font=\sffamily\fontsize{6.2}{7.2}\selectfont, text=cslate, align=center},
  ]

\def\CS{0.40}          \def\PBX{3.68}         \def\GY{0.55}          

\node[ttl, anchor=west] at (0,4.28) {(a) Morton (Z-order)};

\begin{scope}[xshift=0cm, yshift={\GY*1cm}, shift={(0.20cm,0.20cm)}, x=\CS cm, y=\CS cm]
  \foreach \cx/\cy/\r in {    0/0/0,1/0/0,0/1/0,1/1/0,2/0/0,3/0/0,2/1/0,3/1/0,0/2/0,1/2/0,0/3/0,1/3/0,2/2/0,3/2/0,2/3/0,3/3/0,4/0/0,    5/0/1,4/1/1,5/1/1,6/0/1,7/0/1,6/1/1,7/1/1,4/2/1,5/2/1,4/3/1,5/3/1,6/2/1,7/2/1,6/3/1,7/3/1,0/4/1,1/4/1,    0/5/2,1/5/2,2/4/2,3/4/2,2/5/2,3/5/2,0/6/2,1/6/2,0/7/2,1/7/2,2/6/2,3/6/2,2/7/2,3/7/2,    4/4/3,5/4/3,4/5/3,5/5/3,6/4/3,7/4/3,6/5/3,7/5/3,4/6/3,5/6/3,4/7/3,5/7/3,6/6/3,7/6/3,6/7/3,7/7/3}{    \ifcase\r \colorlet{rc}{cblue}\or\colorlet{rc}{cteal}\or\colorlet{rc}{corange}\or\colorlet{rc}{cslate}\fi
    \draw[fill=rc!22, draw=rc!55, line width=0.35pt]
      ({\cx-0.46},{\cy-0.46}) rectangle ({\cx+0.46},{\cy+0.46});}

  \draw[cink!32, line width=0.3pt, line join=round]
    (0,0)--(1,0)--(0,1)--(1,1)--(2,0)--(3,0)--(2,1)--(3,1)--(0,2)--(1,2)--(0,3)--(1,3)--(2,2)--(3,2)
    --(2,3)--(3,3)--(4,0)--(5,0)--(4,1)--(5,1)--(6,0)--(7,0)--(6,1)--(7,1)--(4,2)--(5,2)--(4,3)--(5,3)
    --(6,2)--(7,2)--(6,3)--(7,3)--(0,4)--(1,4)--(0,5)--(1,5)--(2,4)--(3,4)--(2,5)--(3,5)--(0,6)--(1,6)
    --(0,7)--(1,7)--(2,6)--(3,6)--(2,7)--(3,7)--(4,4)--(5,4)--(4,5)--(5,5)--(6,4)--(7,4)--(6,5)--(7,5)
    --(4,6)--(5,6)--(4,7)--(5,7)--(6,6)--(7,6)--(6,7)--(7,7);
  \fill[cink!50] (0,0) circle (1.0pt);

  \foreach \cx/\cy in {0/0,1/0,2/0,3/0,4/0,0/1,1/1,2/1,3/1,0/2,1/2,2/2,3/2,0/3,1/3,2/3,3/3,                       2/4,3/4,4/4,5/4,6/4,7/4}{
    \draw[cink!60, line width=0.45pt]
      ({\cx-0.20},{\cy-0.20})--({\cx+0.20},{\cy+0.20})
      ({\cx-0.20},{\cy+0.20})--({\cx+0.20},{\cy-0.20});}

  \draw[corange!85!black, line width=0.9pt, -{Stealth[length=3.6pt,width=3.2pt]}]
    (7,3) .. controls (5.0,4.02) and (2.0,4.02) .. (0.55,4);

  \draw[cteal!85!black, line width=0.9pt, dash pattern=on 2.6pt off 1.6pt, rounded corners=1pt]
    (-0.62,-0.62) rectangle (7.62,4.62);
\end{scope}

\node[ttl, anchor=west] at (\PBX,4.28) {(b) Hilbert};

\begin{scope}[xshift={\PBX*1cm}, yshift={\GY*1cm}, shift={(0.20cm,0.20cm)}, x=\CS cm, y=\CS cm]
  \foreach \cx/\cy/\r in {    0/0/0,0/1/0,1/1/0,1/0/0,2/0/0,3/0/0,3/1/0,2/1/0,2/2/0,3/2/0,3/3/0,2/3/0,1/3/0,1/2/0,0/2/0,0/3/0,0/4/0,1/4/0,    1/5/1,0/5/1,0/6/1,0/7/1,1/7/1,1/6/1,2/6/1,2/7/1,3/7/1,3/6/1,3/5/1,2/5/1,2/4/1,3/4/1,    4/4/2,5/4/2,5/5/2,4/5/2,4/6/2,4/7/2,5/7/2,5/6/2,6/6/2,6/7/2,7/7/2,7/6/2,7/5/2,6/5/2,6/4/2,7/4/2,    7/3/3,7/2/3,6/2/3,6/3/3,5/3/3,4/3/3,4/2/3,5/2/3,5/1/3,4/1/3,4/0/3,5/0/3,6/0/3,6/1/3,7/1/3,7/0/3}{    \ifcase\r \colorlet{rc}{cblue}\or\colorlet{rc}{cteal}\or\colorlet{rc}{corange}\or\colorlet{rc}{cslate}\fi
    \draw[fill=rc!22, draw=rc!55, line width=0.35pt]
      ({\cx-0.46},{\cy-0.46}) rectangle ({\cx+0.46},{\cy+0.46});}

  \draw[cink!32, line width=0.3pt, line join=round]
    (0,0)--(0,1)--(1,1)--(1,0)--(2,0)--(3,0)--(3,1)--(2,1)--(2,2)--(3,2)--(3,3)--(2,3)--(1,3)--(1,2)
    --(0,2)--(0,3)--(0,4)--(1,4)--(1,5)--(0,5)--(0,6)--(0,7)--(1,7)--(1,6)--(2,6)--(2,7)--(3,7)--(3,6)
    --(3,5)--(2,5)--(2,4)--(3,4)--(4,4)--(5,4)--(5,5)--(4,5)--(4,6)--(4,7)--(5,7)--(5,6)--(6,6)--(6,7)
    --(7,7)--(7,6)--(7,5)--(6,5)--(6,4)--(7,4)--(7,3)--(7,2)--(6,2)--(6,3)--(5,3)--(4,3)--(4,2)--(5,2)
    --(5,1)--(4,1)--(4,0)--(5,0)--(6,0)--(6,1)--(7,1)--(7,0);
  \fill[cink!50] (0,0) circle (1.0pt);

  \foreach \cx/\cy in {0/4,1/4}{
    \draw[cink!60, line width=0.45pt]
      ({\cx-0.20},{\cy-0.20})--({\cx+0.20},{\cy+0.20})
      ({\cx-0.20},{\cy+0.20})--({\cx+0.20},{\cy-0.20});}

  \draw[cteal!85!black, line width=0.9pt, dash pattern=on 2.6pt off 1.6pt, rounded corners=1pt]
    (-0.62,3.38) rectangle (3.62,7.62);
\end{scope}

\node[anchor=north] at (\PBX/2+1.6,\GY-0.20) {  \fontsize{6.2}{7.6}\selectfont\sffamily
  \begin{tabular}{@{}l@{\;}l@{\quad}l@{\;}l@{\quad}l@{\;}l@{\quad}l@{\;}l@{}}
    \sfckeybox{cblue} & r1 & \sfckeybox{cteal} & r2
      & \sfckeybox{corange} & r3 & \sfckeybox{cslate} & r4 \\
  \end{tabular}};
\node[anchor=north] at (\PBX/2+1.6,\GY-0.48) {  \fontsize{6.2}{7.6}\selectfont\sffamily
  \begin{tabular}{@{}l@{\;}l@{\quad}l@{\;}l@{\quad}l@{\;}l@{}}
    \sfckeyaabb  & AABB of rank 2
      & \sfckeycross & foreign cell
      & \sfckeyjump  & long jump \\
  \end{tabular}};

\end{tikzpicture}}

\caption{\textbf{One partitioning rule, two curves: equal atom counts either way, different shard geometry.}
An $8\times8$ grid cut into four shards of equal atom count (cell counts differ, since density does). \textbf{(a)}~Morton's power-of-two jumps attach a remote sliver to a shard (orange arrow): rank~2 fragments into two pieces whose bounding box---what the Level-1 filter tests---encloses $23$ cells it does not own. \textbf{(b)}~Hilbert's consecutive indices stay adjacent, keeping shards connected and boxes tight (mean AABB/owned-cell ratio: $1.42\times$ Morton vs.\ $1.06\times$ Hilbert). Shown in 2D for legibility (the real partition is 3D); consequences at scale in Fig.~\ref{fig:neighbor_ablation}.}
\label{fig:sfc_compare}
\end{figure}
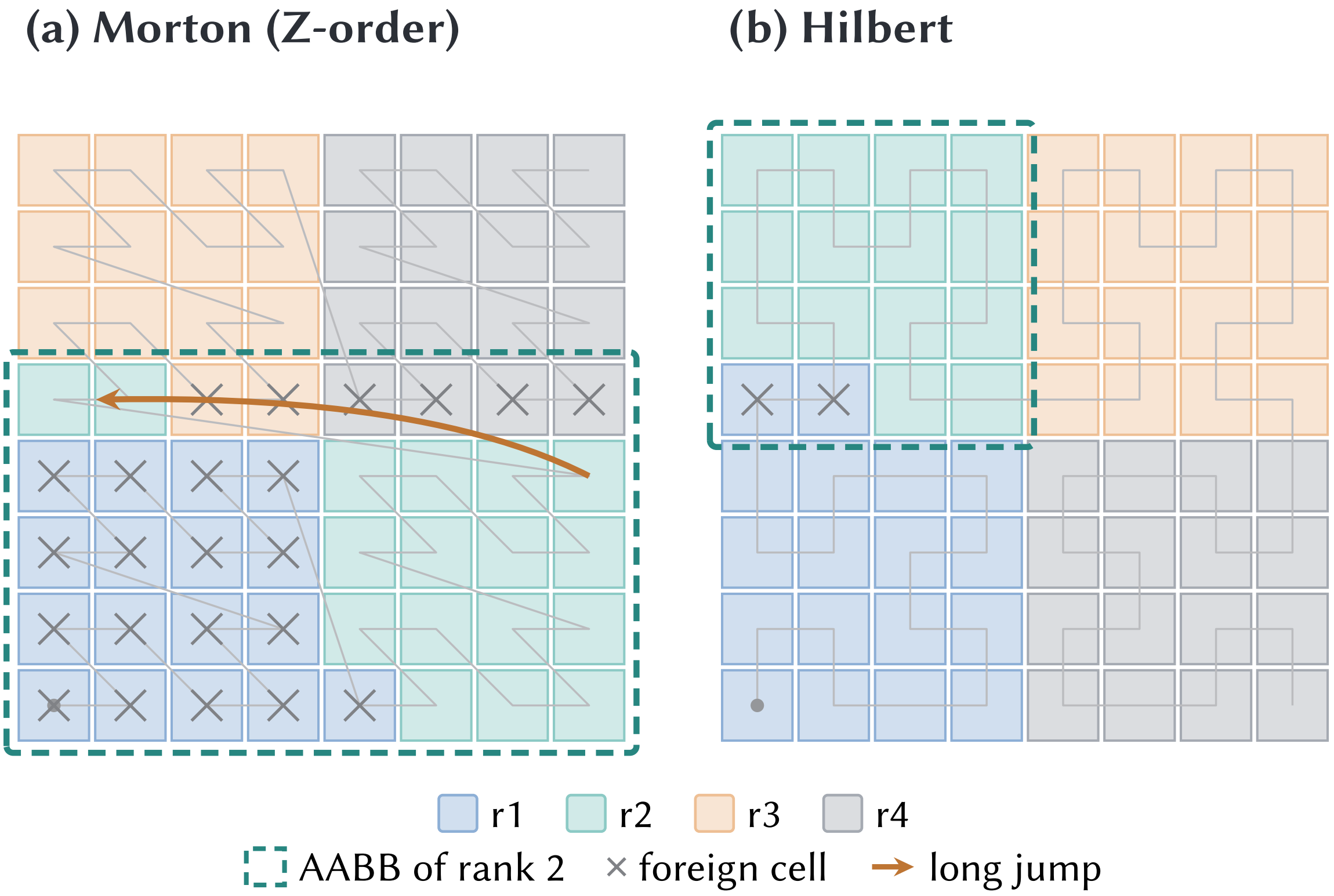

\textbf{Which curve: locality decides the exchange cost.}
Balance, determinism, and static shapes belong to the cut, not the curve, so the choice between the two we implement turns entirely on goal (ii). Morton (Z-order) interleaves the bits of the quantized $x,y,z$---cheap, but it jumps across the domain at every power-of-two boundary, so a contiguous code interval can span \emph{several disconnected} fragments. Hilbert's recursive rotation keeps consecutive indices adjacent, mapping an interval to one connected, near-cubic region. Load balance cannot see the difference; the communication plan can. Figure~\ref{fig:sfc_compare} shows why: the Level-1 filter summarizes each rank by a bounding box, and a fragmented Morton shard's box also covers the gaps between its fragments---enclosing $23$ of $64$ cells it does not own---so the rank requests from far more peers than geometry requires, and the count keeps climbing with the machine where Hilbert's stays sublinear: over $64$--$512$ GPUs Morton rises from $5$ to $27$ and Hilbert from $6$ to $19$ at $25$k atoms/GPU, both measured (Fig.~\ref{fig:neighbor_ablation}c). Neither series saturates, and the slopes differ. Neighbor-construction latency then grows $2.8$--$2.9\times$ under Morton from 64 to 512 GPUs at $25$k atoms/GPU, against $1.6$--$1.7\times$ under Hilbert (Section~\ref{sec:neighbor-scaling}).

We therefore adopt \textbf{Hilbert equal-count partitioning as the default}, keeping Morton as the ablation baseline; its higher encoding cost is paid once per repartition, off the per-step path. One caveat: our two implementations normalize the quantization axes differently, so on strongly anisotropic cells part of the gap reflects resolution, not curve topology---though it persists at 64 GPUs, where a cubic cell makes both coincide (Appendix~\ref{sec:app-partition}).

\textbf{Maintaining the partition under MD dynamics.}
Since positions evolve, a partition fixed at $t=0$ would lose both balance and locality, so we re-establish it every $100$ steps. This re-coding is the only event that changes \emph{which atoms a rank owns}: counts stay fixed, so node shapes never move, only the identities behind them. It also invalidates the neighbor graph built against the old ownership, whose edge shapes turn over on the faster clock of Section~\ref{sec:multilevel_search}. The two cadences compose rather than compete: a repartition is a rebuild that also reassigns atoms, so its cost is the partition-and-broadcast plus one graph re-capture, measured at $407$--$418$\,ms in Section~\ref{sec:experiments} and amortized over $100$ steps. It is also the one event that invalidates the topology cache outright, since the cached edge sets, chunk assignments, and plans are all indexed by ownership; Section~\ref{sec:multilevel_search} states the invariant, and Algorithm~\ref{alg:md-step} makes the invalidation unconditional rather than inferred.

\subsection{Multi-Level Neighbor Search}
\label{sec:multilevel_search}

\definecolor{cblue}{HTML}{2F6DB5}
\definecolor{cteal}{HTML}{2E9E96}
\definecolor{corange}{HTML}{E08A3C}
\definecolor{cslate}{HTML}{5B6573}
\definecolor{cink}{HTML}{2B2F36}
\definecolor{cgreen}{HTML}{6DA34D}
\definecolor{cplum}{HTML}{8E6BB0}

\providecommand{\scard}[3]{  \begin{scope}
    \clip[rounded corners=5pt] (0,0) rectangle (\SW,\SH);
    \fill[white] (0,0) rectangle (\SW,\SH);
    \fill[#1!10] (0,\SH-\HB) rectangle (\SW,\SH);
  \end{scope}
  \draw[rounded corners=5pt, black!18, line width=0.55pt] (0,0) rectangle (\SW,\SH);
  \draw[#1!32, line width=0.5pt] (0,\SH-\HB) -- (\SW,\SH-\HB);
  \node[circle, fill=#1!80!black, text=white, font=\sffamily\bfseries\footnotesize, inner sep=1.5pt] at (0.4,\SH-\HB/2) {#2};
  \node[anchor=west, font=\sffamily\bfseries\footnotesize, text=cink] at (0.72,\SH-\HB/2) {#3};
}

\begin{figure*}[t]
\centering
\resizebox{\textwidth}{!}{\begin{tikzpicture}[
  font=\sffamily,
  every node/.style={inner sep=1.2pt},
  glab/.style={font=\sffamily\fontsize{6.8}{7.6}\selectfont, text=cslate, align=center},
  chip/.style={rounded corners=2.5pt, font=\sffamily\fontsize{8.4}{9.6}\selectfont,
               inner sep=2.5pt},
  flow/.style={-{Stealth[length=6pt,width=5pt]}, line width=1pt, cslate!80},
  x=1cm,y=1cm,
]
\def\OW{4.7}\def\OH{3.16}
\def\Oax{0.2}\pgfmathsetmacro{\Obx}{0.2+\OW+1.05}\pgfmathsetmacro{\Ocx}{0.2+2*(\OW+1.05)}
\pgfmathsetmacro{\fullW}{\Ocx+\OW}
\pgfmathsetmacro{\Ovmid}{\OH/2}
\pgfmathsetmacro{\OabA}{(\Oax+\OW+\Obx)/2}\pgfmathsetmacro{\OabB}{(\Obx+\OW+\Ocx)/2}
\def\SD{-1.12}       \def\pyC{0.34}       

\draw[flow] (\Oax+\OW+0.06,\Ovmid) -- (\Obx-0.06,\Ovmid);
\draw[flow] (\Obx+\OW+0.06,\Ovmid) -- (\Ocx-0.06,\Ovmid);
\node[glab, text=cslate!85, text width=1.08cm] at (\OabA,\Ovmid+0.42) {sparse\\peer set};
\node[glab, text=cslate!85, text width=1.08cm] at (\OabB,\Ovmid+0.42) {per-chunk\\candidates};

\begin{scope}[shift={(\Oax,0)}]
  \def\SW{4.7}\def\SH{3.16}\def\HB{0.42}
  \scard{cblue}{a}{Inter-device AABB filtering}
  \begin{scope}[yshift={\SD*1cm}]
    \def\dx{0.4}\def\dxx{2.05}\def\dy{1.95}\def\dyy{3.85}
    \pgfmathsetmacro{\dmx}{(\dx+\dxx)/2}\pgfmathsetmacro{\dmy}{(\dy+\dyy)/2}
    \fill[cblue!7]  (\dx,\dmy) rectangle (\dmx,\dyy);
    \fill[cteal!7]  (\dmx,\dmy) rectangle (\dxx,\dyy);
    \fill[cgreen!7] (\dx,\dy) rectangle (\dmx,\dmy);
    \fill[corange!7](\dmx,\dy) rectangle (\dxx,\dmy);
    \draw[black!15,line width=0.4pt] (\dx,\dy) rectangle (\dxx,\dyy);
    \draw[black!12,line width=0.3pt] (\dmx,\dy)--(\dmx,\dyy) (\dx,\dmy)--(\dxx,\dmy);
    \node[glab,text=black!35] at (\dx+0.14,\dyy-0.13){D0};
    \node[glab,text=black!35] at (\dxx-0.14,\dyy-0.13){D1};
    \node[glab,text=black!35] at (\dx+0.14,\dy+0.13){D2};
    \node[glab,text=black!35] at (\dxx-0.14,\dy+0.13){D3};
    \foreach \p in {(0.62,3.2),(0.8,3.32),(0.7,3.46)}{\fill[cslate!75] \p circle (1.7pt);}
    \draw[cblue!75,dash pattern=on 2pt off 1.6pt,line width=0.65pt,rounded corners=1.5pt] (0.55,3.1) rectangle (0.92,3.56);
    \foreach \p in {(1.5,3.22),(1.7,3.32),(1.58,3.46)}{\fill[cslate!75] \p circle (1.7pt);}
    \draw[cteal!75,dash pattern=on 2pt off 1.6pt,line width=0.65pt,rounded corners=1.5pt] (1.42,3.12) rectangle (1.82,3.56);
    \foreach \p in {(0.62,2.3),(0.82,2.4),(0.7,2.55)}{\fill[cslate!75] \p circle (1.7pt);}
    \draw[cgreen!75,dash pattern=on 2pt off 1.6pt,line width=0.65pt,rounded corners=1.5pt] (0.55,2.2) rectangle (0.92,2.64);
    \foreach \p in {(1.5,2.32),(1.7,2.42),(1.58,2.56),(1.74,2.5)}{\fill[cslate!75] \p circle (1.7pt);}
    \draw[corange!75,dash pattern=on 2pt off 1.6pt,line width=0.65pt,rounded corners=1.5pt] (1.42,2.22) rectangle (1.84,2.66);
    \draw[cblue!40,line width=0.9pt,line join=round] (0.73,3.32)--(1.6,3.32)--(0.72,2.42)--(1.62,2.42);
    \def\gx{2.65}\def\gy{2.05}\def\gw{1.55}\def\gh{1.55}
    \foreach \i in {0,...,4}\foreach \j in {0,...,4}{
       \pgfmathtruncatemacro{\act}{ifthenelse(abs(\i-\j)<2,1,0)}
       \ifnum\act=1 \fill[cblue!34] ({\gx+\gw*\j/5},{\gy+\gh*(4-\i)/5}) rectangle ({\gx+\gw*(\j+1)/5},{\gy+\gh*(4-\i+1)/5});\fi}
    \foreach \i in {0,...,5}{\draw[black!16,line width=0.3pt] (\gx,{\gy+\gh*\i/5})--(\gx+\gw,{\gy+\gh*\i/5});
       \draw[black!16,line width=0.3pt] ({\gx+\gw*\i/5},\gy)--({\gx+\gw*\i/5},\gy+\gh);}
    \draw[cblue!70,line width=0.9pt] (\gx,\gy) rectangle (\gx+\gw,\gy+\gh);
    \node[glab,rotate=90] at (\gx-0.22,{\gy+\gh/2}) {device $i$};
    \node[glab] at ({\gx+\gw/2},\gy+\gh+0.18) {device $j$};
  \end{scope}
  \node[chip, fill=cblue!12, draw=cblue!45, text=cblue!60!black] at (\SW/2,\pyC) {\ifdefined\ifarxiv SFC-sparse active peers\else SFC-bounded active peers\fi};
\end{scope}

\begin{scope}[shift={(\Obx,0)}]
  \def\SW{4.7}\def\SH{3.16}\def\HB{0.42}
  \scard{cteal}{b}{Recursive query chunking}
  \begin{scope}[yshift={\SD*1cm}]
    \def\sx{0.4}\def\sxx{4.3}\def\sy{1.80}\def\syy{3.85}
    \pgfmathsetmacro{\smx}{(\sx+\sxx)/2}\pgfmathsetmacro{\smy}{(\sy+\syy)/2}
    \fill[cblue!6]  (\sx,\smy) rectangle (\smx,\syy);
    \fill[cteal!6]  (\smx,\smy) rectangle (\sxx,\syy);
    \fill[cgreen!6] (\sx,\sy) rectangle (\smx,\smy);
    \fill[corange!6](\smx,\sy) rectangle (\sxx,\smy);
    \draw[black!15,line width=0.4pt] (\sx,\sy) rectangle (\sxx,\syy);
    \draw[cslate!50,dash pattern=on 2.5pt off 2pt,line width=0.65pt] (2.3,1.88)--(2.3,3.72);
    \draw[cslate!45,dash pattern=on 2pt off 1.8pt,line width=0.55pt] (0.55,2.66)--(2.3,2.66);
    \draw[cslate!45,dash pattern=on 2pt off 1.8pt,line width=0.55pt] (2.3,2.72)--(4.12,2.72);
    \foreach \p in {(1.05,1.98),(1.3,2.06),(1.15,2.26),(1.38,2.28)}{\fill[cblue!82] \p circle (2.2pt);}
    \draw[cblue!75,dash pattern=on 2.5pt off 2pt,line width=0.75pt,rounded corners=2pt] (0.92,1.86) rectangle (1.5,2.4);
    \node[glab,text=cblue!60!black] at (1.75,1.9) {$Q_1$};
    \foreach \p in {(1.1,2.95),(1.35,3.05),(1.2,3.25),(1.42,3.28)}{\fill[cteal!82] \p circle (2.2pt);}
    \draw[cteal!75,dash pattern=on 2.5pt off 2pt,line width=0.75pt,rounded corners=2pt] (0.96,2.82) rectangle (1.55,3.4);
    \node[glab,text=cteal!55!black] at (1.24,3.55) {$Q_2$};
    \foreach \p in {(2.75,1.98),(3.0,2.08),(2.82,2.28),(3.05,2.31),(2.9,2.43)}{\fill[cgreen!82] \p circle (2.2pt);}
    \draw[cgreen!75,dash pattern=on 2.5pt off 2pt,line width=0.75pt,rounded corners=2pt] (2.6,1.86) rectangle (3.2,2.55);
    \node[glab,text=cgreen!45!black] at (3.45,1.94) {$Q_3$};
    \foreach \p in {(2.82,3.0),(3.06,3.1),(2.9,3.3),(3.12,3.32)}{\fill[cplum!82] \p circle (2.2pt);}
    \draw[cplum!75,dash pattern=on 2.5pt off 2pt,line width=0.75pt,rounded corners=2pt] (2.68,2.86) rectangle (3.26,3.44);
    \node[glab,text=cplum!55!black] at (2.97,3.55) {$Q_4$};
    \foreach \px/\py in {4.0/3.55,4.05/2.04,0.62/3.55}{
       \fill[corange!85] (\px,\py) circle (2.5pt);
       \draw[corange!75,dash pattern=on 2pt off 1.8pt,line width=0.65pt] ({\px-0.18},{\py-0.18}) rectangle ({\px+0.18},{\py+0.18});}
  \end{scope}
  \node[chip, fill=cteal!12, draw=cteal!45, text=cteal!55!black] at (\SW/2,\pyC) {tight per-chunk AABB: $C \!\to\! C_i$};
\end{scope}

\begin{scope}[shift={(\Ocx,0)}]
  \def\SW{4.7}\def\SH{3.16}\def\HB{0.42}
  \scard{corange}{c}{Chunk-local range search}
  \begin{scope}[yshift={\SD*1cm}]
    \def\ix{0.4}\def\ixx{1.95}\def\iy{1.95}\def\iyy{3.8}
    \draw[black!14,line width=0.4pt,rounded corners=2.5pt] (\ix,\iy) rectangle (\ixx,\iyy);
    \node[glab,text=cslate!70] at ({(\ix+\ixx)/2},\iyy-0.16) {exact $\le r_c$};
    \coordinate (q) at (1.1,2.78);
    \draw[corange!60,dotted,line width=0.85pt] (q) circle (0.5);
    \node[glab,text=corange!55!black] at (1.45,2.42) {$r_c$};
    \foreach \p in {(0.82,2.92),(1.3,2.98),(0.95,2.55),(1.3,2.58)}{
       \draw[corange!55,line width=0.5pt] (q)--\p; \fill[black!30] \p circle (1.9pt);}
    \foreach \p in {(0.6,3.42),(1.62,2.28)}{\fill[black!12] \p circle (1.9pt);}
    \fill[cslate!80] (q) circle (2.4pt);
    \def\gx{2.55}\def\gy{1.95}\def\gw{1.6}\def\gh{1.7}
    \foreach \i/\c in {0/cblue,1/cteal,2/cgreen,3/cplum}{
       \fill[\c!26] ({\gx+\gw*\i/4},{\gy+\gh*(3-\i)/4}) rectangle ({\gx+\gw*(\i+1)/4},{\gy+\gh*(3-\i+1)/4});}
    \foreach \i in {0,...,4}{\draw[black!15,line width=0.3pt] (\gx,{\gy+\gh*\i/4})--(\gx+\gw,{\gy+\gh*\i/4});
       \draw[black!15,line width=0.3pt] ({\gx+\gw*\i/4},\gy)--({\gx+\gw*\i/4},\gy+\gh);}
    \draw[corange!70,line width=0.9pt] (\gx,\gy) rectangle (\gx+\gw,\gy+\gh);
    \node[glab,rotate=90] at (\gx-0.22,{\gy+\gh/2}) {chunk $Q_i$};
    \node[glab] at ({\gx+\gw/2},\gy+\gh+0.18) {candidates $C_i$};
  \end{scope}
  \node[chip, fill=corange!12, draw=corange!45, text=corange!60!black] at (\SW/2,\pyC) {exact, at \ifdefined\ifarxiv $\textstyle\sum_i Q_iC_i \ll Q_{\mathrm{loc}}C_{\mathrm{cand}}$\else $\textstyle\sum_i Q_iC_i \ll QC$\fi};
\end{scope}

\end{tikzpicture}}
\caption{\textbf{Multi-level neighbor search.} \textbf{(a)}~Per-device AABBs and a $K\!\times\!K$ overlap test bound each device's active peers---the sparse peer set the communication layer reuses. \textbf{(b)}~Recursive bisection into compact chunks $Q_i$, each screened by its own tight box, cuts candidates from $C$ to $C_i\!\ll\!C$ (detail: Fig.~\ref{fig:multilevel_search_detail}). \textbf{(c)}~The exact cutoff search then scores only block-diagonal $Q_i$--$C_i$ pairs, at $\sum_i Q_iC_i$ rather than \ifdefined\ifarxiv $Q_{\mathrm{loc}}C_{\mathrm{cand}}$\else $QC$\fi.}
\label{fig:multilevel_search}
\end{figure*}
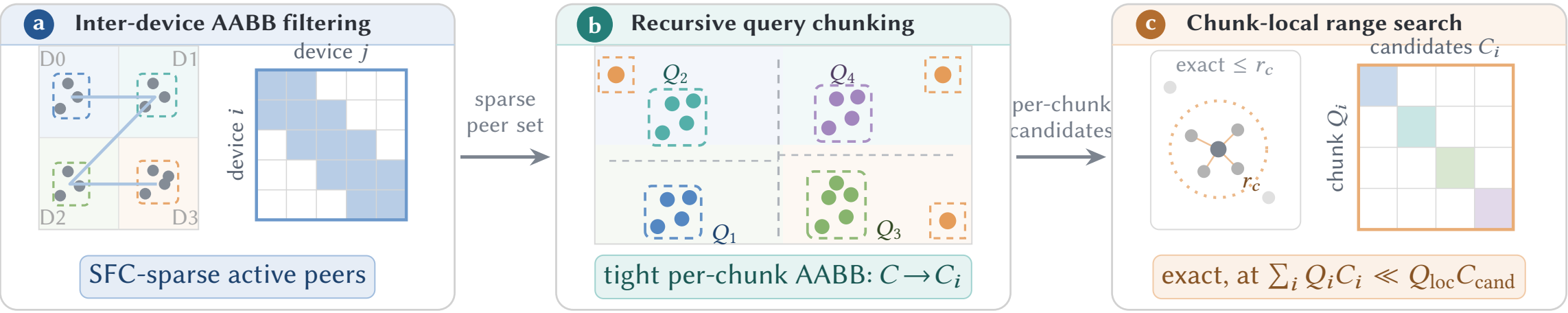

Neighbor construction is the stage that scales worst: each device owns a subset of atoms but must find their neighbors among nearby remote regions. Figure~\ref{fig:multilevel_search} shows our answer---three filters pruning the candidate set progressively, across devices, then within a device, then exactly. Exactness survives the pruning: every filter is \emph{conservative}, discarding only atoms provably beyond the cutoff, and Level~3 is an \emph{exact} brute-force cutoff-radius search via FAISS~\cite{johnson2019billion-scale}, not an approximate nearest-neighbor query. Nor do we impose a fixed-$k$ truncation of our own: where the backend's $k$-selection is unsaturated (Section~\ref{sec:discussion}), the edge set is the complete radius graph, at a fraction of the cost of an all-pairs search.

At Level~1, each of the $K$ devices summarizes its atoms by a coarse axis-aligned bounding box (AABB), and a $K\times K$ test marks $(i,j)$ active whenever atoms on device $j$ \emph{may} fall within the cutoff of device~$i$'s queries. Nominally $O(K^2)$, it runs on coarse boxes from one small metadata all-gather, and this is where the partition pays off: SFC locality confines each device's neighbors to a set of adjacent subdomains that grows sublinearly, keeping every row sparse as $K$ grows. Candidates then move as a selective-subgroup variable-length All-to-All, whose sparse peer set the communication plan of Section~\ref{sec:reduce_volume} reuses. Under periodic boundaries a shard wraps the cell, so each device publishes one AABB per occupied image: at most $27$ including the identity image, with the corresponding counts, $8+27\times8$ doubles and under $2$\,KiB per rank. Because that descriptor is small and fixed in size, one metadata all-gather lets every rank reconstruct the complete rank$\times$rank$\times$image request matrix locally, with no second exchange of request masks.

\textbf{Level-1 granularity.} The three realizations of that filter differ in what they publish, and the comparison is instructive because the scheme that prunes hardest loses. \textbf{All-gather} performs no geometric filtering: every device broadcasts its full PBC-expanded candidate set to every other device. It builds no plan and issues one regular collective, which makes it the correctness/performance baseline and competitive at small $K$, but each device's candidate population then grows with the \emph{global} atom count---the memory and scalability wall the design exists to remove. \textbf{Rank-box} (ours) publishes one query AABB per device plus up to $27$ periodic-image candidate AABBs per source: fixed small metadata, no request-mask exchange. \textbf{Spatial-bin} replaces the per-device boxes with a regular grid sharing a global origin and publishes the AABB and count of every occupied query and candidate bin, capped at $8192$ bins per side and auto-coarsened if the cap is hit; surviving bins double as query chunks for Level~3. Spatial-bin therefore prunes the candidate side hardest---about $0.5$M entries per rank, essentially independent of $K$, against $1.9$M for Hilbert rank-box at 512 GPUs---and is nonetheless slower at every scale we measured, because its metadata is $\approx1$\,MiB per rank rather than under $2$\,KiB, it needs a second all-gather to distribute the request mask, and it hands Level~3 roughly four times as many query groups to schedule.

Two further readings support attributing the gap to the scheme rather than to the partition it runs on. Spatial-bin's latency varies by under $5\%$ between the Hilbert and Morton curves, where rank-box's varies by $2.0\times$: replacing per-device boxes with a shared grid does make shard geometry irrelevant, exactly as designed, and makes the scheme pay for the grid instead. And where the per-rank box is genuinely inflated the grid does help: on Morton, spatial-bin brings the peer degree down to $19$, matching Hilbert rank-box, and at $50$k atoms/GPU it is $\approx\!10\%$ faster than Morton rank-box. That is the only configuration in which it wins: at $25$k atoms/GPU on Morton the two are level, and against Hilbert rank-box it is behind throughout: the latency the smaller box buys back never covers the cost of the plan that produced it. A filter's cost includes publishing what it filtered with. Because the sub-stage timers are MAX-reduced independently (Section~\ref{sec:app-protocol}), we report this as consistent with the telemetry rather than as a measured stage decomposition.

Boundary atoms and outliers still inflate a device's AABB and the candidates it retains, so Level~2 bisects local queries at the median of the longest axis into compact chunks and screens each against its own dilated box (Fig.~\ref{fig:multilevel_search_detail})---conservative, since any atom within the cutoff of a chunk's query lies inside that box. The exact search then runs per chunk over only its survivors, as one batched GPU call. With $Q_{\mathrm{loc}}$ local queries and $C_{\mathrm{cand}}$ candidates after Level~1, an unchunked search costs $\propto Q_{\mathrm{loc}}C_{\mathrm{cand}}$ against $\sum_i Q_iC_i$ chunked, where $Q_i$ and $C_i$ are the per-chunk counts; since $\sum_i Q_i=Q_{\mathrm{loc}}$, the gain is entirely $C_i\ll C_{\mathrm{cand}}$, not fewer queries (Appendix~\ref{sec:app-neighbor}). The chunking itself adds an $O(Q\log Q)$ median bisection and one AABB test per chunk, both small against the pairwise work they remove.

\input{figures/multilevel_search_l2detail_tikz}

\textbf{Cache exactness.}
A step's graph splits into a \emph{topology}---edge sets, periodic-image mapping, and the exchange/\allowbreak communication plans, all indexed by current ownership---and a \emph{geometry}, the positions and displacements that feed it. Only geometry must change every step. Over one cache epoch of ten steps ($5$\,fs at our timestep) a $300$\,K hydrogen moves $\approx\!0.1$\,\AA{} rms, inside the $\delta/2=0.25$\,\AA{} trigger that the $0.5$\,\AA{} skin allows, so the topology survives. Over the hundred steps between re-partitions it moves $\approx\!1$\,\AA{} rms, far inside the $\approx\!67$\,\AA{} subdomain at $\sim\!30$k atoms/rank ($\rho=0.1$\,atoms\,\AA$^{-3}$), so ownership survives too.

Reuse is only legitimate if it is exact, so the scheme is stated as an invariant. Two search radii are in play, and which one runs depends on whether the cache is enabled. With the cache off---the interval-$1$ configuration used as the reference throughout---every filter runs at $r_c$ and each step's edge set is built from scratch. With the cache on, the rebuild instead constructs an untruncated candidate \emph{superset} at $r_c+\delta$ ($\delta=0.5$\,\AA{} by default), and every intervening hit step recomputes minimum-image geometry over that superset and activates only pairs with $r\le r_c$. Activation is a mask over the fixed superset: we never compact the edge list within an epoch, so edge-indexed tensors keep the superset's extent. The two therefore produce the same active edge set, not merely a similar one: the interval-$1$ and interval-$10$ runs agree on the active edge list bit for bit, while their cached supersets and plans differ in length by construction. The model carries two cutoffs, $5$\,\AA{} for the three short-range layers and $8$\,\AA{} for the long-range one, and each is searched in its own stage under the same rule: the stage for cutoff $r_c$ builds its superset at $r_c+\delta$ and activates $r\le r_c$ against it.

The $\delta$ is what the displacement trigger is sized against. A rebuild fires once any rank's own maximum displacement since the last build exceeds $\delta/2$, so two atoms moving directly toward each other close their separation by less than $\delta$ between builds, and no pair beyond $r_c+\delta$ at build time can enter $r_c$ unnoticed---one trigger therefore covers both stages, since the smaller $\delta/2$ margin, not the cutoff, bounds the crossing. The decision must be unanimous: every rank agrees via \texttt{all\allowbreak\_reduce(MAX)} over local triggers, so all ranks re-enter the same cache epoch together and the exchange plans stay mutually consistent. On a hit we re-expand positions under PBC, recompute edge geometry, and re-apply the $r\le r_c$ test over the superset, reusing its indices and plan; the indices are detached from autograd while the recomputed geometry stays in it, so forces are not taken through stale positions. A repartition invalidates the cache outright, since ownership, chunk assignment, and the plan all change with it (Algorithm~\ref{alg:md-step}). Left as a separate caveat is the $k$-selection clamp of Section~\ref{sec:discussion}: the skin bounds which pairs the candidate set must contain, not how many neighbors the backend will return for a single query. This is reuse, not approximation---the skin is sized so no pair crosses undetected between rebuilds---and interval-$1$ and interval-$10$ runs give bitwise-identical topology hashes and forces agreeing to $7.2\times10^{-7}$\,eV/\AA.

Even so, every step must still exchange remote Key, source-embedding, and Value features, and at scale this---not floating-point throughput---is the dominant obstacle.

\subsection{Communication Optimization}
\label{sec:reduce_volume}

A remote exchange rarely costs just the payload's bandwidth: per-call handshakes, full-width equivariant transfers, separate launches, and on-path allocation can dominate instead, and irregularly. Three mechanisms strip that overhead; two more then hide what remains.

\textbf{Removing What Is Not Payload.}
\emph{Plan reuse} (Fig.~\ref{fig:all2all_plan}): blocks sharing a cutoff share a peer set, so the routing plan---peers, send/receive lengths, offsets, index maps, buffer addresses---is built once per cutoff and reused by every later \texttt{all2allv} there. With $R_r$ blocks at cutoff $r$, building the plan once turns $R_r$ handshake-plus-payload exchanges into one handshake and $R_r$ payload-only transfers:
\begin{equation}
R_r\times(\text{handshake}+\text{payload})\;\longrightarrow\;\text{handshake}+R_r\times\text{payload},
\end{equation}
cutting per-block handshakes to one per cutoff---two per step, at $5$ and $8$\,\AA{}, rather than four---and the saving grows with the number of blocks sharing a cutoff. It is realized again on every cache-hit step, since each plan is indexed by topology rather than by step number.

\input{figures/comm_plan_tikz}

\emph{Sender-side projection} (Fig.~\ref{fig:reduce_comm}): of the three per-atom quantities attention consumes, only the Value is aggregated as an equivariant feature and must travel at full width $(l_{\max}+1)^2C$; the source embedding is an invariant scalar of width $C$, and the Key is used solely inside scalar query--key dot products, so it need not be shipped equivariant. Projecting the Key on the sender to $H$ per-head vectors of width $d_k$ ($Hd_k$ scalars) leaves the per-atom payload at $C+Hd_k+(l_{\max}+1)^2C$, against $C+2(l_{\max}+1)^2C$ if the Key were sent in its equivariant form. With the deployed $l_{\max}=2$, $C=256$, $H=128$, $d_k=8$ (Appendix~\ref{sec:app-config}), we have $(l_{\max}+1)^2C=2{,}304$ and $Hd_k=1{,}024$, so the payload falls from $4{,}864$ to $3{,}584$ scalars per atom---$\times4$ bytes in TF32/FP32, i.e.\ $19{,}456\to14{,}336$ bytes---a $26.3\%$ reduction realized entirely on the Key. The saving grows with model order: at fixed $C$ the equivariant Key width scales as $(l_{\max}+1)^2$ while its projected width $Hd_k$ is constant, so higher-$l_{\max}$ models---where communication is most acute---benefit more.

\input{figures/skv_fusion_tikz}

\emph{Fusion and pooling}: the three tensors then travel as one physical \texttt{all2allv} from a pre-registered buffer (Figs.~\ref{fig:reduce_comm}--\ref{fig:mempool}), so launches per exchange fall from three to one, and the received SKV data is consumed directly from the fused buffer with the operators' autograd dependencies rewritten so the backward pass traverses the fused path in a well-defined way. Send and receive buffers are carved from a pool whose Memory-Region registration happens in advance during plan construction, so steady-state transfers repeat neither buffer creation nor dynamic registration. What remains exposed is close to the payload itself, which can only be hidden.

\definecolor{cblue}{HTML}{2F6DB5}
\definecolor{cteal}{HTML}{2E9E96}
\definecolor{corange}{HTML}{E08A3C}
\definecolor{cslate}{HTML}{5B6573}
\definecolor{cink}{HTML}{2B2F36}
\definecolor{cgreen}{HTML}{6DA34D}

\newcommand{\mpcard}[3]{  \begin{scope}
    \clip[rounded corners=5pt] (0,0) rectangle (\SW,\SH);
    \fill[white] (0,0) rectangle (\SW,\SH);
    \fill[#1!9] (0,\SH-\HB) rectangle (\SW,\SH);
  \end{scope}
  \draw[rounded corners=5pt, black!18, line width=0.6pt] (0,0) rectangle (\SW,\SH);
  \draw[#1!35, line width=0.5pt] (0,\SH-\HB) -- (\SW,\SH-\HB);
  \node[anchor=west, font=\sffamily\bfseries\normalsize, text=#1!85!black] at (0.22,\SH-\HB/2) {#2};
  \node[anchor=west, font=\sffamily\bfseries\small, text=cink] at (0.62,\SH-\HB/2) {#3};
}
\newcommand{\mpseg}[6]{  \draw[fill=#4, draw=black!35, line width=0.3pt, rounded corners=1pt] (#1,{#3-0.15}) rectangle (#2,{#3+0.15});
  \node[font=\sffamily\fontsize{5.8}{6.4}\selectfont, text=#6] at ({(#1+#2)/2},#3) {#5};
}

\begin{figure*}[t]
\centering
\resizebox{\textwidth}{!}{\begin{tikzpicture}[
  font=\sffamily,
  every node/.style={inner sep=1.2pt},
  slab/.style={font=\sffamily\fontsize{7.8}{9}\selectfont, text=cslate, align=center},
  ssub/.style={font=\sffamily\fontsize{7}{8}\selectfont, text=cslate, align=center},
  glab/.style={font=\sffamily\fontsize{6.6}{7.4}\selectfont, text=cslate, align=center},
  chip/.style={rounded corners=2.5pt, font=\sffamily\fontsize{8}{9.5}\selectfont, inner sep=3.5pt},
  flow/.style={-{Stealth[length=6pt,width=5pt]}, line width=1pt, cslate!80},
  x=1cm,y=1cm,
]
\def\SW{5.0}\def\SH{5.85}\def\HB{0.6}
\def\Ax{0.2}\pgfmathsetmacro{\Bx}{\Ax+\SW+0.5}\pgfmathsetmacro{\Cx}{\Ax+2*(\SW+0.5)}
\pgfmathsetmacro{\fullW}{\Cx+\SW}\pgfmathsetmacro{\mid}{\SH/2}

\node[anchor=base, font=\sffamily\bfseries\large, text=cink] at (\fullW/2,\SH+0.6)
   {Pooled buffer management with one-time registration};
\draw[black!15, line width=0.5pt] (0,\SH+0.36) -- (\fullW,\SH+0.36);

\draw[flow] (\Ax+\SW+0.06,\mid) -- (\Bx-0.06,\mid);
\draw[flow] (\Bx+\SW+0.06,\mid) -- (\Cx-0.06,\mid);

\begin{scope}[shift={(\Ax,0)}]
  \mpcard{corange}{a}{Naive: per-round setup}
  \foreach \r/\yy in {1/4.55,2/3.95,3/3.35}{
     \node[ssub, anchor=east] at (0.78,\yy) {R\r};
     \mpseg{0.9}{1.5}{\yy}{corange!75}{alloc}{white}
     \mpseg{1.5}{2.3}{\yy}{corange!55}{MR reg}{white}
     \mpseg{2.3}{3.55}{\yy}{cblue!28}{transfer}{cblue!55!black}
     \mpseg{3.55}{4.1}{\yy}{cslate!30}{free}{cslate!70!black}}
  \node[ssub] at (\SW/2,2.78) {$\vdots$\ \ repeated every round};
  \foreach \fx in {0.95,1.35,1.9,2.45,3.0}{\fill[corange!22] (\fx,2.16) rectangle (\fx+0.22,2.36);}
  \node[glab, text=corange!60!black] at (\SW/2,1.95) {short-lived buffers $\Rightarrow$ fragmentation};
  \node[slab, text width=4.6cm] at (\SW/2,1.28) {alloc \& MR registration\\on the critical path};
  \node[chip, fill=corange!12, draw=corange!45, text=corange!60!black] at (\SW/2,0.66) {repeated setup, fragmented};
\end{scope}

\begin{scope}[shift={(\Bx,0)}]
  \mpcard{cslate}{b}{One-time pool initialization}
  \node[circle, fill=cslate!70, text=white, font=\sffamily\bfseries\fontsize{6.6}{7.4}\selectfont, inner sep=1.3pt] at (0.5,4.62) {1};
  \node[ssub, anchor=west] at (0.78,4.62) {allocate contiguous pool (once)};
  \foreach \i in {0,...,5}{\fill[cteal!22] ({0.7+0.68*\i},3.95) rectangle ({0.7+0.68*\i+0.6},4.32);}
  \draw[cteal!55, line width=0.7pt, rounded corners=1.5pt] (0.7,3.95) rectangle (4.78,4.32);
  \node[glab] at (\SW/2,3.74) {one contiguous device-memory pool};
  \draw[flow, line width=0.7pt] (\SW/2,3.6) -- (\SW/2,3.3);
  \node[circle, fill=cslate!70, text=white, font=\sffamily\bfseries\fontsize{6.6}{7.4}\selectfont, inner sep=1.3pt] at (0.5,3.06) {2};
  \node[ssub, anchor=west] at (0.78,3.06) {register pinned MR (once)};
  \draw[rounded corners=3pt, cteal!50, fill=cteal!10, line width=0.7pt] (0.7,1.98) rectangle (4.78,2.74);
  \node[ssub, text=cteal!50!black, align=center] at (\SW/2,2.36) {pre-registered pinned MR\\$+$ symmetric memory};
  \node[slab, text width=4.5cm] at (\SW/2,1.32) {done outside the steady-state path};
  \node[chip, fill=cteal!12, draw=cteal!45, text=cteal!55!black] at (\SW/2,0.66) {pooled, pre-registered, reusable};
\end{scope}

\begin{scope}[shift={(\Cx,0)}]
  \mpcard{cblue}{c}{Fast-path reuse}
  \draw[rounded corners=2pt, cteal!50, fill=cteal!12, line width=0.6pt] (0.85,4.74) rectangle (4.15,5.04);
  \node[glab, text=cteal!50!black] at (2.5,4.89) {pre-registered pool};
  \foreach \r/\yy in {1/4.2,2/3.62,3/3.04}{
     \node[ssub, anchor=east] at (0.78,\yy) {R\r};
     \mpseg{0.9}{1.55}{\yy}{cteal!32}{borrow}{cteal!50!black}
     \mpseg{1.55}{2.95}{\yy}{cblue!28}{transfer}{cblue!55!black}
     \mpseg{2.95}{3.6}{\yy}{cslate!22}{return}{cslate!70!black}
     \draw[cteal!45, line width=0.4pt, -{Stealth[length=2pt]}] (1.2,4.69) -- (1.2,\yy+0.16);}
  \node[ssub] at (\SW/2,2.5) {$\vdots$\ \ no alloc / registration};
  \node[slab, text width=4.5cm] at (\SW/2,1.32) {borrow and return buffers from the pool};
  \node[chip, fill=cblue!12, draw=cblue!45, text=cblue!60!black] at (\SW/2,0.66) {no registration latency};
\end{scope}

\draw[black!15, line width=0.5pt] (\Ax,-0.25) -- (\fullW,-0.25);
\def\ly{-0.74}
\draw[fill=corange!72, draw=corange!90!black, rounded corners=1pt] (\Ax+0.12,\ly-0.12) rectangle (\Ax+0.42,\ly+0.12);
\node[ssub, anchor=west] at (\Ax+0.52,\ly) {alloc $+$ MR registration (setup)};
\draw[fill=cblue!28, draw=cblue!60, rounded corners=1pt] (\Ax+4.55,\ly-0.12) rectangle (\Ax+4.85,\ly+0.12);
\node[ssub, anchor=west] at (\Ax+4.95,\ly) {payload transfer};
\draw[fill=cteal!32, draw=cteal!60, rounded corners=1pt] (\Ax+7.3,\ly-0.12) rectangle (\Ax+7.6,\ly+0.12);
\node[ssub, anchor=west] at (\Ax+7.7,\ly) {borrow / return};
\def\cbx{10.6}
\node[ssub, anchor=east] at (\cbx-0.16,-0.52) {naive};
\mpseg{\cbx}{\cbx+0.7}{-0.52}{corange!72}{}{white}
\mpseg{\cbx+0.7}{\cbx+1.6}{-0.52}{corange!55}{}{white}
\mpseg{\cbx+1.6}{\cbx+3.0}{-0.52}{cblue!28}{}{cblue!55!black}
\mpseg{\cbx+3.0}{\cbx+3.6}{-0.52}{cslate!30}{}{cslate}
\node[ssub, anchor=east] at (\cbx-0.16,-1.02) {ours};
\mpseg{\cbx}{\cbx+0.45}{-1.02}{cteal!32}{}{cteal}
\mpseg{\cbx+0.45}{\cbx+1.85}{-1.02}{cblue!28}{}{cblue!55!black}
\mpseg{\cbx+1.85}{\cbx+2.25}{-1.02}{cslate!22}{}{cslate}
\draw[{Stealth[length=3pt]}-{Stealth[length=3pt]}, cgreen!55!black, line width=0.6pt] (\cbx+2.25,-1.32) -- (\cbx+3.6,-1.32);
\node[glab, text=cgreen!40!black, anchor=north] at ({\cbx+2.92},-1.36) {setup off critical path};

\end{tikzpicture}}
\caption{\textbf{Pooled buffer management with one-time registration.}
\textbf{a},~Allocating buffers and performing Memory-Region (MR) registration on the critical path of every communication round incurs repeated setup cost and leaves scattered, short-lived buffers that fragment device memory.
\textbf{b},~We instead allocate a single contiguous device-memory pool once and pre-register it as a pinned MR (together with symmetric-memory registration), entirely outside the steady-state path.
\textbf{c},~During communication, buffers are borrowed from and returned to the pool with no further allocation or registration, removing dynamic-registration latency from every transfer and leaving only the useful payload on the critical path.}
\label{fig:mempool}
\end{figure*}

\textbf{Overlapping the Exchange.}
\label{sec:overlap}
Attention splits into a \emph{local branch}, whose source atoms already reside on this GPU, and a \emph{remote branch}, fetched from peers; running them concurrently hides the smaller of the two times. The bound follows directly. Let $T_c$ be the time to exchange remote features, $T_\ell$ the local branch's execution time, and $T_r$ the remote branch's time once the data has arrived. Ignoring terms common to both schedules---the initial QKV projection and the final merge---the sequential schedule costs
\begin{equation}
T_{\mathrm{seq}}=T_c+T_\ell+T_r,
\end{equation}
whereas launching communication and local computation concurrently exposes only
\begin{equation}
T_{\mathrm{ovlp}}=\max(T_c,T_\ell)+T_r .
\end{equation}
The saving is therefore
\begin{equation}
\Delta T=T_{\mathrm{seq}}-T_{\mathrm{ovlp}}=\min(T_c,T_\ell),
\qquad
\eta_{\mathrm{hide}}=\frac{\Delta T}{T_c}=\min\!\left(1,\frac{T_\ell}{T_c}\right),
\end{equation}
so the exchange is fully hidden exactly when the local branch is at least as long as the transfer, and otherwise hidden in proportion. Note that $T_{\mathrm{ovlp}}$ is a conservative upper bound: in practice the remote branch can begin as soon as the transfer finishes, rather than only after $T_\ell$ completes, so a second level of overlap is available whenever communication lands early.

The split pays only if $T_\ell$ is a large share of the total, and that follows from geometry alone, not from tuning. For a cubic subdomain of side $L$ and cutoff $r_c$, the expected fraction of an atom's in-cutoff neighbors that are locally owned is
\begin{equation}
f_{\mathrm{local}}=1-\frac{9r_c}{8L}+\mathcal{O}\!\left((r_c/L)^2\right),
\end{equation}
via the box covariogram of a uniform-density cube (verified against Monte Carlo sampling). The coefficient follows from the cube's geometry. For a cube of side $L$, the width along a unit direction $\mathbf{n}$ is $L(|\mathbf{n}_x|+|\mathbf{n}_y|+|\mathbf{n}_z|)$, so a point uniformly placed in the cube leaves it under a displacement $t\mathbf{n}$ with probability $t(|\mathbf{n}_x|+|\mathbf{n}_y|+|\mathbf{n}_z|)/L$; averaging over the sphere gives $\tfrac32 t/L$, and averaging $t$ over a ball of radius $r_c$ gives $\tfrac34 r_c$, so the two averages multiply to $9r_c/(8L)$. At the weak-scaling loads we study ($10$k--$60$k atoms/GPU), $L\approx46$--$84$\,\AA{} at $\rho=0.1$\,atoms\,\AA$^{-3}$, so $f_{\mathrm{local}}\ge80\%$ even for the $8$\,\AA{} long-range layer and $\ge87\%$ for the $5$\,\AA{} short-range layers; at the $8$k-atom/GPU strong-scaling points $L$ falls to $\approx\!40$\,\AA{} and the $8$\,\AA{} bound to $\approx\!78\%$. The local branch therefore carries the larger share of the neighbor work at the loads that dominate the weak-scaling study, which is what the overlap exploits, and the fraction depends on the ratio $r_c/L$: at fixed density a rank holding more atoms has a larger $L$, so the bound improves with per-GPU load rather than degrading with device count.

There is one implementation trap. A naive forward implementation issues the remote exchange and the local branch on two asynchronous streams and lets the scheduler interleave them. On SIMT hardware this is not enough: once a compute-heavy local branch is enqueued it occupies the scheduling queue and can delay the communication stream's own launch, so the realized overlap window shrinks below the logical $\min(T_c,T_\ell)$ of the bound above. We instead submit the exchange first, before any local compute is enqueued, and record a CUDA event when the transfer lands; only the remote branch is gated on that event. Separating \emph{launch order} from \emph{data readiness} is what makes the logical bound attainable.

\textbf{Placing the Launch in Both Passes.}
Figure~\ref{fig:main_architecture}(d) sketches the schedule; Figs.~\ref{fig:overlap_design} and~\ref{fig:bwd_overlap} give it in full. Backward does not follow the same recipe. Autograd traverses in reverse, so a forward-only scheme launches the exchange late and leaves a serial tail that no event placement can repair---the consequence, in scheduling terms, of the second force-pass requirement of Section~\ref{sec:bg-forces}. We therefore split the interface into two asymmetric operators: \texttt{pre} launches communication in the forward pass and is empty in backward, while \texttt{post} is a forward no-op that launches it in backward. Because \texttt{post} sits later in the forward graph, its backward runs \emph{first}, issuing the reverse exchange early, with the remote-backward branch waiting on the resulting event.

The local/remote split follows communication dependency, not a strict partition of the source atoms: the remote branch still consumes local source atoms, so the two branches share part of the same local features. Implementing them as independent operators with separately materialized inputs would duplicate those features in the autograd graph, inflating activation residency and complicating gradient flow. We instead let both branches consume a single unified representation of the local source data, differing only in the communication-dependent indices they access, and rewrite the backward pass to accumulate through the shared dependency exactly once, preserving the overlap schedule with no additional activation memory.

\definecolor{cblue}{HTML}{2F6DB5}
\definecolor{cteal}{HTML}{2E9E96}
\definecolor{corange}{HTML}{E08A3C}
\definecolor{cslate}{HTML}{5B6573}
\definecolor{cink}{HTML}{2B2F36}
\definecolor{cgreen}{HTML}{6DA34D}
\definecolor{cred}{HTML}{C0504D}

\newcommand{\ovcard}[3]{  \begin{scope}
    \clip[rounded corners=5pt] (0,0) rectangle (\SW,\SH);
    \fill[white] (0,0) rectangle (\SW,\SH);
    \fill[#1!9] (0,\SH-\HB) rectangle (\SW,\SH);
  \end{scope}
  \draw[rounded corners=5pt, black!18, line width=0.6pt] (0,0) rectangle (\SW,\SH);
  \draw[#1!35, line width=0.5pt] (0,\SH-\HB) -- (\SW,\SH-\HB);
  \node[anchor=west, font=\sffamily\bfseries\normalsize, text=#1!85!black] at (0.22,\SH-\HB/2) {#2};
  \node[anchor=west, font=\sffamily\bfseries\small, text=cink] at (0.62,\SH-\HB/2) {#3};
}
\newcommand{\ovbar}[6]{  \draw[fill=#4, draw=black!35, line width=0.3pt, rounded corners=1.5pt] (#1,{#3-0.18}) rectangle (#2,{#3+0.18});
  \node[font=\sffamily\fontsize{6.4}{7}\selectfont, text=#6] at ({(#1+#2)/2},#3) {#5};
}

\begin{figure*}[t]
\centering
\resizebox{\textwidth}{!}{\begin{tikzpicture}[
  font=\sffamily,
  every node/.style={inner sep=1.2pt},
  lab/.style={font=\sffamily\fontsize{7}{8}\selectfont, text=cslate, anchor=east},
  ssub/.style={font=\sffamily\fontsize{7}{8}\selectfont, text=cslate, align=center},
  glab/.style={font=\sffamily\fontsize{6.6}{7.4}\selectfont, text=cslate, align=center},
  chip/.style={rounded corners=2.5pt, font=\sffamily\fontsize{7.4}{8.6}\selectfont, inner sep=3pt},
  x=1cm,y=1cm,
]
\def\SW{15.6}\def\HB{0.55}
\def\Wbc{7.55}
\def\rowBC{0.9}\pgfmathsetmacro{\rowA}{\rowBC+3.3+0.35}
\pgfmathsetmacro{\fullW}{\SW}
\pgfmathsetmacro{\Tity}{\rowA+3.5+0.46}

\node[anchor=base, font=\sffamily\bfseries\large, text=cink] at (\fullW/2,\Tity)
   {Forward-pass communication--computation overlap};
\draw[black!15, line width=0.5pt] (0,\Tity-0.24) -- (\fullW,\Tity-0.24);

\begin{scope}[shift={(0,\rowA)}]
  \def\SW{15.6}\def\SH{3.5}\def\HB{0.55}
  \ovcard{cblue}{a}{Ideal mechanism: communication on a separate stream}
  \node[lab] at (1.7,2.05) {comm stream};
  \node[lab] at (1.7,1.3) {compute stream};
  \ovbar{1.85}{2.2}{2.05}{corange!70}{}{white}
  \ovbar{2.2}{6.3}{2.05}{cteal!30}{remote feature exchange (in flight)}{cteal!50!black}
  \node[glab, text=corange!60!black] at (2.55,2.46) {launch async};
  \draw[cslate!55, dash pattern=on 2pt off 1.6pt, line width=0.6pt] (6.3,0.95) -- (6.3,2.42);
  \node[glab, text=cslate!70] at (6.3,2.56) {remote data arrives};
  \ovbar{1.85}{5.65}{1.3}{cblue!28}{local branch (local compute)}{cblue!55!black}
  \draw[fill=cslate!8, draw=cslate!40, dash pattern=on 1.6pt off 1.6pt] (5.65,1.12) rectangle (6.3,1.48);
  \node[glab, text=cslate!60] at (5.98,0.99) {wait};
  \ovbar{6.3}{8.1}{1.3}{cblue!55}{remote branch}{white}
  \draw[{Stealth[length=3pt]}-{Stealth[length=3pt]}, cgreen!55!black, line width=0.7pt] (2.2,0.82) -- (5.65,0.82);
  \node[glab, text=cgreen!40!black] at (3.92,0.64) {overlap window};
  \draw[-{Stealth[length=4pt]}, black!40, line width=0.5pt] (1.85,0.42) -- (8.4,0.42);
  \node[glab, anchor=west] at (8.45,0.42) {time};
  \draw[rounded corners=3pt, cblue!28, fill=cblue!4, line width=0.5pt] (9.6,0.55) rectangle (15.35,2.66);
  \node[anchor=west, font=\sffamily\bfseries\fontsize{7.4}{8.4}\selectfont, text=cblue!45!black] at (9.8,2.42) {exposed cost};
  \node[anchor=west, font=\sffamily\fontsize{7.6}{9.2}\selectfont, text=cink] at (9.8,2.0) {$T_{\mathrm{seq}}=T_{\mathrm{comm}}+T_{\mathrm{local}}+T_{\mathrm{remote}}$};
  \node[anchor=west, font=\sffamily\fontsize{7.6}{9.2}\selectfont, text=cink] at (9.8,1.55) {$T_{\mathrm{ovlp}}=\max(T_{\mathrm{comm}},T_{\mathrm{local}})+T_{\mathrm{remote}}$};
  \node[anchor=west, font=\sffamily\fontsize{7.6}{9.2}\selectfont, text=cgreen!45!black] at (9.8,1.1) {$\Delta T=\min(T_{\mathrm{comm}},T_{\mathrm{local}})$};
  \node[anchor=west, font=\sffamily\fontsize{7.6}{9.2}\selectfont, text=cgreen!45!black] at (9.8,0.78) {$\eta_{\mathrm{hide}}=\min(1,\,T_{\mathrm{local}}/T_{\mathrm{comm}})$};
\end{scope}

\begin{scope}[shift={(0,\rowBC)}]
  \def\SW{7.55}\def\SH{3.3}\def\HB{0.55}
  \ovcard{corange}{b}{Naive order: local kernels first}
  \node[lab] at (1.7,2.0) {comm stream};
  \node[lab] at (1.7,1.32) {compute stream};
  \ovbar{1.85}{4.7}{1.32}{cblue!28}{heavy local kernel}{cblue!55!black}
  \ovbar{6.2}{7.2}{1.32}{cblue!55}{remote}{white}
  \draw[draw=cslate!35, dash pattern=on 1.6pt off 1.6pt, line width=0.5pt] (1.85,1.82) rectangle (4.7,2.18);
  \node[glab, text=cslate!55] at (3.27,2.0) {comm queued (SMs busy)};
  \ovbar{4.7}{6.2}{2.0}{cteal!30}{in flight}{cteal!50!black}
  \node[glab, anchor=east, text=cslate!70] at (1.7,0.78) {SM use};
  \fill[cred!55] (1.85,0.66) rectangle (4.7,0.9);
  \fill[cgreen!30] (4.7,0.66) rectangle (7.2,0.9);
  \node[glab, text=white] at (3.27,0.78) {saturated by local};
  \node[glab, text=cslate!60] at (8.0,0.6) {};
  \node[chip, fill=corange!12, draw=corange!45, text=corange!60!black] at (\SW/2,0.3) {comm starved $\Rightarrow$ overlap collapses};
\end{scope}

\begin{scope}[shift={(8.05,\rowBC)}]
  \def\SW{7.55}\def\SH{3.3}\def\HB{0.55}
  \ovcard{cblue}{c}{Correct order: async exchange first}
  \node[lab] at (1.7,2.0) {comm stream};
  \node[lab] at (1.7,1.32) {compute stream};
  \ovbar{1.85}{2.15}{2.0}{corange!70}{}{white}
  \ovbar{2.15}{5.1}{2.0}{cteal!30}{remote exchange (in flight)}{cteal!50!black}
  \ovbar{1.85}{4.6}{1.32}{cblue!28}{local branch}{cblue!55!black}
  \ovbar{5.1}{6.6}{1.32}{cblue!55}{remote}{white}
  \draw[{Stealth[length=3pt]}-{Stealth[length=3pt]}, cgreen!55!black, line width=0.7pt] (2.15,2.34) -- (4.6,2.34);
  \node[glab, text=cgreen!40!black] at (3.37,2.52) {overlap window};
  \node[glab, anchor=east, text=cslate!70] at (1.7,0.78) {SM use};
  \fill[cgreen!35] (1.85,0.66) rectangle (4.6,0.9);
  \fill[cgreen!30] (4.6,0.66) rectangle (6.6,0.9);
  \node[glab, text=cgreen!45!black] at (3.22,0.78) {compute $+$ copy-engine};
  \node[chip, fill=cblue!12, draw=cblue!45, text=cblue!60!black] at (\SW/2,0.3) {comm hidden behind local compute};
\end{scope}

\node[anchor=west, font=\sffamily\fontsize{7}{8}\selectfont, text=cslate] at (0.1,0.2) {};
\draw[fill=cblue!28, draw=black!35, rounded corners=1pt] (0.15,0.08) rectangle (0.5,0.32);
\node[ssub, anchor=west] at (0.6,0.2) {local compute};
\draw[fill=cblue!55, draw=black!35, rounded corners=1pt] (2.5,0.08) rectangle (2.85,0.32);
\node[ssub, anchor=west] at (2.95,0.2) {remote branch};
\draw[fill=cteal!30, draw=black!35, rounded corners=1pt] (5.0,0.08) rectangle (5.35,0.32);
\node[ssub, anchor=west] at (5.45,0.2) {communication in flight};
\draw[fill=corange!70, draw=black!35, rounded corners=1pt] (8.1,0.08) rectangle (8.45,0.32);
\node[ssub, anchor=west] at (8.55,0.2) {async launch};
\draw[draw=cslate!45, dash pattern=on 1.6pt off 1.6pt] (10.4,0.08) rectangle (10.75,0.32);
\node[ssub, anchor=west] at (10.85,0.2) {queued / wait};
\draw[fill=cred!55, draw=black!35] (12.7,0.08) rectangle (13.05,0.32);
\node[ssub, anchor=west] at (13.15,0.2) {SM saturated};

\end{tikzpicture}}
\caption{\textbf{Forward-pass communication--computation overlap.}
\textbf{a},~The attention operator is split into a local branch (source atoms already resident on the GPU) and a remote branch; the remote feature exchange is launched asynchronously on the communication stream while the local branch executes on the compute stream, and the remote branch waits on the completion event only immediately before it consumes the received data; the panel's cost model gives the resulting exposed cost.
\textbf{b},~Launch order is decisive: if the heavy local kernels are enqueued first, the exchange is issued late behind them and the overlap window collapses.
\textbf{c},~Issuing the asynchronous exchange first gets the communication kernels resident before the local branch fills the device, preserving a large overlap window.}
\label{fig:overlap_design}
\end{figure*}

\definecolor{cblue}{HTML}{2F6DB5}
\definecolor{cteal}{HTML}{2E9E96}
\definecolor{corange}{HTML}{E08A3C}
\definecolor{cslate}{HTML}{5B6573}
\definecolor{cink}{HTML}{2B2F36}
\definecolor{cgreen}{HTML}{6DA34D}
\definecolor{cred}{HTML}{C0504D}

\newcommand{\bovtwocard}[3]{  \begin{scope}
    \clip[rounded corners=5pt] (0,0) rectangle (\SW,\SH);
    \fill[white] (0,0) rectangle (\SW,\SH);
    \fill[#1!9] (0,\SH-\HB) rectangle (\SW,\SH);
  \end{scope}
  \draw[rounded corners=5pt, black!18, line width=0.6pt] (0,0) rectangle (\SW,\SH);
  \draw[#1!35, line width=0.5pt] (0,\SH-\HB) -- (\SW,\SH-\HB);
  \node[anchor=west, font=\sffamily\bfseries\normalsize, text=#1!85!black] at (0.22,\SH-\HB/2) {#2};
  \node[anchor=west, font=\sffamily\bfseries\small, text=cink] at (0.62,\SH-\HB/2) {#3};
}

\newcommand{\bovtwobar}[6]{  \draw[fill=#4, draw=black!35, line width=0.3pt, rounded corners=1.5pt] (#1,{#3-0.17}) rectangle (#2,{#3+0.17});
  \node[font=\sffamily\fontsize{6.2}{6.8}\selectfont, text=#6, align=center] at ({(#1+#2)/2},#3) {#5};
}

\begin{figure*}[t]
\centering
\resizebox{\textwidth}{!}{\begin{tikzpicture}[
  font=\sffamily,
  every node/.style={inner sep=1.2pt},
  lab/.style={font=\sffamily\fontsize{6.8}{7.6}\selectfont, text=cslate, anchor=east},
  ssub/.style={font=\sffamily\fontsize{7}{8}\selectfont, text=cslate, align=center},
  glab/.style={font=\sffamily\fontsize{6.4}{7.2}\selectfont, text=cslate, align=center},
  chip/.style={rounded corners=2.5pt, font=\sffamily\fontsize{7.2}{8.4}\selectfont, inner sep=3pt},
  opb/.style={rounded corners=3pt, draw=cslate!55, fill=white, align=center,
              font=\sffamily\fontsize{7.2}{8.4}\selectfont, text=cink, inner sep=2.5pt},
  x=1cm,y=1cm,
]
\def\SW{15.6}\def\HB{0.55}\def\PH{2.95}
\def\rowB{0.9}\pgfmathsetmacro{\rowA}{\rowB+\PH+0.32}
\pgfmathsetmacro{\opy}{\rowA+\PH+0.95}
\pgfmathsetmacro{\fullW}{\SW}\pgfmathsetmacro{\Tity}{\opy+1.05}
\def\divx{7.95}

\node[anchor=base, font=\sffamily\bfseries\large, text=cink] at (\fullW/2,\Tity)
   {Autograd-compatible overlap via paired \texttt{pre}/\texttt{post} operators};
\draw[black!15, line width=0.5pt] (0,\Tity-0.24) -- (\fullW,\Tity-0.24);

\draw[rounded corners=4pt, black!15, fill=cslate!4, line width=0.5pt] (0.2,\opy-0.55) rectangle (15.4,\opy+0.62);
\node[lab, anchor=west, text=cslate!80] at (0.4,\opy+0.38) {operator chain (forward order)};
\node[opb] (pre) at (3.4,\opy) {\texttt{pre}};
\node[opb, minimum width=2.6cm] (att) at (6.7,\opy) {equivariant attention\\(local $\parallel$ remote)};
\node[opb] (post) at (10.0,\opy) {\texttt{post}};
\node[glab, anchor=west, text=cslate!55] at (10.7,\opy) {$\cdots$};
\node[glab, anchor=east, text=cslate!55] at (2.7,\opy) {$\cdots$};
\foreach \a/\b in {pre/att, att/post}{\draw[-{Stealth[length=4pt]}, cslate!70, line width=0.7pt] (\a) -- (\b);}
\node[glab, text=corange!60!black] at (3.4,\opy-0.4) {fwd: launch comm $\cdot$ bwd: empty};
\node[glab, text=cblue!60!black] at (10.0,\opy-0.4) {fwd: no-op $\cdot$ bwd: launch comm};
\node[glab, anchor=west, text=cslate!75] at (12.0,\opy+0.16) {backward traverses in reverse,};
\node[glab, anchor=west, text=cslate!75] at (12.0,\opy-0.14) {so \texttt{post.backward} fires first};
\draw[{Stealth[length=4pt]}-, cslate!50, line width=0.7pt, dash pattern=on 2pt off 1.6pt] (10.5,\opy+0.28) -- (12.0,\opy+0.46);

\begin{scope}[shift={(0,\rowA)}]
  \def\SW{15.6}\def\SH{2.95}\def\HB{0.55}
  \bovtwocard{corange}{a}{Naive forward-only overlap}
  \draw[black!15, dash pattern=on 2pt off 2pt, line width=0.5pt] (\divx,0.35) -- (\divx,2.25);
  \node[glab, text=cslate!75] at (4.4,2.18) {forward pass};
  \node[glab, text=cslate!75] at (11.6,2.18) {backward pass};
  \node[lab] at (1.65,1.78) {comm};
  \node[lab] at (1.65,1.18) {compute};

  \bovtwobar{1.8}{2.1}{1.78}{corange!70}{}{white}
  \bovtwobar{2.1}{4.5}{1.78}{cteal!30}{pre launch exchange}{cteal!50!black}
  \bovtwobar{1.8}{4.05}{1.18}{cblue!28}{local compute}{cblue!55!black}
  \bovtwobar{4.5}{5.7}{1.18}{cblue!55}{remote}{white}
  \draw[{Stealth[length=2.5pt]}-{Stealth[length=2.5pt]}, cgreen!55!black, line width=0.6pt] (2.1,0.78) -- (4.05,0.78);
  \node[glab, text=cgreen!45!black] at (3.07,0.6) {overlap window};

  \bovtwobar{8.3}{8.6}{1.78}{corange!70}{}{white}
  \bovtwobar{8.6}{10.6}{1.78}{cred!28}{exchange remote}{cred!60!black}
  \bovtwobar{10.6}{11.85}{1.18}{cblue!55}{remote-grad}{white}
  \bovtwobar{11.85}{14.7}{1.18}{cblue!28}{local-grad compute}{cblue!55!black}
  \draw[-{Stealth[length=3pt]}, cred!65, line width=0.7pt] (10.6,1.55) -- (10.6,1.35);
  \draw[-{Stealth[length=3pt]}, cred!65, line width=0.7pt] (11.85,1.18) -- (12.05,1.18);
  \draw[{Stealth[length=2.5pt]}-{Stealth[length=2.5pt]}, cred!70, line width=0.7pt] (8.6,0.78) -- (14.7,0.78);
  \node[glab, text=cred!65!black] at (11.65,0.6) {serial backward tail};
\end{scope}

\begin{scope}[shift={(0,\rowB)}]
  \def\SW{15.6}\def\SH{2.95}\def\HB{0.55}
  \bovtwocard{cblue}{b}{Paired \texttt{pre}/\texttt{post}: overlap in both directions}
  \draw[black!15, dash pattern=on 2pt off 2pt, line width=0.5pt] (\divx,0.35) -- (\divx,2.25);
  \node[glab, text=cslate!75] at (4.4,2.18) {forward pass --- \texttt{pre} launches};
  \node[glab, text=cslate!75] at (11.6,2.18) {backward pass --- \texttt{post.backward} launches};
  \node[lab] at (1.65,1.78) {comm};
  \node[lab] at (1.65,1.18) {compute};

  \bovtwobar{1.8}{2.1}{1.78}{corange!70}{}{white}
  \bovtwobar{2.1}{4.5}{1.78}{cteal!30}{pre launch exchange}{cteal!50!black}
  \bovtwobar{1.8}{4.05}{1.18}{cblue!28}{local compute}{cblue!55!black}
  \bovtwobar{4.5}{5.7}{1.18}{cblue!55}{remote}{white}
  \draw[{Stealth[length=2.5pt]}-{Stealth[length=2.5pt]}, cgreen!55!black, line width=0.6pt] (2.1,0.78) -- (4.05,0.78);
  \node[glab, text=cgreen!45!black] at (3.07,0.6) {overlap window};

  \bovtwobar{8.3}{8.6}{1.78}{corange!70}{}{white}
  \bovtwobar{8.6}{11.0}{1.78}{cteal!30}{post launch exchange}{cteal!50!black}
  \bovtwobar{8.3}{10.55}{1.18}{cblue!28}{local-grad compute}{cblue!55!black}
  \bovtwobar{11.0}{12.25}{1.18}{cblue!55}{scatter-sum}{white}
  \draw[-{Stealth[length=3pt]}, cslate!55, line width=0.7pt, dash pattern=on 2pt off 1.6pt] (11.0,1.55) -- (11.0,1.35);
  \node[glab, text=cslate!70] at (11.62,1.55) {wait};
  \draw[{Stealth[length=2.5pt]}-{Stealth[length=2.5pt]}, cgreen!55!black, line width=0.6pt] (8.6,0.78) -- (10.55,0.78);
  \node[glab, text=cgreen!45!black] at (9.57,0.6) {overlap window};
\end{scope}

\def\ly{0.2}
\draw[fill=corange!70, draw=black!35, rounded corners=1pt] (0.15,\ly-0.12) rectangle (0.5,\ly+0.12);
\node[ssub, anchor=west] at (0.6,\ly) {async launch};
\draw[fill=cteal!30, draw=black!35, rounded corners=1pt] (2.5,\ly-0.12) rectangle (2.85,\ly+0.12);
\node[ssub, anchor=west] at (2.95,\ly) {remote exchange};
\draw[fill=cblue!28, draw=black!35, rounded corners=1pt] (5.4,\ly-0.12) rectangle (5.75,\ly+0.12);
\node[ssub, anchor=west] at (5.85,\ly) {local compute};
\draw[fill=cblue!55, draw=black!35, rounded corners=1pt] (7.9,\ly-0.12) rectangle (8.25,\ly+0.12);
\node[ssub, anchor=west] at (8.35,\ly) {remote branch};
\draw[fill=cred!28, draw=black!35, rounded corners=1pt] (10.4,\ly-0.12) rectangle (10.75,\ly+0.12);
\node[ssub, anchor=west] at (10.85,\ly) {exposed serial tail};

\end{tikzpicture}}
\caption{\textbf{Autograd-compatible overlap via paired \texttt{pre}/\texttt{post} operators.}
\textbf{a},~A naive forward-only overlap works in the forward pass, where \texttt{pre} launches the remote exchange before local computation. In backward, however, communication is not relocated early enough: the remote exchange, remote-gradient computation, and local-gradient computation appear as a serial tail.
\textbf{b},~With paired operators, \texttt{pre} launches communication in the forward pass, while \texttt{post} is a forward no-op whose backward method launches the reverse remote exchange. Because \texttt{post.backward} is reached early under reverse traversal, the exchange goes out as soon as its payload---the gradient with respect to the received remote features---exists, and overlaps with unblocked local-gradient computation. Writing the forward halo gather as $\mathbf{x}_{\text{halo}}=G\,\mathbf{x}_{\text{owned}}$ makes the bar after the wait explicit: the owner applies $G^{\!\top}$, accumulating $\nabla\mathbf{x}_{\text{owned}} \mathrel{+}= G^{\!\top}\nabla\mathbf{x}_{\text{halo}}$ over every rank and periodic image that requested the atom.}
\label{fig:bwd_overlap}
\end{figure*}

\subsection{A Decentralized On-GPU MD Engine}
\label{sec:md_engine}

\definecolor{cblue}{HTML}{2F6DB5}
\definecolor{cteal}{HTML}{2E9E96}
\definecolor{corange}{HTML}{E08A3C}
\definecolor{cslate}{HTML}{5B6573}
\definecolor{cink}{HTML}{2B2F36}
\definecolor{cgreen}{HTML}{6DA34D}
\definecolor{cred}{HTML}{C0504D}

\begin{figure*}[t]
\centering
\resizebox{\textwidth}{!}{\begin{tikzpicture}[
  font=\sffamily,
  proc/.style   ={rounded corners=3pt, draw=black!35, line width=0.5pt, align=center,
                  inner sep=3pt, font=\sffamily\footnotesize, text width=3.05cm, minimum height=0.95cm},
  r0/.style     ={proc, fill=corange!20},
  r0b/.style    ={proc, fill=corange!14, dash pattern=on 2pt off 1.6pt},
  rks/.style    ={proc, fill=cblue!12},
  loc/.style    ={proc, fill=cgreen!16},
  bigarr/.style ={-{Latex[length=2.8mm]}, line width=2.2pt, draw=cred},
  thinarr/.style={-{Latex[length=2.2mm]}, line width=0.8pt, draw=cslate, dash pattern=on 2pt off 1.6pt},
  looparr/.style={-{Latex[length=2.2mm]}, line width=1.0pt, draw=cgreen!60!black},
  elab/.style   ={font=\sffamily\fontsize{6.6}{7.6}\selectfont, align=center},
  badr/.style   ={rounded corners=2pt, fill=cred!12, draw=cred!50, line width=0.5pt, inner sep=2.5pt,
                  font=\sffamily\bfseries\fontsize{6.8}{7.8}\selectfont, text=cred!80!black},
  badg/.style   ={rounded corners=2pt, fill=cgreen!16, draw=cgreen!55!black, line width=0.5pt, inner sep=2.5pt,
                  font=\sffamily\bfseries\fontsize{6.8}{7.8}\selectfont, text=cgreen!40!black},
  hot/.style    ={font=\sffamily\bfseries\fontsize{6}{7}\selectfont, text=cred!85!black},
  hdr/.style    ={font=\sffamily\bfseries\small, text=cink},
]

\def\W{7.6} \def\H{6.45} \def\GAP{0.9} \def\cx{2.55} \def\PYS{2.35} \def\RH{1.9}
\def\Wtot{16.1} 

\begin{scope}[yshift=\PYS cm]
  \draw[rounded corners=6pt, black!16, line width=0.7pt] (0,0) rectangle (\W,\H);
  \draw[black!12, line width=0.5pt] (0,\H-0.72) -- (\W,\H-0.72);
  \node[hdr, anchor=west] at (0.22,\H-0.38) {\textbf{a}\;\; Conventional MD engine};
  \node[badr, anchor=east] at (\W-0.18,\H-0.38) {$O(N)$ / step};

  \node[r0]  (a0) at (\cx,\H-1.55) {\textbf{rank 0:} SFC encode $+$ atom partition};
  \node[rks] (ar) at (\cx,\H-3.55) {ranks $0\ldots K{-}1$: distributed inference $\Rightarrow$ forces};
  \node[r0]  (ai) at (\cx,\H-5.55) {\textbf{rank 0:} integrate $\Rightarrow$ update positions};

  \node[hot, anchor=west] at (a0.east) {\,serial hub};
  \node[hot, anchor=west] at (ai.east) {\,serial hub};

  \draw[bigarr] (a0) -- (ar) node[elab, text=cred, midway, anchor=west, xshift=5pt, text width=2.3cm] {scatter positions\\ $1 \to K$};
  \draw[bigarr] (ar) -- (ai) node[elab, text=cred, midway, anchor=west, xshift=5pt, text width=2.3cm] {reduce forces\\ $K \to 1$};
  \draw[thinarr] (ai.west) .. controls (0.42,\H-5.55) and (0.42,\H-1.55) .. (a0.west)
        node[elab, text=cslate, midway, anchor=east, xshift=-2pt, text width=1.0cm] {next step};
\end{scope}

\begin{scope}[xshift={(\W+\GAP)*1cm}, yshift=\PYS cm]
  \draw[rounded corners=6pt, black!16, line width=0.7pt] (0,0) rectangle (\W,\H);
  \draw[black!12, line width=0.5pt] (0,\H-0.72) -- (\W,\H-0.72);
  \node[hdr, anchor=west] at (0.22,\H-0.38) {\textbf{b}\;\; Decentralized on-GPU engine};
  \node[badg, anchor=east] at (\W-0.18,\H-0.38) {fixed size / step};

  \node[r0b] (b0) at (\cx,\H-1.55) {\textbf{rank 0:} SFC encode $+$ partition \textit{(once per re-partition)}};
  \node[rks] (br) at (\cx,\H-3.55) {ranks $0\ldots K{-}1$: distributed inference $\Rightarrow$ forces};
  \node[loc] (bl) at (\cx,\H-5.55) {\textbf{each rank:} local integrate on GPU};
  \node[hot, text=cgreen!40!black, anchor=west] at (bl.east) {\,parallel};

  \draw[thinarr] (b0) -- (br) node[elab, text=cslate, midway, anchor=west, xshift=5pt, text width=2.3cm] {scatter once /\\ re-partition $1\to K$};
  \draw[looparr] (br) -- (bl);
  \draw[looparr] (bl.east) .. controls +(1.1,0.35) and +(1.1,-0.35) .. (br.east)
        node[elab, text=cgreen!45!black, midway, anchor=west, xshift=2pt, text width=2.0cm] {$\times 100$ steps ($\approx$50\,fs): local update $+$ fixed-size sync};
  \draw[thinarr] (bl.west) .. controls (0.42,\H-5.55) and (0.42,\H-1.55) .. (b0.west)
        node[elab, text=cslate, midway, anchor=east, xshift=-2pt, text width=1.05cm] {gather every $100$ steps};
\end{scope}

\begin{scope}
  \def\cA{2.35} \def\cB{8.0} \def\cC{13.4}   \fill[cred!6,   rounded corners=3pt] (5.05,0.02) rectangle (10.55,\RH);
  \fill[cgreen!7, rounded corners=3pt] (10.75,0.02) rectangle (\Wtot,\RH);
  \draw[rounded corners=3pt, black!14, line width=0.6pt] (0,0.02) rectangle (\Wtot,\RH);
  \node[font=\sffamily\bfseries\fontsize{7.4}{8.6}\selectfont, text=cred!80!black]   at (\cB,\RH-0.22) {Conventional};
  \node[font=\sffamily\bfseries\fontsize{7.4}{8.6}\selectfont, text=cgreen!42!black] at (\cC,\RH-0.22) {Ours};
  \draw[black!12, line width=0.5pt] (0,\RH-0.44) -- (\Wtot,\RH-0.44);
  \foreach \y/\lab/\cv/\ov in {      1.18/{Inter-rank comm / step}/{$O(N)$: $1\!\to\!K$ scatter $+$ $K\!\to\!1$ reduce}/{fixed size: sync signal},
      0.70/{Rank-0 work / step}/{integrates \emph{every} atom (serial hub)}/{idle between re-partitions},
      0.24/{CPU\,$\leftrightarrow$\,GPU transfer / step}/{every step}/{none --- fully on-GPU}  }{    \node[anchor=west, font=\sffamily\fontsize{7}{8.4}\selectfont, text=cink]            at (0.18,\y) {\lab};
    \node[font=\sffamily\fontsize{7}{8.4}\selectfont, text=cred!80!black, align=center]   at (\cB,\y) {\cv};
    \node[font=\sffamily\fontsize{7}{8.4}\selectfont, text=cgreen!40!black, align=center] at (\cC,\y) {$\checkmark$\,\,\ov};
  }
\end{scope}

\end{tikzpicture}}
\caption{\textbf{Conventional versus decentralized molecular-dynamics engine.}
\textbf{a},~A standard distributed loop is rank-0 centric: each step scatters atom positions from rank~0 to all ranks ($1\!\to\!K$), runs distributed inference, reduces all forces back to rank~0 ($K\!\to\!1$), and lets rank~0 integrate and re-encode. The two collective hot spots through rank~0 form a communication bottleneck, and the serial curve encoding and integration on rank~0 form a compute bottleneck---both worsening with system size.
\textbf{b},~Our engine keeps every operation on the GPU. After the initial scatter, each rank integrates its own atoms locally using the globally computed forces and emits only a fixed-size synchronization signal, eliminating the per-step $K\!\to\!1$ reduction and the central integration. A global gather and re-partition are performed only every $100$ steps ($\approx\!50$\,fs at the $0.5$\,fs step), which is safe because atoms move far less than the neighbor-search margin over that interval (Section~\ref{sec:morton_partition}).
The scorecard contrasts the two engines per step: the decentralized design removes the $O(N)$ collective bottlenecks, the rank-0 serial-compute bottleneck, and the per-step CPU--GPU transfers entirely.}
\label{fig:md_engine}
\end{figure*}

One global collective per step survives everything above---not in the attention operator but in the integration loop wrapping it. Conventionally rank~0 orchestrates every step (Fig.~\ref{fig:md_engine}a): it \emph{scatters} positions and \emph{reduces} forces through its single network endpoint, which $K\!-\!1$ ranks contend for every step. Both per-step collectives funnel through one endpoint: the force reduction is a many-to-one incast and the position scatter its one-to-many mirror, so $K-1$ ranks contend for a single network interface---congestion rather than aggregate bandwidth---and the volume crossing it grows with total system size instead of staying local. The deeper cost sits between those two collectives: the loop re-encodes the partition on one device every step, quantizing and radix-sorting the full coordinate set, and that sort is most of the partition-and-broadcast bundle it sits in ($\approx\!70.7$\,ms of the bundle's $\approx\!87$\,ms at $10.24$M atoms). Re-encoding re-derives \emph{ownership}, and since the edge sets and exchange plan of Section~\ref{sec:multilevel_search} are indexed by ownership, it invalidates the topology cache on every step---erasing exactly what the cache exists to preserve. A centralized driver therefore cannot hold a rebuild off for even the ten steps the cache needs, let alone reach the hundred steps between re-partitions, however fast its collectives run. Integrating on-device additionally removes the host--device coordinate and force transfers a host-driven integrator performs each step.

Neither centralization is needed every step: the subdomain size bounds \emph{ownership}, and the skin bounds which edges can cross, on the two timescales of Section~\ref{sec:multilevel_search}. We therefore integrate per rank, on the GPU (Fig.~\ref{fig:md_engine}b; one full step is Algorithm~\ref{alg:md-step}). Rank~0 encodes and scatters once at initialization; from then on each rank advances the atoms it owns from its own positions and the globally computed forces, emitting only a fixed handful of scalars---so the per-step force reduction and the centralized integrator collapse into a constant-payload barrier. A global gather and re-partition fire only every $100$ steps ($\approx\!50$\,fs at our timestep)---the shape-changing event of Section~\ref{sec:morton_partition}---leaving the steady-state trajectory to run as replayed static graphs. Section~\ref{sec:md-validation} measures that bundle: $\approx\!87$--$88$\,ms for the partition-and-broadcast half ($\approx\!16$--$18$\,ms of it cross-rank communication) plus $\approx\!320$--$330$\,ms for the rebuild it triggers, $407$--$418$\,ms in total, amortized to roughly $4$\,ms of per-step cost against an $\approx\!4.55$\,s full step.

Algorithm~\ref{alg:md-step} spells this out in full, executed identically (SPMD) by every rank $p$: the half-kick/drift, the unanimous rebuild-or-reuse decision of Section~\ref{sec:multilevel_search}, the fixed-schedule re-partition of Section~\ref{sec:morton_partition} that unconditionally forces a rebuild on top of it, the overlapped forward--backward pass of Section~\ref{sec:reduce_volume}, and the closing half-kick. Lines 12--15 are the only branch that can change tensor shape---forced by the unanimous rebuild decision at line 5, taken from the trigger at line 4, or unconditionally by a repartition event (lines 6--11)---while lines 1--5 and 16--19 run identically whether or not the step rebuilds, including the mask re-application of line 16, which sees the same $(\mathcal{E},\Pi)$ shape on a hit as a rebuild just produced. The shape-changing branch is explicitly fenced; everything outside it is data-independent in extent, which is what makes the captured graph reusable.

\begin{algorithm}
\caption{Decentralized MD step (rank $p$; identical on every rank)}
\label{alg:md-step}
\begin{algorithmic}[1]
\Require owned positions $\mathbf{r}$, velocities $\mathbf{v}$, forces $\mathbf{F}$ from the previous step, build-time reference positions $\mathbf{r}_{\mathrm{ref}}$, cached topology $(\mathcal{E}, \Pi)$, step $t$, timestep $\Delta t$, rebuild interval $I$, displacement skin $\delta$
\State $\mathbf{v} \gets \mathbf{v} + \tfrac{1}{2}\Delta t\,\mathbf{F}/m$ \Comment{half-kick, force from previous step}
\State $\mathbf{r} \gets \mathbf{r} + \Delta t\,\mathbf{v}$ \Comment{drift}
\State $d \gets \max$ displacement of an owned atom since $\mathbf{r}_{\mathrm{ref}}$
\State $\mathit{trigger} \gets [\,t \bmod I = 0\,] \lor [\,d > \delta/2\,]$
\State $\mathit{rebuild} \gets \texttt{all\_reduce}(\mathrm{MAX}, \mathit{trigger})$ \Comment{unanimous, Sec.~\ref{sec:multilevel_search}}
\If{$t \bmod 100 = 0$} \Comment{fixed schedule, not data-dependent}
    \State $\mathbf{r}^{\text{all}} \gets \Call{Gather}{\mathbf{r}, \text{rank } 0}$
    \State $\sigma \gets \Call{HilbertSort}{\mathbf{r}^{\text{all}}}$, cut into $K$ equal-count contiguous segments
    \State $\Call{Broadcast}{\sigma}$; migrate owned positions and velocities to match this rank's new segment
    \State $\mathit{rebuild} \gets \mathrm{true}$ \Comment{which atoms this rank owns changed: cached topology and plan invalid, regardless of line 5}
\EndIf
\If{$\mathit{rebuild}$}
    \State $(\mathcal{E}, \Pi) \gets \Call{ThreeLevelSearch}{\mathbf{r}, r_c\!+\!\delta}$ \Comment{candidate superset; AABB $\to$ chunk $\to$ exact FAISS; also resets $\mathbf{r}_{\mathrm{ref}}$}
    \State re-capture the static CUDA Graph for this $(\mathcal{E}, \Pi)$ shape
\EndIf
\State re-expand $\mathbf{r}$ under PBC; recompute edge geometry from $\mathcal{E}$ and activate the pairs with $r \le r_c$ \Comment{both paths; indices detached, geometry stays in the autograd graph}
\State $\mathbf{F} \gets \Call{ReplayGraph}{\mathbf{r}, \mathcal{E}, \Pi}$ \Comment{fwd + autograd bwd; All-to-All overlapped both passes, Sec.~\ref{sec:reduce_volume}}
\State $\mathbf{v} \gets \mathbf{v} + \tfrac{1}{2}\Delta t\,\mathbf{F}/m$ \Comment{half-kick, force from this step}
\State $\texttt{all\_reduce}(\mathrm{SUM}, E, T, \mathrm{virial})$ \Comment{constant payload, independent of $N$}
\end{algorithmic}
\end{algorithm}

\section{Experimental Evaluation}
\label{sec:experiments}

\subsection{Experimental Setup}
\label{sec:exp-setup}

Experiments run on $64$ compute nodes, each with $8$ NVIDIA H100 GPUs (80\,GB
HBM3) on NVLink~4.0 and linked by a 400\,Gbps InfiniBand NDR fabric; the
largest run uses all $512$. The stack is CUDA 13.0, PyTorch 2.11, NCCL 2.28,
Triton 3.7 and FAISS 1.14.2. A second cluster of $256$ A100 GPUs ($40$\,GB
HBM2e, PCIe, no NVLink) carries the strong-scaling sweep and the two MD
trajectory studies of Section~\ref{sec:md-validation}; it was swapped in for
the H100/\allowbreak NVLink/\allowbreak InfiniBand fabric for availability, and
Section~\ref{sec:discussion} scopes what a comparison across the two
interconnects can mean. Every timing reported below is MAX-reduced across
ranks, so it measures the distributed critical path rather than the rank-$0$
view.

All runs use a variant of the open-source \textbf{UBio-MolFM}
potential~\cite{huang2026ubiomolfm}, itself built on E2Former~V2~\cite{huang2025e2former,
huang2026e2formerv2}---three short-range layers at $5$\,\AA{} then one
long-range layer at $8$\,\AA{}, a $23$\,\AA{} receptive field per atom,
$23.5$M parameters (full configuration in Appendix~\ref{sec:app-config}). This
paper's contribution is the model's distributed execution, not its chemistry:
its DFT-level accuracy on biomolecular systems is established separately, and
Table~\ref{tab:toy_infer_ablation} validates that our execution path preserves
it. Module benchmarks use a 10k-atom toy system and bulk water to the
single-node limit; the scaling study grows bulk water with the device count, at
$10$k, $25$k and $60$k atoms per GPU, to $30.7$M atoms on $512$ GPUs; the MD
studies simulate a $10.24$M-atom aqueous KCl solution at $40$k atoms per GPU on
$256$ devices and a $10.06$M-atom Norovirus capsid in solvent at $39.3$k.
Sections~\ref{sec:neighbor-scaling}--\ref{sec:e2e_scaling}
use TF32 mixed precision, with the accuracy trade-off characterized below;
Section~\ref{sec:md-validation}'s trajectory validation runs with TF32 off, and
the TF32 A/B of Appendix~\ref{sec:app-protocol} runs on this section's
$1.28$M-atom strong-scaling system, where its off arm reproduces the $64$-GPU
step time to $0.9\%$.

\begin{table}[t]
\centering
\caption{Run configurations. ``n/s'' marks a value the experimental text does
not state. TF32 is the mixed-precision mode of Section~\ref{sec:single_gpu};
recomputation is activation recomputation; the cache interval is the rebuild
period in steps. The H100 cluster is the 64-node, 8-GPU-per-node
NVLink/InfiniBand system above, the A100 cluster the $256$-GPU PCIe system; the
two are never mixed within a single comparison.}
\label{tab:run-config}
\small
\begin{tabular}{lrrllll}
    \toprule
    Configuration & GPUs & Atoms/GPU & TF32 & Recomp. & Cache & Cluster \\
    \midrule
    Single-GPU ablation & $1$ & $10$k & off/on & off/on & n/s & H100 \\
    Neighbor, $25$k & $64$--$512$ & $25$k & on & n/s & $1$/$10$ & H100 \\
    Neighbor, $50$k & $64$--$512$ & $50$k & on & n/s & $1$/$10$ & H100 \\
    Overlap A/B & $32$--$512$ & $25$k & on & off & $10$ & H100 \\
    Weak scaling, $10$k & $32$--$512$ & $10$k & on & off & $10$ & H100 \\
    Weak scaling, $25$k & $32$--$512$ & $25$k & on & off & $10$ & H100 \\
        Weak scaling, $60$k & $32$--$512$ & $60$k & on & on & $10$ & H100 \\
    Full-step cache A/B & $512$ & $60$k & on & on & $1$/$10$ & H100 \\
        Strong scaling & $32$--$256$ & $40$k$\to$$5$k & off & on & n/s & A100 \\
    KCl trajectory & $256$ & $40$k & off & n/s & $10$ & A100 \\
        Capsid pair & $256$ & $39.3$k & off & n/s & $10$ & A100 \\
    \bottomrule
\end{tabular}
\end{table}

Table~\ref{tab:run-config} collects the configurations the section reports.
The cluster column is uniform within every comparison but not across the
section: the weak-scaling, overlap, neighbor-scaling and single-GPU results are
H100 numbers, while the strong-scaling sweep and the trajectory studies are A100
numbers with no NVLink. The recomputation column marks the runs that carry
it---the $60$k weak-scaling load and the strong-scaling sweep, whose
$32$/$64$-GPU points exceed its cluster's $40$\,GB budget otherwise.

\textbf{Timing protocol.} Every timing in this section is a steady-state
number: step~$0$ is excluded from every series, since it carries warm-up,
\texttt{torch.\allowbreak compile} tracing, autotuning and first-touch
allocation. The single-GPU ablation averages the last $50$ of $100$
repetitions; the end-to-end curves average steps $300$--$399$ of a $500$-step
run, a window that spans exactly ten interval-$10$ rebuild periods so the mean
is the amortized per-step cost rather than a sample biased by where rebuilds
land; the strong-scaling points average the last five logged windows of their
run. Sub-stage timers are each MAX-reduced independently, and their maxima may
originate from different ranks, so we do not decompose a stage additively---an
attribution resting on those timers is \emph{consistent with} the telemetry, not
a measured stage decomposition. One official run per point, so no error bars are
drawn and differences of a few percent are not treated as separable.
Appendix~\ref{sec:app-protocol} gives the protocol in full.

\subsection{Single-GPU Acceleration}
We first quantify the single-GPU stack of Section~\ref{sec:engine} with a
waterfall ablation on a fixed 10k-atom toy system
(Table~\ref{tab:toy_infer_ablation}).

\begin{table}[t]
    \centering
    \caption{Ablation on a 10k-atom toy system on H100. Config abbreviates T=Triton, C=\texttt{torch.compile}, F=TF32, R=activation recomputation, applied cumulatively left to right. Precision metrics are relative to the base (no T/C/F/R).}
    \small
    \begin{tabular}{lrrrr}
        \toprule
        Config & Latency(ms) & Mem(MiB) & $|\Delta E|$(eV) & $\max|\Delta F|$ \\
        \midrule
        base       & 363.4 & 67918 & $0$                 & $0$                 \\
        $+$T       & 234.9 & 21078 & $3.05\times10^{-5}$ & $2.38\times10^{-7}$ \\
        $+$T+C     & 141.1 & 15370 & $3.05\times10^{-5}$ & $1.01\times10^{-6}$ \\
        $+$T+C+F   & 95.7  & 15370 & $8.93\times10^{-1}$ & $1.43\times10^{-3}$ \\
        $+$T+C+R   & 165.0 & 7368  & $3.05\times10^{-5}$ & $1.19\times10^{-6}$ \\
        $+$T+C+F+R & 114.4 & 7368  & $8.96\times10^{-1}$ & $1.37\times10^{-3}$ \\
        \bottomrule
    \end{tabular}
    \label{tab:toy_infer_ablation}
\end{table}

The largest gains come from restructuring the sparse execution path itself:
Triton and \texttt{torch.\allowbreak compile} together take latency
$363.4\!\rightarrow\!141.1$\,ms and peak memory $66.3\!\rightarrow\!15.0$\,GiB
(Table~\ref{tab:toy_infer_ablation}), consistent with fusion removing both
execution overhead and intermediate materialization. TF32 and recomputation are
complementary but non-equivalent \emph{optional} modes: TF32 buys $1.47\times$
throughput at a $\sim\!3\times10^4\times$ larger energy error
($3.05\times10^{-5}\!\rightarrow\!8.93\times10^{-1}$\,eV) and $6.5\times$ the
NVE drift rate---but not the temperature, and not the water or solvation
structure, whose first peaks hold to $0.5\%$ while the ion--ion pairs move on
the scale of the scatter between two halves of one run
(Appendix~\ref{sec:app-protocol}); recomputation buys memory headroom at
moderate latency cost and leaves that error untouched.

The attention kernel itself we isolate against a vanilla sparse baseline and
dense SDPA in Appendix~\ref{sec:app-kernels}. What that comparison can mean is
fixed by the workload: molecular attention's MFU ceiling is set by sparsity,
not the GPU's dense peak, so the informative comparison is \emph{within} sparse
implementations.

\subsection{Neighbor Search at Scale}
\label{sec:neighbor-scaling}

\textbf{Setup.} This stage is the design's primary scaling risk. We measure
curve, exchange granularity, and cache, separately and composed, from 64 to 512
GPUs at fixed atoms per GPU ($12.8$M atoms at $25$k/GPU, $25.6$M at $50$k),
$\rho=0.1$\,atoms\,\AA$^{-3}$, a common $10$k query-chunk limit. Reported
\textbf{neighbor-pipeline} time excludes the partition and the forward/backward
pass, so these reductions do not transfer one-to-one to a full step
(Section~\ref{sec:e2e_scaling} measures that; protocol in
Appendix~\ref{sec:app-protocol}).

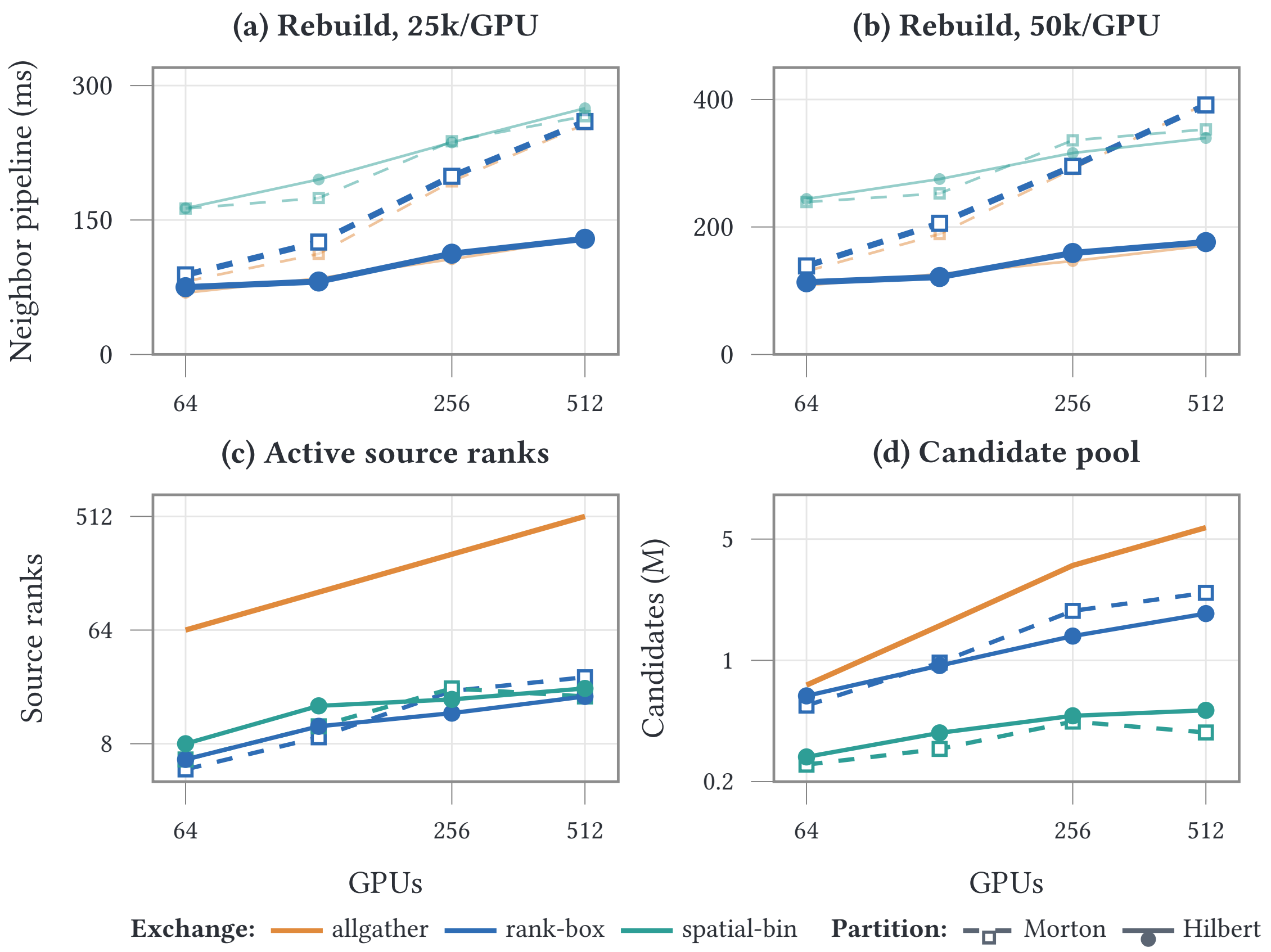
\begin{figure}[t]
    \centering
    \definecolor{cblue}{HTML}{2F6DB5}
\definecolor{cteal}{HTML}{2E9E96}
\definecolor{corange}{HTML}{E08A3C}
\definecolor{cslate}{HTML}{5B6573}
\definecolor{cink}{HTML}{2B2F36}

\providecommand{\nbkeyline}[1]{  \tikz[baseline=-0.55ex]{\draw[#1, line width=0.9pt] (0,0) -- (0.34,0);}}
\providecommand{\nbkeymark}[2]{  \tikz[baseline=-0.55ex]{\draw[#1, line width=0.9pt] (0,0) -- (0.34,0);
    \node[#2] at (0.17,0) {};}}

\pgfplotsset{
  nbsty2/.style={
    width=3.05cm, height=1.88cm, scale only axis,
    scaled ticks=false,
    tick align=outside, tick pos=left,
    grid=major, grid style={black!10, line width=0.3pt},
    axis line style={black!45, line width=0.5pt},
    label style={font=\scriptsize, text=cink},
    tick label style={font=\tiny, text=cink},
    title style={font=\scriptsize\bfseries, text=cink, yshift=-4.5pt},
    every axis plot/.append style={line width=0.8pt, mark size=1.2pt},
    xmode=log, log basis x=2, xmin=54, xmax=609,
    xtick={64,256,512}, xticklabels={64,256,512},
  },
  nbmorton/.style={dashed, mark=square*, mark options={solid, fill=white, line width=0.6pt}},
  nbhilbert/.style={solid, mark=*, mark options={solid}},
  nbprimary/.style={line width=1.15pt, mark size=1.3pt},
  nbref/.style={opacity=0.5, line width=0.5pt, mark size=0.85pt},
}

\resizebox{\linewidth}{!}{\begin{tikzpicture}
\begin{groupplot}[
    group style={group size=2 by 2, horizontal sep=1.02cm, vertical sep=0.92cm},
    nbsty2,
  ]

\nextgroupplot[
    title={(a) Rebuild, 25k/GPU},
    ylabel={Neighbor pipeline (ms)},
    ymin=0, ymax=320, ytick={0,150,300},
  ]
\addplot[corange, nbmorton, nbref] coordinates {(64,81.376)(128,112.266)(256,193.101)(512,256.891)};
\addplot[cblue,   nbmorton, nbprimary] coordinates {(64,88.767)(128,125.397)(256,198.790)(512,259.826)};
\addplot[cteal,   nbmorton, nbref] coordinates {(64,162.773)(128,174.541)(256,237.850)(512,265.802)};
\addplot[corange, nbhilbert, nbref] coordinates {(64,68.806)(128,84.975)(256,106.277)(512,129.735)};
\addplot[cblue,   nbhilbert, nbprimary] coordinates {(64,75.037)(128,81.376)(256,112.404)(512,129.025)};
\addplot[cteal,   nbhilbert, nbref] coordinates {(64,163.181)(128,195.254)(256,236.570)(512,274.669)};

\nextgroupplot[
    title={(b) Rebuild, 50k/GPU},
    ymin=0, ymax=450, ytick={0,200,400},
  ]
\addplot[corange, nbmorton, nbref] coordinates {(64,130.532)(128,188.927)(256,291.030)(512,393.641)};
\addplot[cblue,   nbmorton, nbprimary] coordinates {(64,138.871)(128,205.755)(256,295.227)(512,391.468)};
\addplot[cteal,   nbmorton, nbref] coordinates {(64,239.115)(128,252.648)(256,335.962)(512,353.105)};
\addplot[corange, nbhilbert, nbref] coordinates {(64,108.376)(128,126.253)(256,146.522)(512,170.845)};
\addplot[cblue,   nbhilbert, nbprimary] coordinates {(64,113.236)(128,121.480)(256,159.195)(512,176.276)};
\addplot[cteal,   nbhilbert, nbref] coordinates {(64,243.982)(128,275.163)(256,316.080)(512,339.495)};

\nextgroupplot[
    title={(c) Active source ranks},
    xlabel={GPUs}, ylabel={Source ranks},
    ymode=log, log basis y=2, ymin=4, ymax=760,
    ytick={8,64,512}, yticklabels={8,64,512},
  ]
\addplot[corange, solid, mark=none, line width=1.0pt]
  coordinates {(64,64)(128,128)(256,256)(512,512)};
\addplot[cblue,   nbmorton] coordinates {(64,5)(128,9)(256,21)(512,27)};
\addplot[cteal,   nbmorton] coordinates {(64,6)(128,11)(256,22)(512,19)};
\addplot[cblue,   nbhilbert] coordinates {(64,6)(128,11)(256,14)(512,19)};
\addplot[cteal,   nbhilbert] coordinates {(64,8)(128,16)(256,18)(512,22)};

\nextgroupplot[
    title={(d) Candidate pool},
    xlabel={GPUs}, ylabel={Candidates (M)},
    ymode=log, ymin=0.2, ymax=9, log ticks with fixed point,
    ytick={0.2,1,5}, yticklabels={0.2,1,5},
  ]
\addplot[corange, solid, mark=none, line width=1.0pt]
  coordinates {(64,0.7200)(128,1.5827)(256,3.5234)(512,5.8243)};
\addplot[cblue,   nbmorton] coordinates {(64,0.5485)(128,0.9707)(256,1.9292)(512,2.4502)};
\addplot[cteal,   nbmorton] coordinates {(64,0.2506)(128,0.3087)(256,0.4436)(512,0.3841)};
\addplot[cblue,   nbhilbert] coordinates {(64,0.6240)(128,0.9333)(256,1.3801)(512,1.8605)};
\addplot[cteal,   nbhilbert] coordinates {(64,0.2776)(128,0.3820)(256,0.4791)(512,0.5151)};

\end{groupplot}

\node[anchor=north, yshift=-0.78cm, text=cink]
  at ($(group c1r2.south)!0.5!(group c2r2.south)$) {  \tiny
  \textbf{Exchange:} \nbkeyline{corange} allgather
                     \nbkeyline{cblue} rank-box
                     \nbkeyline{cteal} spatial-bin
  \quad
  \textbf{Partition:} \nbkeymark{cslate,dashed}{draw=cslate, fill=white, rectangle, inner sep=1.0pt, line width=0.6pt} Morton
                     \nbkeymark{cslate}{fill=cslate, circle, inner sep=1.0pt} Hilbert};
\end{tikzpicture}}
    \caption{\textbf{Weak-scaling ablation of the distributed neighbor search on H100}, rebuilding every step; color is the Level-1 exchange scheme, line style the partition curve. \textbf{(a,b)} Neighbor-pipeline latency at $25$k/$50$k atoms per GPU---Morton bends upward ($2.8$--$2.9\times$ across the two panels), Hilbert rises gently ($1.6$--$1.7\times$); rank-box (default) is bold, allgather (within $11\%$ of it throughout) is lightened for clutter, spatial-bin for being the slowest scheme on Hilbert instead, as detailed next. \textbf{(c)} Source ranks requested by the heaviest rank, all schemes at equal weight (all-gather $=K$ by construction, drawn once for (c,d)). \textbf{(d)} Candidate population: spatial-bin prunes hardest---$2.6$--$11.3\times$ below all-gather, $2.2$--$3.6\times$ below rank-box---yet is slowest on Hilbert at every scale: volume alone does not predict cost.}
    \label{fig:neighbor_ablation}
\end{figure}

\textbf{The curve sets the slope.}
At identical per-rank atom counts and rank-box granularity, neighbor cost
grows $2.8$--$2.9\times$ under Morton against $1.6$--$1.7\times$ under Hilbert
over the range (Fig.~\ref{fig:neighbor_ablation}a,b)---the slope, not the
constant, separates a stage that scales from one that does not. Panels (c,d)
give the mechanism: a fragmented Z-order shard's box encloses the gaps between
its pieces, so Morton's peer degree and candidate pool keep climbing fast---$5$
to $27$ source ranks over $64$--$512$ GPUs at $25$k atoms/GPU---where Hilbert's
connected shard rises from $6$ only to $19$
(Section~\ref{sec:morton_partition}). Both grow; the exponents differ, near
$0.5$ for Hilbert against a steeper Morton. Some of the
penalty lands downstream of the exchange: handed \emph{identical} all-gather
pools, Morton is still $2.0\times$ slower in the exact search itself.

\textbf{Pruning hardest loses.}
Spatial-bin holds the smallest candidate pool (panel d) yet runs $1.9$--$2.4\times$
slower than rank-box on Hilbert at every scale, because pruning must itself be
published. All-gather fails at the other extreme: latency indistinguishable
from rank-box's, but a pool growing with the global atom count and no sparse
peer set for Section~\ref{sec:reduce_volume}'s plan to reuse.
Appendix~\ref{sec:app-neighbor} accounts for both. We adopt \textbf{Hilbert
$+$ rank-box}, which at this scale costs $3\%$ against all-gather rather than
saving; the two differ by about $9\%$ at most across the Hilbert points, a margin we do not
treat as separable given the single run per point, and rank-box is the default
for its small, fixed-size metadata and reusable peer set, not for latency.

\begin{figure*}[t]
    \centering
    \definecolor{cblue}{HTML}{2F6DB5}
\definecolor{cteal}{HTML}{2E9E96}
\definecolor{corange}{HTML}{E08A3C}
\definecolor{cslate}{HTML}{5B6573}
\definecolor{cink}{HTML}{2B2F36}
\definecolor{cgreen}{HTML}{6DA34D}

\providecommand{\mkey}[2]{  \tikz[baseline=-0.55ex]{\draw[#1, line width=0.9pt] (0,0) -- (0.34,0);
    \node[#2] at (0.17,0) {};}}

\ifdefined\ifarxiv
  \def\coePanelW{3.95cm}  \def\coePanelH{2.09cm}  \pgfplotsset{/pgfplots/group/coegroup/.style={      group size=3 by 3, horizontal sep=1.1cm, vertical sep=1.15cm}}  \def\coeLegendAt{($(group c1r3.north east)+(0.62cm,0cm)$)}\else
  \def\coePanelW{3.35cm}  \def\coePanelH{1.99cm}  \pgfplotsset{/pgfplots/group/coegroup/.style={      group size=4 by 2, horizontal sep=1.1cm, vertical sep=1.15cm}}  \def\coeLegendAt{($(group c3r2.north east)+(0.68cm,0.25cm)$)}\fi

\pgfplotsset{
  msty/.style={
    width=\coePanelW, height=\coePanelH, scale only axis,
    tick align=outside, tick pos=left,
    grid=major, grid style={black!10, line width=0.3pt},
    axis line style={black!45, line width=0.5pt},
    label style={font=\scriptsize, text=cink},
    tick label style={font=\tiny, text=cink},
    title style={font=\scriptsize\bfseries, text=cink, yshift=-5pt},
    every axis plot/.append style={line width=0.95pt, mark size=1.3pt},
  },
}

\begin{tikzpicture}
\begin{groupplot}[
    group style={coegroup},
    msty,
  ]

\nextgroupplot[
    title={(a) Graph cache, 50k/GPU},
    ylabel={Pipeline (ms/step)},
    xmode=log, log basis x=2, xmin=54, xmax=609,
    xtick={64,256,512}, xticklabels={64,256,512},
    ymode=log, ymin=3, ymax=1000, log ticks with fixed point,
    ytick={3,30,300}, yticklabels={3,30,300},
  ]
\addplot[cslate, dashed, mark=square*, mark options={solid, fill=white}]
  coordinates {(64,130.532)(128,188.927)(256,291.030)(512,393.641)};
\addplot[cblue, mark=*]
  coordinates {(64,113.236)(128,121.480)(256,159.195)(512,176.276)};
\addplot[cgreen, mark=triangle*, mark size=1.9pt]
  coordinates {(64,15.018)(128,16.295)(256,20.936)(512,25.066)};
\addplot[corange, densely dotted, mark=diamond*, mark size=1.9pt]
  coordinates {(64,4.106)(128,4.608)(256,5.574)(512,8.264)};

\nextgroupplot[
    title={(b) Cumulative, 512 GPUs},
    ylabel={ms/step},
    ybar, bar width=9pt,
    xmin=0.45, xmax=4.55, ymin=0, ymax=460,
    ytick={0,200,400},
    xtick={1,2,3,4},
    xticklabels={{\tiny Mor.\\a-g},{\tiny Hil.\\a-g},{\tiny Hil.\\r-b},{\tiny $+$c.}},
    xticklabel style={font=\tiny, align=center, text=cink, yshift=-1pt},
    enlarge x limits=false,
  ]
\addplot[draw=cslate!70, fill=cslate!30, bar shift=0pt]         coordinates {(1,393.641)};
\addplot[draw=corange!80!black, fill=corange!45, bar shift=0pt] coordinates {(2,170.845)};
\addplot[draw=cblue!80!black, fill=cblue!40, bar shift=0pt]     coordinates {(3,176.276)};
\addplot[draw=cgreen!80!black, fill=cgreen!55, bar shift=0pt]   coordinates {(4,25.066)};
\node[anchor=south, font=\tiny\bfseries, text=cgreen!58!black, inner sep=1pt]
  at (axis cs:3.78,58) {$15.7\times$};

\nextgroupplot[
    title={(c) Overlap on vs.\ off},
    ylabel={Full step (s)},
    xmode=log, log basis x=2, xmin=27.2, xmax=609,
    xtick={32,128,512}, xticklabels={32,128,512},
    ymin=0.5, ymax=0.78, ytick={0.5,0.7}, yticklabels={0.5,0.7},
  ]
\addplot[cslate, dashed, mark=square*, mark options={solid, fill=white}]
  coordinates {(32,0.5395)(64,0.6093)(128,0.6326)(256,0.6597)(512,0.7403)};
\addplot[cblue, mark=*]
  coordinates {(32,0.5361)(64,0.5504)(128,0.5743)(256,0.5989)(512,0.6328)};
\node[anchor=east, font=\tiny\bfseries, text=cblue!80!black, inner sep=1pt,
      fill=white, fill opacity=0.92, text opacity=1]
  at (axis cs:512,0.66) {$-14.5\%$};

\nextgroupplot[
    title={(d) Reduction by step type},
    ylabel={Reduction (\%)},
    ybar, bar width=4.5pt,
    symbolic x coords={32,64,128,256,512},
    xtick=data, xticklabel style={font=\tiny},
    ymin=0, ymax=20, ytick={0,10,20},
    enlarge x limits=0.14,
  ]
\addplot[draw=cblue!80!black, fill=cblue!40]   coordinates {(32,0.77)(64,10.3)(128,10.3)(256,10.9)(512,16.5)};
\addplot[draw=corange!80!black, fill=corange!45] coordinates {(32,0.0)(64,6.2)(128,4.0)(256,2.0)(512,5.9)};

\nextgroupplot[
    title={(e) End-to-end step time},
    ylabel={Full step (s)},
    xmode=log, log basis x=2, xmin=27.2, xmax=609,
    xtick={32,128,512}, xticklabels={32,128,512},
    ymode=log, ymin=0.2, ymax=2.0, log ticks with fixed point,
    ytick={0.2,1,2}, yticklabels={0.2,1,2},
  ]
\addplot[corange, mark=diamond*, mark size=1.9pt]
  coordinates {(32,0.2289)(64,0.2371)(128,0.2466)(256,0.2682)(512,0.2963)};
\addplot[cblue, mark=*, mark size=1.3pt]
  coordinates {(32,0.5361)(64,0.5504)(128,0.5743)(256,0.5989)(512,0.6328)};
\addplot[cgreen, mark=triangle*, mark size=1.9pt]
  coordinates {(32,1.5111)(64,1.5458)(128,1.5857)(256,1.6077)(512,1.6837)};

\nextgroupplot[
    title={(f) Weak-scaling efficiency},
    ylabel={Efficiency (\%)},
    xmode=log, log basis x=2, xmin=27.2, xmax=609,
    xtick={32,128,512}, xticklabels={32,128,512},
    ymin=68, ymax=104, ytick={70,100}, yticklabels={70,100},
  ]
\addplot[black!35, line width=0.5pt, forget plot, mark=none]
  coordinates {(32,100)(512,100)};
\addplot[corange, mark=diamond*, mark size=1.9pt]
  coordinates {(32,100)(64,96.51)(128,92.82)(256,85.34)(512,77.25)};
\addplot[cblue, mark=*, mark size=1.3pt]
  coordinates {(32,100)(64,97.41)(128,93.35)(256,89.51)(512,84.72)};
\addplot[cgreen, mark=triangle*, mark size=1.9pt]
  coordinates {(32,100)(64,97.75)(128,95.30)(256,93.99)(512,89.75)};

\nextgroupplot[
    title={(g) Aggregate throughput},
    ylabel={M atoms/s},
    xmode=log, log basis x=2, xmin=27.2, xmax=609,
    xtick={32,128,512}, xticklabels={32,128,512},
    ymode=log, ymin=1.1, ymax=26, log ticks with fixed point,
    ytick={2,10}, yticklabels={2,10},
  ]
\addplot[corange!45, densely dashed, line width=0.5pt, forget plot, mark=none]
  coordinates {(32,1.3982)(64,2.7964)(128,5.5928)(256,11.1856)(512,22.3712)};
\addplot[cblue!45, densely dashed, line width=0.5pt, forget plot, mark=none]
  coordinates {(32,1.4924)(64,2.9848)(128,5.9696)(256,11.9392)(512,23.8784)};
\addplot[cgreen!45, densely dashed, line width=0.5pt, forget plot, mark=none]
  coordinates {(32,1.2706)(64,2.5412)(128,5.0824)(256,10.1648)(512,20.3296)};
\addplot[corange, mark=diamond*, mark size=1.9pt]
  coordinates {(32,1.3982)(64,2.6988)(128,5.1913)(256,9.5460)(512,17.2819)};
\addplot[cblue, mark=*, mark size=1.3pt]
  coordinates {(32,1.4924)(64,2.9074)(128,5.5728)(256,10.6857)(512,20.2283)};
\addplot[cgreen, mark=triangle*, mark size=1.9pt]
  coordinates {(32,1.2706)(64,2.4843)(128,4.8435)(256,9.5534)(512,18.2463)};

\end{groupplot}

\node[anchor=north west, text=cink, font=\tiny, inner sep=0pt]
  at \coeLegendAt {  \renewcommand{\arraystretch}{1.0}
  \begin{tabular}{@{}l@{\;}l@{\quad}l@{\;}l@{}}
    \multicolumn{4}{@{}l@{}}{\textbf{(a,b)}} \\
    \mkey{cslate,dashed}{draw=cslate,fill=white,rectangle,inner sep=0.9pt,line width=0.6pt} & Mort.$+$a.g. &
      \mkey{cblue}{fill=cblue,circle,inner sep=0.9pt} & Hil.$+$r.b. \\
    \mkey{cgreen}{fill=cgreen,regular polygon,regular polygon sides=3,inner sep=0.8pt} & $+$cache &
      \mkey{corange,densely dotted}{fill=corange,diamond,inner sep=0.8pt} & hit-only \\
    \noalign{\vskip 1pt}\hline\noalign{\vskip 1.5pt}
    \multicolumn{4}{@{}l@{}}{\textbf{(c,d)}} \\
    \mkey{cslate,dashed}{draw=cslate,fill=white,rectangle,inner sep=0.9pt,line width=0.6pt} & off &
      \mkey{cblue}{fill=cblue,circle,inner sep=0.9pt} & on \\
    \tikz{\draw[draw=cblue!80!black,fill=cblue!40](0,-0.04)rectangle(0.16,0.12);} & hit &
      \tikz{\draw[draw=corange!80!black,fill=corange!45](0,-0.04)rectangle(0.16,0.12);} & rebuild \\
    \noalign{\vskip 1pt}\hline\noalign{\vskip 1.5pt}
    \multicolumn{4}{@{}l@{}}{\textbf{(e-g)}} \\
    \mkey{corange}{fill=corange,diamond,inner sep=0.8pt} & 10k &
      \mkey{cblue}{fill=cblue,circle,inner sep=0.9pt} & 25k \\
    \mkey{cgreen}{fill=cgreen,regular polygon,regular polygon sides=3,inner sep=0.8pt} & 60k$^*$ \\
    \multicolumn{4}{@{}p{2.55cm}@{}}{$^*$recompute on (HBM limit)} \\
  \end{tabular}};
\end{tikzpicture}
    \caption{H100 results across varying GPU counts. \textbf{(a,b) Neighbor-graph cache}: (a) per-step cost at $50$k atoms/GPU---cache hits (orange) an order of magnitude cheaper than rebuilds; (b) cumulative reduction composing curve $+$ exchange $+$ cache at 512 GPUs, the cache bar an upper estimate (Appendix~\ref{sec:app-protocol}). \textbf{(c,d) Overlap} on the full step at $25$k atoms/GPU: (c) per-step time on vs.\ off, gap widening with scale; (d) reduction by step type, larger on cache hits. \textbf{(e-g) End-to-end weak scaling}, 32--512 GPUs, three per-GPU loads: (e) step time, rising $1.11$--$1.29\times$ over $16\times$; (f) parallel efficiency vs.\ 32 GPUs; (g) aggregate throughput, each load against its own $32$-GPU-anchored linear reference (dashed)---a curve's gap to its line is (f)'s efficiency. Heaviest load: $30.7$M atoms on 512 GPUs. $^*$At $60$k atoms/GPU (e-g), recomputation is on since the load exceeds the $80$\,GB HBM budget otherwise; $10$k/$25$k run without it.}
    \label{fig:neighbor_cache}
    \label{fig:overlap_scaling}
    \label{fig:e2e_weak_scaling}
\end{figure*}
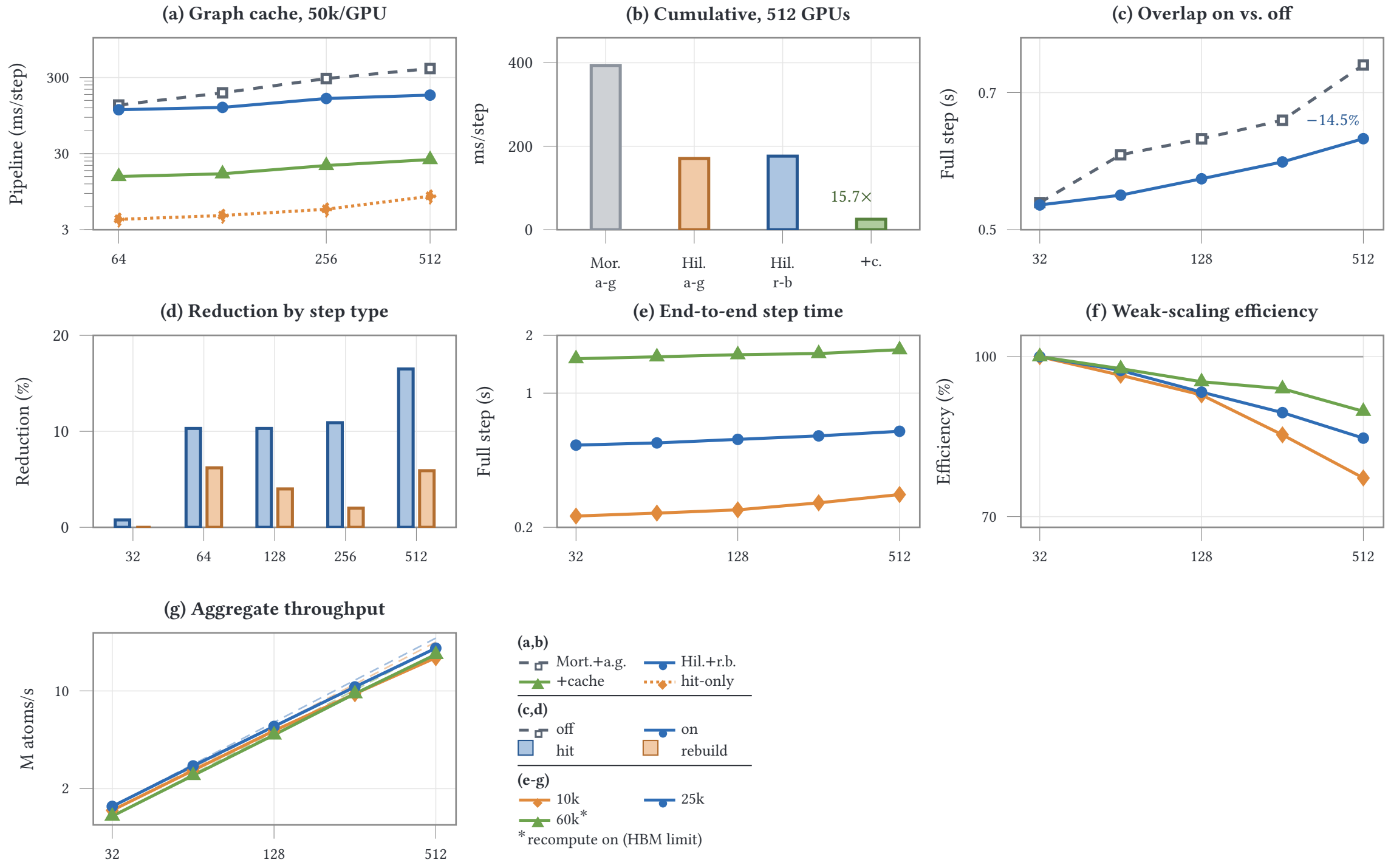

\textbf{The cache, and the three composed.}
A hit costs an order of magnitude less than a rebuild
(Fig.~\ref{fig:neighbor_cache}a, interval $1$ against $10$), and is worth
\emph{more} at larger scale: it skips plan construction, metadata collectives,
and candidate transport, exactly the part that grows with $K$, while the
geometry refresh it retains reuses the fixed halo plan and recomputes distances
locally. The reuse is exact: the interval-$1$ and interval-$10$ runs give bitwise-identical topology hashes and forces agreeing to $7.2\times10^{-7}$\,eV/\AA{} (Section~\ref{sec:multilevel_search}). Composed at
512 GPUs and $25.6$M atoms (Fig.~\ref{fig:neighbor_cache}b), a
Morton$+$all-gather baseline of $393.6$\,ms falls to $170.8$\,ms on the curve,
holds at $176.3$\,ms on granularity (kept for footprint, not latency), and
reaches $\mathbf{25.1}$\,ms with the cache, $15.7\times$ overall---that last
figure composed from separately measured rebuild and hit steps rather than
timed on one interval-$10$ run (Appendix~\ref{sec:app-protocol}). Again the
slope matters as much: the baseline grows $3.0\times$ over the range, capping
weak-scaling efficiency at $33\%$ from this stage alone, against $1.7\times$
and $60\%$ for the full configuration, at an absolute cost small enough that
the stage no longer limits the system. That $1.7\times$ is not flat, and the
telemetry names the next target: the metadata all-gather, still rising with
$K$. Table~\ref{tab:neighbor_composition} gives the per-mechanism numbers
behind the composition, including each row's own $64$-to-$512$-GPU growth---a
scaling behavior Fig.~\ref{fig:neighbor_cache}(b) does not show, since it plots
only the 512-GPU endpoint.

\begin{table}[t]
    \centering
    \caption{Composing the three neighbor-search mechanisms (a-g=all-gather, r-b=rank-box) at 512 GPUs, $50$k atoms/GPU; each row adds one mechanism to the row above. ``Growth'' is the 512-/64-GPU cost ratio, so $1.0$ is ideal weak scaling. Cached rows are a composed interval-$10$ amortized cost, not a timed interval-$10$ run; the composition and its known bias are stated below. Neighbor pipeline only. Rank-box costs $3\%$ against all-gather at this scale rather than saving: the two differ by under $9\%$ at every Hilbert point in the observed runs, a margin we do not treat as separable given the single run per point (Section~\ref{sec:neighbor-scaling}), and rank-box is the default for its small, fixed-size metadata and reusable peer set, not for latency. At $25$k atoms/GPU the baseline grows $3.2\times$ over the same range.}
    \label{tab:neighbor_composition}
    \small
    \begin{tabular}{llrrr}
        \toprule
        Partition & Exchange & ms & Reduc. & Growth \\
        \midrule
        Morton    & a-g (base) & $393.6$        & $1.0\times$           & $3.0\times$ \\
        Hilbert   & a-g        & $170.8$         & $2.3\times$           & $1.6\times$ \\
        Hilbert   & r-b        & $176.3$         & $2.2\times$           & $1.6\times$ \\
        Hilbert   & r-b$+$cache & $\mathbf{25.1}$ & $\mathbf{15.7\times}$ & $1.7\times$ \\
        \bottomrule
    \end{tabular}
\end{table}

\textbf{How much weight the composed figure carries.} The cached row is
$(t_{\mathrm{rebuild}}+9\,t_{\mathrm{hit}})/10$, where $t_{\mathrm{hit}}$ comes
from an interval-$10$ run and $t_{\mathrm{rebuild}}$ from a \emph{separate}
interval-$1$ run of the same configuration. They are not interchangeable: the
interval-$1$ arm searches at each layer's bare cutoff, while an interval-$10$
rebuild builds the $r_c\!+\!\delta$ superset the following nine steps
consume---$1.33\times$ the candidate volume at the $5$\,\AA{} layers and
$1.20\times$ at the $8$\,\AA{} one---so $t_{\mathrm{rebuild}}$ is measured on a
strictly cheaper search than the one it stands in for, and these runs contain no
steady-state interval-$10$ rebuild to substitute. Where both quantities do exist
in one place---Table~\ref{tab:neighbor_cache_fullinfer}, whose rebuilds fall on
steps $10$ and $20$---a steady-state interval-$10$ rebuild runs
$0.98$--$1.62\times$ its interval-$1$ counterpart. The composition therefore
understates the amortized cost by up to $1.4\times$, and the composed
$15.7\times$ is an \emph{upper estimate} rather than a measured speedup. It does
not affect the cache-hit latencies of Fig.~\ref{fig:neighbor_cache}(a),
measured directly, or the full-step $1.87\times$ of
Section~\ref{sec:e2e_scaling}, an A/B on the complete pipeline.
Appendix~\ref{sec:app-protocol} carries the full accounting.

Measured directly on the full pipeline at $20$k atoms/GPU instead of composed,
the cache leaves the same $13$--$16\%$ of the rebuild-every-step cost, row by
row in Table~\ref{tab:neighbor_cache_fullinfer}. Each row's exchange scheme is
tuned for its own curve, so the two rows are \emph{not} a controlled
exchange-scheme comparison.

\begin{table}[t]
    \centering
    \caption{Neighbor-stage cost measured during full-inference-mode runs at $20$k atoms per GPU (the timer is the neighbor pipeline, not end-to-end step wall time), the largest per-GPU configuration in this cache-validation sweep that runs without activation recomputation. These runs execute the complete model---forward, autograd forces, \texttt{torch.compile}, TF32, and remote-communication overlap---over a 20-step window after step~0 holding exactly 2 interval-$10$ rebuilds and 18 hits, so the amortized column is measured, not composed---though that $2$:$18$ split makes it equal to $(\text{Rebuild}+9\times\text{Cache hit})/10$. Each row's candidate-exchange scheme is tuned for its own curve, so the two rows are \emph{not} a controlled exchange-scheme comparison; that remains the neighbor-only matrix of Fig.~\ref{fig:neighbor_ablation}. The load-bearing column is $t_{\mathrm{hit}}/t_{\mathrm{rebuild}}$, which licenses composing the amortized cost for Table~\ref{tab:neighbor_composition}; the last column is the resulting $12.6$--$16.0\%$ of the rebuild-every-step cost.}
    \small
    \begin{tabular}{lrrrrrr}
        \toprule
        Configuration          & GPUs & Rebuild (ms) & Cache hit (ms) & $t_{\mathrm{hit}}/t_{\mathrm{rebuild}}$ & Amortized (ms) & Amortized/rebuild \\
        \midrule
        Morton $+$ spatial-bin & 64   & $162.8$      & $4.8$          & $0.029$ & $20.6$ & $0.126$ \\
                               & 128  & $184.2$      & $5.6$         & $0.031$ & $23.5$ & $0.128$ \\
                               & 256  & $204.0$      & $8.0$          & $0.039$ & $27.6$ & $0.135$ \\
                               & 512  & $316.2$      & $14.4$         & $0.046$ & $44.6$ & $0.141$ \\
        \midrule
        Hilbert $+$ multi-box  & 64   & $91.5$       & $4.4$          & $0.048$ & $13.1$ & $0.143$ \\
                               & 128  & $122.4$      & $5.1$         & $0.041$ & $16.8$ & $0.137$ \\
                               & 256  & $182.8$      & $8.3$          & $0.045$ & $25.7$ & $0.141$ \\
                               & 512  & $230.0$      & $15.3$         & $0.067$ & $36.8$ & $0.160$ \\
        \bottomrule
    \end{tabular}
    \label{tab:neighbor_cache_fullinfer}
\end{table}

\subsection{Communication Optimization at Scale}
\label{sec:overlap-results}
What is the communication optimization of Section~\ref{sec:reduce_volume} worth
once embedded in the complete step? We run a controlled A/B on a single flag,
under the $25$k-atoms-per-GPU configuration of Section~\ref{sec:e2e_scaling}:
the two runs differ only in whether the remote exchange runs on a side stream
awaited late or synchronously before the forward pass. Because one flag
separates the arms, the difference is attributable to the schedule, and the A/B
inherits the end-to-end protocol of Section~\ref{sec:e2e_scaling}. Figure~\ref{fig:overlap_scaling}(c)
shows a result whose shape matters more than its magnitude: at 32 GPUs the flag
changes step time by under $1\%$, and beyond that the time it recovers grows
with the device count---the signature of a cost proportional to inter-node
traffic, not a fixed overhead. Because overlap acts on exactly the term that
grows, its effect on scaling exceeds its per-point size: it removes $52\%$ of
the scale-induced increase over the measured range, lifting 512-GPU efficiency
from $72.9\%$ to the $84.7\%$ of Section~\ref{sec:e2e_scaling}.
Neighbor-construction time stays close between the two arms at every scale,
confirming the flag only reschedules the exchange. Panel (d) resolves this by
step type: the reduction is roughly $3\times$ larger on cache-hit steps---almost
entirely model forward plus the exchange it depends on---than on rebuild steps,
which also carry graph construction the forward-pass overlap cannot hide; cache
and overlap are therefore complementary, not redundant, and rebuild remains the
costlier step type even with overlap on. Nsight timelines corroborate the
mechanism qualitatively, showing the layer-wise exchange largely masked behind
the local branch, but we do not count them as a second measurement: they are an
aggregate, run-level view on a different denominator than the per-step timers
the A/B compares.

\subsection{End-to-End Weak Scaling}
\label{sec:e2e_scaling}
Integrating all modules, we measure the full inference step from 32 to 512
GPUs---PBC expansion, neighbor construction, remote-atom exchange, the E2Former
forward pass, autograd forces---with \texttt{torch.\allowbreak compile}, graph
capture, TF32 and overlap enabled and the cache at interval $10$. We fix the
per-GPU atom count and grow the system with the device count at three loads
spanning the usable memory range: $10$k, $25$k, and $60$k atoms per GPU; the
$60$k load exceeds the $80$\,GB HBM budget under the otherwise-identical
settings, so activation recomputation (Section~\ref{sec:single_gpu}) is enabled
for it alone, trading the moderate latency cost characterized there for the
memory headroom this load needs. Every point is the mean over steps $300$--$399$
of a $500$-step run: the window excludes the \texttt{torch.\allowbreak compile}
first step and spans exactly ten rebuild periods at interval $10$, which makes
the mean the amortized per-step cost rather than a sample biased by where
rebuilds happen to land, and one official run per point means no error bars
(protocol in Appendix~\ref{sec:app-protocol}).

Figure~\ref{fig:e2e_weak_scaling}(e,f) shows that the integrated step is close
to flat under weak scaling, though not flat. Growing $16\times$ in device count
and in total system size, the per-step time rises $1.29\times$ at $10$k atoms
per GPU, $1.18\times$ at $25$k, and $1.11\times$ at $60$k, for parallel
efficiencies at 512 GPUs of $77.3\%$, $84.7\%$, and $89.7\%$ respectively. The
ordering is the
informative part: efficiency \emph{improves} with the per-GPU load, which is
the signature of a fixed residual overhead being amortized against a larger
local compute window rather than of a cost that grows with the device count.
One caveat on the $60$k point specifically: it is the only load carrying
activation recomputation, which enlarges the local compute window for a reason
unrelated to atom count, so it is not a like-for-like continuation of the
$10$k$\rightarrow$$25$k trend. The mechanism is cleanest between $10$k and
$25$k, which differ only in load. Aggregate throughput tracks each load's own
linear reference (Fig.~\ref{fig:e2e_weak_scaling}(g)). The largest
configuration is $60$k atoms per GPU on 512 GPUs---$30.7$M atoms at $1.684$\,s
per step, or $18.2$M atoms per second.

Two mechanisms can be attributed directly at this level, both measured as
controlled A/B pairs of the complete pipeline. Overlap's own contribution is
already quantified in Section~\ref{sec:overlap-results}; the second is the
neighbor cache. Rebuilding the graph every step instead of every tenth costs $1.87\times$ on the full step at 512 GPUs and $60$k atoms/GPU over steps $300$--$399$---smaller than the neighbor-only ratio of Section~\ref{sec:neighbor-scaling}, as expected: the forward/backward
compute that dominates a full step is identical whether or not that step
rebuilds, so embedding the stage-only gap in the full step can only shrink it,
never widen it. The same A/B at $25$k atoms/GPU costs $1.49\times$ over steps $300$--$399$: a lighter step spends proportionally less of itself in the neighbour stage, so the same rebuild cost buys a smaller saving. Both match the design argument: the equal-count cut removes
load-balancing stalls, the peer set grows sublinearly in device count, and the
overlap hides the residual latency---so the
compute-to-communication ratio degrades gracefully rather than collapsing into
the communication-bound regime that limits a naive distribution of long-range
Transformers. Reaching ns/day---against the $\approx\!8$\,ps/day of
Section~\ref{sec:density-heterogeneity}'s A100 run---is left to future work: it
depends on per-GPU memory headroom and fabric bandwidth at the target size,
neither of which this study varies.

\subsection{MD Trajectory Validation}
\label{sec:md-validation}
Every result above measures throughput and scaling on repeated, independent
steps, not whether a live trajectory behaves as designed once atoms move and
ownership churns. We close that gap on a separate $256$-GPU A100 cluster
($40$\,GB HBM2e, PCIe, no NVLink, swapped in for
Section~\ref{sec:exp-setup}'s H100/\allowbreak NVLink/\allowbreak InfiniBand
fabric for availability): a $150$\,mmol/L aqueous KCl solution, $10.24$M atoms
($40$k/GPU average), $0.5$\,fs steps, $20{,}000$ steps of NVT equilibration at
$300$\,K followed by $20{,}000$ of NVE at the same temperature to expose drift
a thermostat would mask---$10$\,ps per segment, TF32 off throughout, with
cache, sync and skin at Sections~\ref{sec:multilevel_search}
and~\ref{sec:md_engine}'s defaults (Appendix~\ref{sec:app-protocol}). The
trajectory rebuilds the neighbor graph every $10$ steps and re-partitions every
$100$, with the $0.5$\,\AA{} displacement skin triggering an early rebuild past
half that margin ($0.25$\,\AA{}).

\begin{figure}[t]
    \centering
    \definecolor{cblue}{HTML}{2F6DB5}
\definecolor{cslate}{HTML}{5B6573}
\definecolor{cink}{HTML}{2B2F36}

\providecommand{\mkyline}[1]{\tikz[baseline=-0.55ex]{\draw[#1, line width=1.1pt] (0,0) -- (0.36,0);}}

\input{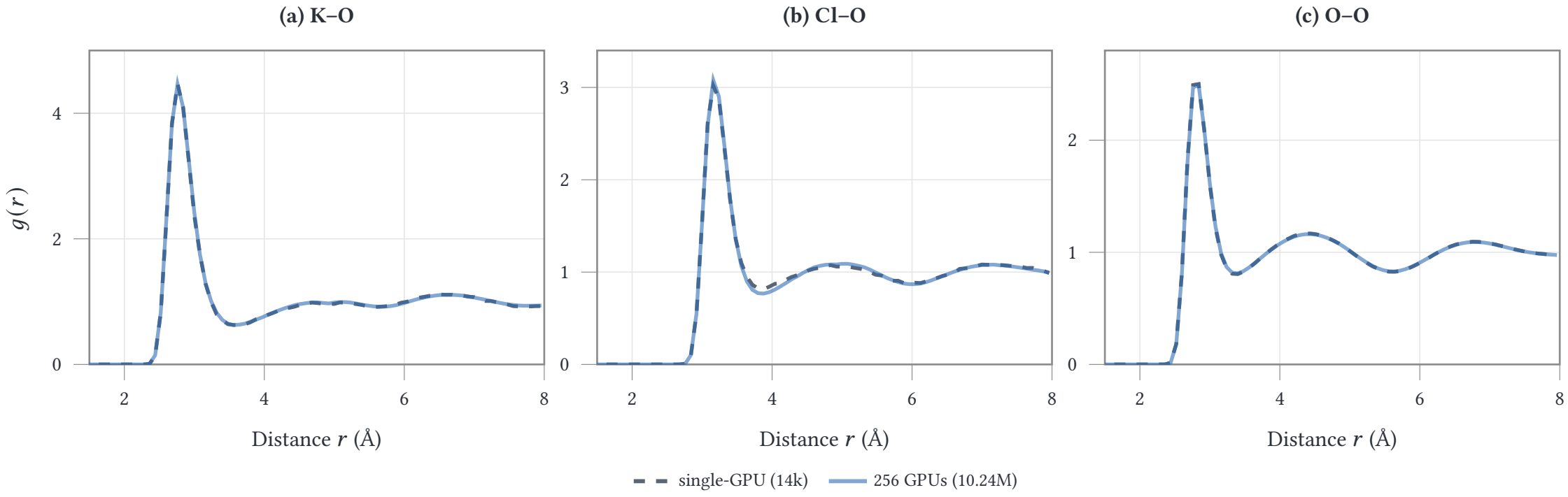}

\pgfplotsset{
  rdfsty/.style={
    width=\figKclPanelW, height=\figKclPanelH, scale only axis,
    tick align=outside, tick pos=left,
    grid=major, grid style={black!10, line width=0.3pt},
    axis line style={black!45, line width=0.5pt},
    label style={font=\scriptsize, text=cink},
    tick label style={font=\tiny, text=cink},
    title style={font=\scriptsize\bfseries, text=cink, yshift=-3pt},
    xlabel={Distance $r$ (\AA)}, xtick={2,4,6,8},
    xmin=1.5, xmax=8, ymin=0,
    every axis plot/.append style={line width=1.0pt},
  },
}

\begin{tikzpicture}
\begin{groupplot}[
    group style={group size=3 by 1, horizontal sep=0.5cm},
    rdfsty,
  ]

\nextgroupplot[title={(a) K--O}, ymax=5, ylabel={$g(r)$}]
\addplot[cslate, dashed, mark=none] coordinates {(0.04,0.0000)(0.12,0.0000)(0.20,0.0000)(0.28,0.0000)(0.36,0.0000)(0.44,0.0000)(0.52,0.0000)(0.60,0.0000)(0.68,0.0000)(0.76,0.0000)(0.84,0.0000)(0.92,0.0000)(1.00,0.0000)(1.08,0.0000)(1.16,0.0000)(1.24,0.0000)(1.32,0.0000)(1.40,0.0000)(1.48,0.0000)(1.56,0.0000)(1.64,0.0000)(1.72,0.0000)(1.80,0.0000)(1.88,0.0000)(1.96,0.0000)(2.04,0.0000)(2.12,0.0000)(2.20,0.0000)(2.28,0.0000)(2.36,0.0081)(2.44,0.1405)(2.52,0.8263)(2.60,2.2902)(2.68,3.8336)(2.76,4.4665)(2.84,4.0887)(2.92,3.2394)(3.00,2.3615)(3.08,1.7560)(3.16,1.2839)(3.24,0.9939)(3.32,0.7926)(3.40,0.7040)(3.48,0.6404)(3.56,0.6280)(3.64,0.6280)(3.72,0.6376)(3.80,0.6704)(3.88,0.7181)(3.96,0.7446)(4.04,0.7854)(4.12,0.8186)(4.20,0.8545)(4.28,0.8920)(4.36,0.9015)(4.44,0.9223)(4.52,0.9510)(4.60,0.9671)(4.68,0.9854)(4.76,0.9767)(4.84,0.9681)(4.92,0.9641)(5.00,0.9634)(5.08,0.9874)(5.16,0.9813)(5.24,0.9752)(5.32,0.9591)(5.40,0.9459)(5.48,0.9212)(5.56,0.9131)(5.64,0.9189)(5.72,0.9249)(5.80,0.9348)(5.88,0.9551)(5.96,0.9841)(6.04,0.9991)(6.12,1.0352)(6.20,1.0617)(6.28,1.0760)(6.36,1.0926)(6.44,1.1027)(6.52,1.1131)(6.60,1.1090)(6.68,1.1108)(6.76,1.0964)(6.84,1.0942)(6.92,1.0786)(7.00,1.0730)(7.08,1.0417)(7.16,1.0101)(7.24,1.0097)(7.32,0.9903)(7.40,0.9696)(7.48,0.9530)(7.56,0.9276)(7.64,0.9250)(7.72,0.9229)(7.80,0.9228)(7.88,0.9278)(7.96,0.9290)};
\addplot[cblue, opacity=0.6, mark=none] coordinates {(0.04,0.0000)(0.12,0.0000)(0.20,0.0000)(0.28,0.0000)(0.36,0.0000)(0.44,0.0000)(0.52,0.0000)(0.60,0.0000)(0.68,0.0000)(0.76,0.0000)(0.84,0.0000)(0.92,0.0000)(1.00,0.0000)(1.08,0.0000)(1.16,0.0000)(1.24,0.0000)(1.32,0.0000)(1.40,0.0000)(1.48,0.0000)(1.56,0.0000)(1.64,0.0000)(1.72,0.0000)(1.80,0.0000)(1.88,0.0000)(1.96,0.0000)(2.04,0.0000)(2.12,0.0000)(2.20,0.0000)(2.28,0.0002)(2.36,0.0101)(2.44,0.1461)(2.52,0.8279)(2.60,2.3070)(2.68,3.8371)(2.76,4.4624)(2.84,4.0911)(2.92,3.2607)(3.00,2.4158)(3.08,1.7509)(3.16,1.2979)(3.24,1.0070)(3.32,0.8198)(3.40,0.7084)(3.48,0.6496)(3.56,0.6278)(3.64,0.6328)(3.72,0.6486)(3.80,0.6789)(3.88,0.7121)(3.96,0.7453)(4.04,0.7843)(4.12,0.8223)(4.20,0.8572)(4.28,0.8903)(4.36,0.9187)(4.44,0.9457)(4.52,0.9628)(4.60,0.9760)(4.68,0.9861)(4.76,0.9888)(4.84,0.9789)(4.92,0.9712)(5.00,0.9850)(5.08,0.9936)(5.16,0.9926)(5.24,0.9843)(5.32,0.9690)(5.40,0.9501)(5.48,0.9368)(5.56,0.9251)(5.64,0.9173)(5.72,0.9197)(5.80,0.9277)(5.88,0.9430)(5.96,0.9668)(6.04,0.9913)(6.12,1.0183)(6.20,1.0436)(6.28,1.0650)(6.36,1.0832)(6.44,1.0960)(6.52,1.1063)(6.60,1.1108)(6.68,1.1088)(6.76,1.1033)(6.84,1.0959)(6.92,1.0821)(7.00,1.0649)(7.08,1.0472)(7.16,1.0295)(7.24,1.0109)(7.32,0.9906)(7.40,0.9729)(7.48,0.9574)(7.56,0.9463)(7.64,0.9373)(7.72,0.9343)(7.80,0.9354)(7.88,0.9361)(7.96,0.9380)};

\nextgroupplot[title={(b) Cl--O}, ymax=3.4]
\addplot[cslate, dashed, mark=none] coordinates {(0.04,0.0000)(0.12,0.0000)(0.20,0.0000)(0.28,0.0000)(0.36,0.0000)(0.44,0.0000)(0.52,0.0000)(0.60,0.0000)(0.68,0.0000)(0.76,0.0000)(0.84,0.0000)(0.92,0.0000)(1.00,0.0000)(1.08,0.0000)(1.16,0.0000)(1.24,0.0000)(1.32,0.0000)(1.40,0.0000)(1.48,0.0000)(1.56,0.0000)(1.64,0.0000)(1.72,0.0000)(1.80,0.0000)(1.88,0.0000)(1.96,0.0000)(2.04,0.0000)(2.12,0.0000)(2.20,0.0000)(2.28,0.0000)(2.36,0.0000)(2.44,0.0000)(2.52,0.0000)(2.60,0.0000)(2.68,0.0006)(2.76,0.0078)(2.84,0.1013)(2.92,0.5510)(3.00,1.5489)(3.08,2.6220)(3.16,3.0177)(3.24,2.8743)(3.32,2.3360)(3.40,1.7838)(3.48,1.3583)(3.56,1.1004)(3.64,0.9484)(3.72,0.8614)(3.80,0.8238)(3.88,0.8158)(3.96,0.8415)(4.04,0.8729)(4.12,0.8783)(4.20,0.9176)(4.28,0.9470)(4.36,0.9671)(4.44,0.9823)(4.52,1.0183)(4.60,1.0355)(4.68,1.0461)(4.76,1.0688)(4.84,1.0745)(4.92,1.0596)(5.00,1.0591)(5.08,1.0594)(5.16,1.0483)(5.24,1.0380)(5.32,1.0270)(5.40,1.0014)(5.48,0.9709)(5.56,0.9634)(5.64,0.9362)(5.72,0.9057)(5.80,0.9068)(5.88,0.8824)(5.96,0.8806)(6.04,0.8850)(6.12,0.8815)(6.20,0.8990)(6.28,0.9275)(6.36,0.9365)(6.44,0.9595)(6.52,0.9737)(6.60,0.9914)(6.68,1.0346)(6.76,1.0408)(6.84,1.0454)(6.92,1.0670)(7.00,1.0818)(7.08,1.0796)(7.16,1.0752)(7.24,1.0780)(7.32,1.0741)(7.40,1.0674)(7.48,1.0670)(7.56,1.0579)(7.64,1.0441)(7.72,1.0455)(7.80,1.0270)(7.88,1.0134)(7.96,0.9909)};
\addplot[cblue, opacity=0.6, mark=none] coordinates {(0.04,0.0000)(0.12,0.0000)(0.20,0.0000)(0.28,0.0000)(0.36,0.0000)(0.44,0.0000)(0.52,0.0000)(0.60,0.0000)(0.68,0.0000)(0.76,0.0000)(0.84,0.0000)(0.92,0.0000)(1.00,0.0000)(1.08,0.0000)(1.16,0.0000)(1.24,0.0000)(1.32,0.0000)(1.40,0.0000)(1.48,0.0000)(1.56,0.0000)(1.64,0.0000)(1.72,0.0000)(1.80,0.0000)(1.88,0.0000)(1.96,0.0000)(2.04,0.0000)(2.12,0.0000)(2.20,0.0000)(2.28,0.0000)(2.36,0.0000)(2.44,0.0000)(2.52,0.0000)(2.60,0.0000)(2.68,0.0002)(2.76,0.0078)(2.84,0.1019)(2.92,0.5518)(3.00,1.5396)(3.08,2.5911)(3.16,3.0789)(3.24,2.8996)(3.32,2.3593)(3.40,1.7947)(3.48,1.3589)(3.56,1.0690)(3.64,0.8984)(3.72,0.8067)(3.80,0.7715)(3.88,0.7673)(3.96,0.7829)(4.04,0.8104)(4.12,0.8449)(4.20,0.8823)(4.28,0.9225)(4.36,0.9542)(4.44,0.9904)(4.52,1.0221)(4.60,1.0503)(4.68,1.0671)(4.76,1.0785)(4.84,1.0827)(4.92,1.0824)(5.00,1.0895)(5.08,1.0896)(5.16,1.0800)(5.24,1.0670)(5.32,1.0513)(5.40,1.0262)(5.48,0.9954)(5.56,0.9663)(5.64,0.9367)(5.72,0.9128)(5.80,0.8929)(5.88,0.8765)(5.96,0.8691)(6.04,0.8692)(6.12,0.8755)(6.20,0.8916)(6.28,0.9109)(6.36,0.9355)(6.44,0.9562)(6.52,0.9804)(6.60,1.0038)(6.68,1.0238)(6.76,1.0421)(6.84,1.0561)(6.92,1.0671)(7.00,1.0758)(7.08,1.0780)(7.16,1.0810)(7.24,1.0749)(7.32,1.0700)(7.40,1.0642)(7.48,1.0551)(7.56,1.0463)(7.64,1.0374)(7.72,1.0256)(7.80,1.0183)(7.88,1.0077)(7.96,0.9901)};

\nextgroupplot[title={(c) O--O}, ymax=2.8]
\addplot[cslate, dashed, mark=none] coordinates {(0.04,0.0000)(0.12,0.0000)(0.20,0.0000)(0.28,0.0000)(0.36,0.0000)(0.44,0.0000)(0.52,0.0000)(0.60,0.0000)(0.68,0.0000)(0.76,0.0000)(0.84,0.0000)(0.92,0.0000)(1.00,0.0000)(1.08,0.0000)(1.16,0.0000)(1.24,0.0000)(1.32,0.0000)(1.40,0.0000)(1.48,0.0000)(1.56,0.0000)(1.64,0.0000)(1.72,0.0000)(1.80,0.0000)(1.88,0.0000)(1.96,0.0000)(2.04,0.0000)(2.12,0.0000)(2.20,0.0000)(2.28,0.0000)(2.36,0.0004)(2.44,0.0155)(2.52,0.1829)(2.60,0.8166)(2.68,1.8185)(2.76,2.4921)(2.84,2.5010)(2.92,2.0887)(3.00,1.6022)(3.08,1.2182)(3.16,0.9808)(3.24,0.8571)(3.32,0.8085)(3.40,0.8055)(3.48,0.8276)(3.56,0.8617)(3.64,0.9008)(3.72,0.9444)(3.80,0.9874)(3.88,1.0266)(3.96,1.0633)(4.04,1.0965)(4.12,1.1210)(4.20,1.1454)(4.28,1.1593)(4.36,1.1674)(4.44,1.1679)(4.52,1.1611)(4.60,1.1494)(4.68,1.1299)(4.76,1.1064)(4.84,1.0800)(4.92,1.0474)(5.00,1.0109)(5.08,0.9743)(5.16,0.9387)(5.24,0.9068)(5.32,0.8762)(5.40,0.8530)(5.48,0.8361)(5.56,0.8268)(5.64,0.8260)(5.72,0.8323)(5.80,0.8472)(5.88,0.8676)(5.96,0.8920)(6.04,0.9195)(6.12,0.9477)(6.20,0.9778)(6.28,1.0061)(6.36,1.0313)(6.44,1.0525)(6.52,1.0699)(6.60,1.0822)(6.68,1.0901)(6.76,1.0947)(6.84,1.0931)(6.92,1.0884)(7.00,1.0794)(7.08,1.0700)(7.16,1.0575)(7.24,1.0450)(7.32,1.0317)(7.40,1.0201)(7.48,1.0094)(7.56,0.9999)(7.64,0.9919)(7.72,0.9876)(7.80,0.9829)(7.88,0.9791)(7.96,0.9756)};
\addplot[cblue, opacity=0.6, mark=none] coordinates {(0.04,0.0000)(0.12,0.0000)(0.20,0.0000)(0.28,0.0000)(0.36,0.0000)(0.44,0.0000)(0.52,0.0000)(0.60,0.0000)(0.68,0.0000)(0.76,0.0000)(0.84,0.0000)(0.92,0.0000)(1.00,0.0000)(1.08,0.0000)(1.16,0.0000)(1.24,0.0000)(1.32,0.0000)(1.40,0.0000)(1.48,0.0000)(1.56,0.0000)(1.64,0.0000)(1.72,0.0000)(1.80,0.0000)(1.88,0.0000)(1.96,0.0000)(2.04,0.0000)(2.12,0.0000)(2.20,0.0000)(2.28,0.0000)(2.36,0.0004)(2.44,0.0154)(2.52,0.1800)(2.60,0.8055)(2.68,1.7981)(2.76,2.4718)(2.84,2.4909)(2.92,2.0926)(3.00,1.6122)(3.08,1.2320)(3.16,0.9931)(3.24,0.8677)(3.32,0.8176)(3.40,0.8121)(3.48,0.8316)(3.56,0.8644)(3.64,0.9035)(3.72,0.9452)(3.80,0.9860)(3.88,1.0251)(3.96,1.0604)(4.04,1.0913)(4.12,1.1176)(4.20,1.1382)(4.28,1.1529)(4.36,1.1612)(4.44,1.1622)(4.52,1.1568)(4.60,1.1450)(4.68,1.1276)(4.76,1.1055)(4.84,1.0788)(4.92,1.0472)(5.00,1.0128)(5.08,0.9770)(5.16,0.9416)(5.24,0.9089)(5.32,0.8801)(5.40,0.8569)(5.48,0.8399)(5.56,0.8307)(5.64,0.8297)(5.72,0.8364)(5.80,0.8505)(5.88,0.8703)(5.96,0.8943)(6.04,0.9209)(6.12,0.9494)(6.20,0.9780)(6.28,1.0050)(6.36,1.0294)(6.44,1.0505)(6.52,1.0672)(6.60,1.0800)(6.68,1.0880)(6.76,1.0917)(6.84,1.0910)(6.92,1.0863)(7.00,1.0784)(7.08,1.0682)(7.16,1.0563)(7.24,1.0439)(7.32,1.0316)(7.40,1.0196)(7.48,1.0092)(7.56,1.0002)(7.64,0.9930)(7.72,0.9874)(7.80,0.9829)(7.88,0.9790)(7.96,0.9754)};
\end{groupplot}
\node[anchor=north, font=\tiny, text=cink, inner sep=1pt]
  at ($(group c1r1.south)!0.5!(group c3r1.south)+(0,-0.98cm)$)
  {\mkyline{cslate,dashed} single-GPU (14k)\quad\mkyline{cblue,opacity=0.6} 256 GPUs (10.24M)};
\end{tikzpicture}
    \caption{\textbf{Structural validation of the distributed trajectory.} K--O, Cl--O, and O--O radial distribution functions, $256$-GPU/$10.24$M-atom run (solid) vs.\ $14$k-atom single-GPU serial reference (dashed, Section~\ref{sec:md-validation}'s protocol at single-GPU scale), same concentration and temperature. First-shell peaks match the $0.08$\,\AA{} bin width in all three; peak heights agree to about $2\%$ (Cl--O, the widest gap) and $0.4\%$ for O--O, the bulk solvent structure.}
    \label{fig:kcl_rdf_compare}
\end{figure}

\textbf{Energy conservation and structure.}
Over the full $10$\,ps of NVE the per-atom drift rate holds steady rather than
growing, at $\approx\!0.42\,\mu$eV/atom/ps ($6\times10^{-9}$ of total energy, a
gauge-dependent ratio), and the final frame shows no broken water molecules and
no anomalously close atom pairs. K--O, Cl--O, and O--O radial distribution
functions agree with the $14$k-atom single-GPU serial reference within the bin
width in peak position and to about $2\%$ in peak height
(Fig.~\ref{fig:kcl_rdf_compare}), confirming the distributed path reproduces a
correct serial integrator's physics. Model accuracy is established separately
(Section~\ref{sec:exp-setup}); these are stability and structure checks on
partitioning, live rebuilding, and cross-rank force reduction, not a same-input
equivalence test.

\textbf{Cache and driver cost under real displacement.}
Section~\ref{sec:e2e_scaling}'s nine-in-ten hit rate comes from static
repetition---an upper bound, since a real trajectory can also trigger a rebuild
by displacement, and a global-sync event forces one outright. The measured hit
rate is $\approx\!79\%$ against that $90\%$ bound, and the $11$-point gap is
itself the accounting: a schedule alone rebuilds on one step in ten, so
displacement roughly doubles the rebuild rate---the two causes are comparable,
not lopsided. Only $1$--$2\%$ of rank-step records are a rank crossing its own
skin (Section~\ref{sec:multilevel_search}), but that counts which rank
\emph{initiated} a rebuild: the \texttt{all\allowbreak\_reduce(MAX)} makes the
decision unanimous, so one early crossing pulls every rank into the same
rebuild, and those induced records carry most of the cost
(Appendix~\ref{sec:app-protocol}). A global-sync event is a separate, larger
bundle: partition, broadcast, and the rebuild it unconditionally forces cost
$\approx\!87$--$88$\,ms for the partition-and-broadcast half
($\approx\!70.7$\,ms rank~$0$ serial quantize-and-sort,
$\approx\!16$--$18$\,ms cross-rank communication) plus
$\approx\!320$--$330$\,ms for the rebuild, $407$--$418$\,ms in total---what our
prior rank-$0$ implementation paid every step. The decentralized engine pays it
once every $100$ steps, roughly $4$\,ms per step against the $\approx\!4.55$\,s
full step (Section~\ref{sec:discussion} scopes this measurement).

\begin{figure}[t]
    \centering
    \definecolor{cblue}{HTML}{2F6DB5}
\definecolor{cteal}{HTML}{2E9E96}
\definecolor{corange}{HTML}{E08A3C}
\definecolor{cslate}{HTML}{5B6573}
\definecolor{cink}{HTML}{2B2F36}
\definecolor{cgreen}{HTML}{6DA34D}
\definecolor{cred}{HTML}{C0504D}

\providecommand{\mkey}[2]{  \tikz[baseline=-0.55ex]{\draw[#1, line width=0.9pt] (0,0) -- (0.34,0);
    \node[#2] at (0.17,0) {};}}

\pgfplotsset{
  ssty/.style={
    width=3.05cm, height=1.55cm, scale only axis,
    tick align=outside, tick pos=left,
    grid=major, grid style={black!10, line width=0.3pt},
    axis line style={black!45, line width=0.5pt},
    label style={font=\scriptsize, text=cink},
    tick label style={font=\tiny, text=cink},
    title style={font=\scriptsize\bfseries, text=cink, yshift=-4pt},
    xmode=log, log basis x=2, xmin=27, xmax=310,
    xtick={32,64,128,256}, xticklabels={32,64,128,256},
    every axis plot/.append style={line width=0.9pt, mark size=1.2pt},
  },
}

\resizebox{\linewidth}{!}{\begin{tikzpicture}
\begin{groupplot}[
    group style={group size=2 by 2, horizontal sep=1.32cm, vertical sep=0.86cm},
    ssty,
  ]

\nextgroupplot[
    title={(a) Strong-scaling step time},
    ylabel={Step (ms)},
    ymode=log, ymin=450, ymax=4500, log ticks with fixed point,
    ytick={500,1000,4000}, yticklabels={500,1000,4000},
  ]
\addplot[cblue, mark=*]
  coordinates {(32,3915.1)(64,1404.5)(96,1092.0)(128,829.3)(160,699.4)(192,706.2)(224,650.4)(256,546.3)};

\nextgroupplot[
    title={(b) Efficiency (ref.\ 64 GPUs)},
    ylabel={Efficiency (\%)},
    xmin=48, xmax=310, xtick={64,128,256}, xticklabels={64,128,256},
    ymin=40, ymax=112, ytick={50,75,100},
  ]
\addplot[black!35, line width=0.5pt, forget plot, mark=none]
  coordinates {(48,100)(310,100)};
\addplot[cblue, mark=*]
  coordinates {(64,100.0)(96,85.7)(128,84.7)(160,80.3)(192,66.3)(224,61.7)(256,64.3)};

\nextgroupplot[
    title={(c) Comm.\ \& rebuild cost},
    xlabel={GPUs (log)}, ylabel={ms / event},
    ymin=0, ymax=225, ytick={0,100,200},
  ]
\addplot[cteal, mark=square*]
  coordinates {(32,24.143)(64,25.938)(96,29.710)(128,32.076)(160,36.314)(192,38.424)(224,40.148)(256,44.081)};
\addplot[cred, mark=diamond*]
  coordinates {(32,205.31)(64,101.284)(96,92.903)(128,88.786)(160,89.033)(192,97.979)(224,107.949)(256,112.433)};

\nextgroupplot[
    title={(d) Fwd-labeled per atom},
    xlabel={GPUs (log)}, ylabel={$\mu$s / atom},
    ymin=52, ymax=122, ytick={60,80,100},
  ]
\addplot[cgreen!70!black, mark=triangle*, mark size=1.5pt]
  coordinates {(32,94.17)(64,68.48)(96,80.41)(128,82.79)(160,85.00)(192,101.22)(224,103.45)(256,104.38)};
\node[cslate, font=\tiny, anchor=south west] at (axis cs:33,93) {near-OOM};
\node[cred!85!black, font=\tiny, anchor=south east] at (axis cs:290,105) {under-utilized};

\end{groupplot}

\node[anchor=north, yshift=-0.60cm, text=cink]
  at ($(group c1r2.south)!0.5!(group c2r2.south)$) {  \tiny
  \textbf{(c)} \mkey{cteal}{fill=cteal,rectangle,inner sep=0.8pt} repartition\quad
  \mkey{cred}{fill=cred,diamond,inner sep=0.8pt} rebuild};
\end{tikzpicture}}
    \caption{\textbf{Strong scaling on a fixed $1.28$M-atom KCl-in-water system}, NVE, $2{,}000$ steps. \textbf{(a)} Step time, activation recomputation on (forward-pass detail and recomputation off are in Table~\ref{tab:strong-scaling-detail}, Appendix~\ref{sec:app-protocol}). \textbf{(b)} Efficiency$(K) = \frac{T(64)\cdot 64}{T(K)\cdot K}\times100\%$, referenced to $64$ GPUs, not $32$: that point sits near recomputation's own memory ceiling, allocator garbage collection inflating its step time. \textbf{(c)} Repartition (rank-$0$ sort $+$ broadcast) and rebuild cost per event. Amortized, repartition stays under $0.1\%$ of step time; neighbor-cache management overall (rebuild $+$ hit handling) runs $2.2\!\to\!6.8\%$ from $64$--$256$ GPUs---$4.6$ points against (b)'s $36$-point drop. \textbf{(d)} Per-atom forward time (utilization proxy, not profiled)---bottoms at $64$ GPUs' $20$k atoms/GPU, rising both ways, consistent with allocator pressure at $32$ and too small a batch at $256$.}
    \label{fig:strong_scaling}
\end{figure}
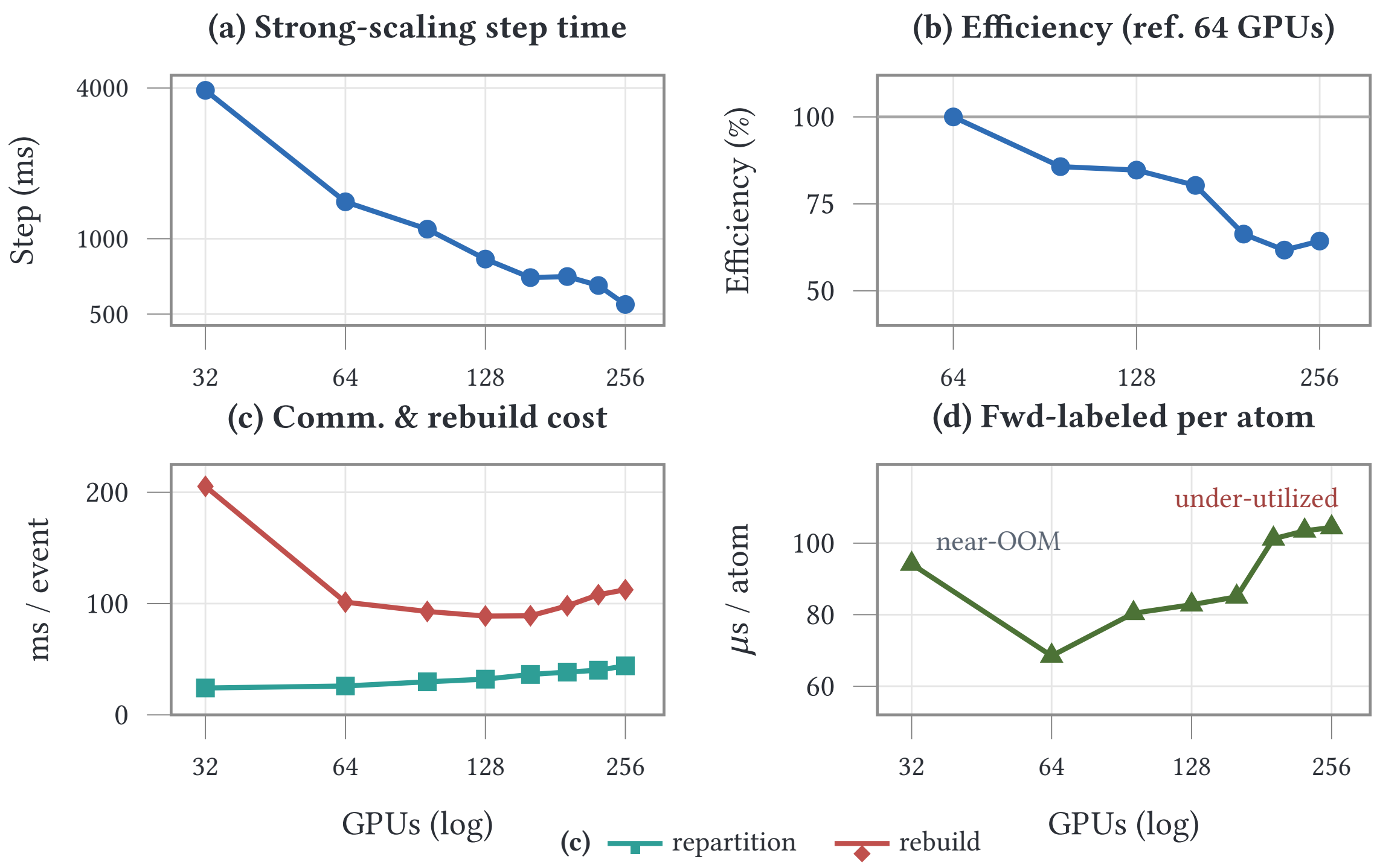

\textbf{Strong scaling.}
Where every scaling result above grew the system with the device count, a
separate sweep holds the $1.28$M-atom system of Fig.~\ref{fig:strong_scaling}
fixed and grows only the count, from $32$ to $256$ GPUs, so the per-GPU load
falls from $40$k to $5$k atoms. Activation recomputation
(Section~\ref{sec:single_gpu}) is not an optional latency/memory trade at this
end of the range: without it, $32$ and $64$ GPUs, at $40$k and $20$k atoms
each, exceed this cluster's $40$\,GB budget and do not run at all, so every
point plotted has it on. Recomputation off is therefore undefined at those two points; where both fit it is $4.7$--$18.6\%$ faster, flattening at the same knee, and the margin narrows with device count ($18.6\%$ at $96$ GPUs, $4.7\%$ at $256$), so an efficiency figure is only comparable against a shared reference point.
Referenced to $64$ GPUs rather than $32$, for the reason
Fig.~\ref{fig:strong_scaling}b gives, efficiency declines to $64\%$ at $256$ GPUs ($61.7\%$ at $224$), non-monotonically, the fall steepening past $160$ GPUs, where each GPU
holds only $8$k atoms. Around that knee the returns stall rather than plateau: $192$ GPUs is within $1\%$ of $160$'s step time, and the step time resumes falling at $224$ and $256$. Section~\ref{sec:negative-results} records this as a finding.

\begin{table}[t]
    \centering
    \caption{Strong-scaling step time with recomputation on/off, and the forward-pass hit/rebuild split (recomputation on).}
    \small
    \begin{tabular}{rrrrr}
        \toprule
        GPUs & Step, on & Step, off & Fwd, hit & Fwd, rebuild \\
        \midrule
        32  & $3915.1$ & OOM      & $3766.7$ & $3776.0$ \\
        64  & $1404.5$ & OOM      & $1369.7$ & $1366.4$ \\
        96  & $1092.0$ & $888.6$  & $1072.2$ & $1070.9$ \\
        128 & $829.3$  & $728.5$  & $827.9$  & $823.6$ \\
        160 & $699.4$  & $615.1$  & $680.0$  & $711.1$ \\
        192 & $706.2$  & $598.2$  & $674.8$  & $655.8$ \\
        224 & $650.4$  & $567.6$  & $591.2$  & $626.0$ \\
        256 & $546.3$  & $520.5$  & $521.9$  & $532.2$ \\
        \bottomrule
    \end{tabular}
    \label{tab:strong-scaling-detail}
\end{table}

Only one of the two candidate mechanisms is large enough to matter.
Neighbor-cache management grows from $2.2\%$ of wall time at $64$ GPUs to
$6.8\%$ at $256$ and repartition stays under $0.1\%$ amortized, which is $4.6$
points against a $36$-point drop (Fig.~\ref{fig:strong_scaling}c). The
deviation sits instead in the forward-labeled phase, which bottoms out per atom
at $64$ GPUs' $20$k atoms/GPU and rises on both sides of it, following the
per-GPU batch size (Fig.~\ref{fig:strong_scaling}d). That is consistent with
small-batch under-utilization, though the same timer spans the remote branch
and its waits, so it does not isolate a cause. The forward-pass hit/rebuild
split of Table~\ref{tab:strong-scaling-detail} shows no consistent ordering
there ($-35$ to $+19$\,ms, with no consistent sign), unlike the hit-slower-than-rebuild gap the
next paragraph reports at the much larger per-step scale of the trajectory
studies.

\textbf{Density heterogeneity.}
\label{sec:density-heterogeneity}
Every system above is near-uniform in density, bulk water or this KCl solution,
which is the easiest case for a cut that equalizes \emph{atom counts}. We
therefore run a matched pair on $256$ GPUs, identical in hyperparameters
($dt{=}0.5$\,fs, rebuild/$10$ steps, sync/$100$, $0.5$\,\AA{} skin) and
differing only in the system: the $10.24$M-atom KCl solution, and a $10.06$M-atom
Norovirus capsid in solvent whose protein-dense regions sit against a nearly
protein-free water bath (a separate NVT pair, not comparable to the NVE run
above). Table~\ref{tab:density-heterogeneity} splits a steady-state window:
the capsid costs $14.5\%$ more per step, almost entirely from forward compute
($+17$--$20\%$ on both paths), while rebuild/hit overhead, hit rate, and the
per-event cost of repartition---the one mechanism here that touches every
rank---are essentially unchanged. At $40.0$k vs.\ $39.3$k atoms per GPU the gap
is not a load-size confound. The margin moves with trajectory phase
($13$--$33\%$ across three windows), so read it as a range; every window puts
forward compute dominant. Its capsid arm is a $100$\,ps NVT trajectory,
$\approx\!8$\,ps/day on these $256$ A100s, that ran with no broken bond, no
anomalously close pair, no energy excursion, and only three compiled variants
across its $200{,}000$ steps, with no graph break and no fallback to eager
(Appendix~\ref{sec:app-protocol}).

\begin{table}[t]
\centering
\caption{Per-step cost under matched hyperparameters on the same A100 cluster, near-uniform KCl vs.\ heterogeneous Norovirus capsid, over steps $3000$--$4000$. The capsid holds \emph{fewer} atoms per GPU, so the penalty is not a load artifact. Rows compose as Step $\approx$ hit rate$\times$(fwd,hit$+$hit ovh.) $+$ ($1{-}$hit rate)$\times$(fwd,rebuild$+$rebuild ovh.) $+$ residual.}
\small
\begin{tabular}{lrrr}
    \toprule
     & KCl ($10.24$M) & Virus ($10.06$M) & Diff. \\
    \midrule
    Atoms per GPU                & $40.0$k           & $39.3$k           & $-1.8\%$ \\
    \midrule
    Step time                    & $4552.3$\,ms      & $5212.6$\,ms      & $+14.5\%$ \\
    \quad forward, cache hit     & $3650.8$\,ms      & $4258.0$\,ms      & $+16.6\%$ \\
    \quad forward, rebuild       & $3369.0$\,ms      & $4030.5$\,ms      & $+19.6\%$ \\
    \quad rebuild overhead       & $334.9$\,ms       & $367.4$\,ms       & $+9.7\%$ \\
    \quad hit overhead           & $33.8$\,ms        & $33.7$\,ms        & $-0.3\%$ \\
    \quad residual (integrate, I/O) & $862.7$\,ms    & $898.2$\,ms       & $+4.1\%$ \\
    Repartition (per event)      & $92.7$\,ms        & $87.5$\,ms        & $-5.6\%$ \\
    Cache hit rate               & $79.3\%$          & $79.5\%$          & -- \\
    \bottomrule
\end{tabular}
\label{tab:density-heterogeneity}
\end{table}

One oddity: cache-hit forward costs $5.6$--$8.4\%$ \emph{more} than rebuild
forward in both columns, despite skipping the search a rebuild pays for---not a
compute effect but cross-rank wait-time drift
(Appendix~\ref{sec:app-protocol}), and taken up as a finding in
Section~\ref{sec:negative-results}.

One mechanism is consistent with the capsid's margin, and scopes
Section~\ref{sec:morton_partition}'s balance claim: the cut fixes atom count
exactly, which is compute balance only insofar as work is per-atom rather than
per-\emph{edge}---and a protein-dense region carries more edges per atom than
the water beside it, so equal counts no longer mean equal work, in the term
where the cost lands. Node shapes survive untouched, fixed by counts alone;
edge shapes and the peer set follow geometry, holding only within a cache
epoch. What degrades is the work behind the invariants, not the invariants. An
edge-weighted cut would rebalance it, at the cost of the exact-count invariant
the captured graph rests on; whether the gap widens with device count is
untested (Section~\ref{sec:discussion}).

\subsection{Negative and Unverified Results}
\label{sec:negative-results}

Several of this section's results are negative or unresolved, and they bound
what the positive ones establish. We collect them here rather than leaving them
scattered as caveats.

\textbf{Strong scaling stops paying at the knee.} At $192$ GPUs the step time is
$706.2$\,ms against $160$ GPUs' $699.4$\,ms---within $1\%$, and inside the
run's own window-to-window spread---so a $20\%$ device increase buys nothing
this run can resolve, at $8$k atoms per GPU and beyond. The knee appears in
both recomputation modes, and it is the axis on which the timescale gap would
have to be closed (Appendix~\ref{sec:app-protocol}).

\textbf{A cache hit is slower than a rebuild, in the forward pass.} Cache-hit
forward costs $5.6$--$8.4\%$ more than rebuild forward in both columns of
Table~\ref{tab:density-heterogeneity}, despite skipping the search a rebuild
pays for. The effect is not compute: in the separate $256$-GPU profiling run of Appendix~\ref{sec:app-protocol}, the model-compute kernel is if anything $1.6\%$ cheaper on hits ($1804.3$ vs.\ $1833.4$\,ms), while the cross-rank NCCL
exchange is $724$\,ms \emph{cheaper} on rebuild ($1235.4$ vs.\ $1959.6$\,ms), and
per-step communication on hit steps spans $735$--$2451$\,ms (a $3.3\times$
spread) while compute holds within $1\%$. The gap is wait time at the attention
All-to-All: a rebuild's ghost-atom exchange re-aligns ranks, and consecutive hit
steps carry no equivalent, so timing differences build up until the next
exchange waits on whichever rank is slowest. It is a property of the loaded
regime, not a universal one---it does not appear at the
Table~\ref{tab:strong-scaling-detail} points, where hit is the faster path at
four of the eight points (Appendix~\ref{sec:app-protocol}).

\textbf{The heterogeneous penalty is a range, not a number.} The capsid costs
$14.5\%$ more per step in the window of Table~\ref{tab:density-heterogeneity},
but the margin moves with trajectory phase, $13$--$33\%$ across three windows.
The single-window figure is one sample of that range. The mechanism we offer---
equal atom counts are not equal edge counts---is consistent with the margin
rather than isolated by it, and whether the gap widens with device count is
untested.

\textbf{TF32 and ion--ion structure are unresolved.} In the $1.28$M-atom, $\approx\!201$\,K TF32 A/B of Appendix~\ref{sec:app-protocol}, TF32 leaves water and solvation structure alone, with first-peak heights agreeing to $0.5\%$ (K--O),
$0.2\%$ (Cl--O) and $0.0\%$ (O--O), and coordination numbers to $0.2\%$. The
three ion--ion first peaks differ by $-3.2\%$ (K--Cl), $+23.8\%$ (K--K) and
$+11.6\%$ (Cl--Cl), but splitting the TF32-\emph{off} run alone into halves
moves the same peaks by $10.3\%$, $14.1\%$ and $12.5\%$, so only K--K clears the
noise floor, and only by $1.7\times$. The effective sample behind a K--K peak is
tens of pairs, not the $10^3$ frames suggest, because ion diffusion
decorrelates far more slowly than the $5$\,ps window. Resolving whether TF32
touches ion--ion structure needs a trajectory long enough to decorrelate the
ions, not a longer averaging window over the same one
(Appendix~\ref{sec:app-protocol}).

\textbf{The $60$k weak-scaling point is not like-for-like.} It is the only load
carrying activation recomputation, which enlarges the local compute window for
a reason unrelated to atom count, so its $89.7\%$ efficiency is not a
continuation of the $10$k$\rightarrow$$25$k trend; the mechanism is cleanest
between those two loads, which differ only in load.

\textbf{There is no external baseline.} Every comparison in this section is
against another configuration of our own system. The closest message-passing
implementation stops at $32$ GPUs (Section~\ref{sec:related}), so it cannot
supply an end-to-end comparator at these scales, and the kernel comparison of
Appendix~\ref{sec:app-kernels} is against a vanilla sparse baseline and dense
SDPA rather than another production MD engine. The speedups we report are
ablation and composition numbers, not a claim of superiority over an external
system.

Read together, these results mark where the evidence stops: the cache's
composed $15.7\times$ is an upper estimate, one official run per point leaves no
error bars, and differences of a few percent---rank-box's $3\%$ cost against
all-gather, the $\pm 35$\,ms forward split of
Table~\ref{tab:strong-scaling-detail}---are not separable by these
measurements.

\section{Related Work}
\label{sec:related}

Prior work partitions a large graph-like state and hides the communication. We inherit more from them than we replace: space-filling-curve decomposition is long-standing practice~\cite{springel2005gadget}, skin-plus-displacement neighbor reuse dates to Verlet~\cite{verlet1967computer}, and distributing an equivariant potential has direct precedent. SevenNet~\cite{park2024sevennet} spatially decomposes a message-passing potential for MD across GPUs, Allegro's scale-out~\cite{musaelian2023scaling} runs a deep equivariant model at biomolecular scale, and a NequIP-framework rebuild~\cite{tan2026nequip} pairs compiler integration with custom Triton and vendor equivariant kernels---the same two levers as our Section~\ref{sec:single_gpu}, aimed at data-parallel training rather than at splitting one system across ranks. MACE takes the algorithmic route instead, using two high-body-order message-passing layers, which also bounds its receptive field~\cite{batatia2022mace}; we keep the depth and pay in the runtime. None of these cited systems combines propagation with scale: strictly-local models decompose without propagation; the message-passing ones stop well short of hundreds of GPUs---$32$ GPUs at over $80\%$ weak-scaling efficiency on silica in the closest scheme~\cite{park2024sevennet}, $8$ in a recent MPI design whose $10^8$-atom run is on CPUs~\cite{xia2025efficient}---where holding shapes, plans, and captured execution invariant as the graph moves becomes binding.

\begin{table*}[t]
\centering
\caption{Where the design sits against systems cited in this paper, with the cells confined to what those citations and our own results establish. ``Model class'' separates potentials whose energy depends on one cutoff ball from those that propagate features across layers. ``Per-layer exch.'' asks whether a distributed step synchronizes remote features once per layer, as propagation requires, rather than once per step as a ghost-atom halo does. Exact forces means forces are the gradient of the evaluated energy rather than a separately regressed output. Live-graph reuse means the neighbor topology, or the exchange plan built from it, survives across steps of a moving trajectory. Static-shape capture means a step is replayed under a captured graph whose allocations and launch geometry are fixed. An em dash marks an item this paper could not establish from the cited work; no cell is a ranking, the systems differ in task, model, and hardware, so the columns are not additive.}
\label{tab:related-comparison}
\footnotesize
\setlength{\tabcolsep}{4pt}
\begin{tabular}{@{}lccccccc@{}}
\toprule
System & Model class & \shortstack{Per-layer\\exch.} & \shortstack{Max\\GPUs} & \shortstack{Max\\atoms} & \shortstack{Exact\\forces} & \shortstack{Live-graph\\reuse} & \shortstack{Static-shape\\capture} \\
\midrule
DPMD~\cite{jia2020pushing,zhang2018deep}        & strictly local    & no  & --- & 100M  & yes & --- & --- \\
LAMMPS~\cite{plimpton1995fast,thompson2022lammps} & strictly local  & no  & --- & ---   & yes & yes & no  \\
Allegro scale-out~\cite{musaelian2023scaling}    & strictly local   & no  & --- & ---   & yes & --- & --- \\
SevenNet~\cite{park2024sevennet}                & message passing   & yes & 32  & ---   & yes & --- & --- \\
Xia et al.~\cite{xia2025efficient}              & message passing   & yes & 8   & $10^8$ (CPU) & --- & --- & --- \\
NequIP rebuild~\cite{tan2026nequip}             & message passing   & --- & --- & ---   & --- & --- & --- \\
MACE~\cite{batatia2022mace}                     & message passing   & --- & --- & ---   & yes & --- & --- \\
UMA~\cite{wood2025uma}                          & message passing   & --- & 1   & ---   & --- & --- & --- \\
\midrule
This work                                       & message passing   & yes & 512 & $30.7$M & yes & yes & yes \\
\bottomrule
\end{tabular}
\end{table*}

Table~\ref{tab:related-comparison} makes the shape of the gap concrete without scoring anyone. Two columns do most of the work. \emph{Per-layer boundary exchange} is a property of the model class, and it is what turns a distributed step from one halo update into a synchronization per Transformer block; the systems marked ``no'' are distributed but either forgo propagation or replicate the whole system on every rank, and both choices avoid the problem rather than solve it, while a dash marks a system that is not distributed in this sense at all. \emph{Live-graph reuse} and \emph{static-shape capture} are properties of the runtime, and they are the two that a moving molecular graph actively attacks: a training system partitions a fixed graph once, and a strictly local MD code can rebuild a halo every step because the halo is a handful of scalars per ghost atom rather than an equivariant payload. The combination in the last row---propagation, exact gradients, reuse, and capture, at hundreds of GPUs---is the one no cited system reports, which is the scoped form of our claim.

\textbf{Spatial domain decomposition in classical molecular dynamics.}
Spatial domain decomposition with ghost-atom communication has been standard in classical MD for three decades~\cite{plimpton1995fast,thompson2022lammps,abraham2015gromacs,phillips2005namd}, routing only boundary updates between adjacent domains, and neutral-territory schemes choose where each interaction is evaluated to minimize imported volume~\cite{bowers2006midpoint,bowers2006desmond}, a decision Anton commits to silicon~\cite{shaw2007anton}. Our equal-count space-filling-curve partition (Section~\ref{sec:morton_partition}) descends from that line, since all bound communication by geometry, not device count, but optimizes for a per-rank shape that does not move between steps, not the smallest import set. Classical short-range force fields are cheap per atom and tolerant of a domain's count fluctuations; nothing there needs shapes bit-identical across steps, as \texttt{torch.compile} guards and a captured CUDA Graph do---and compiling for dynamic shapes~\cite{ansel2024pytorch2} removes the re-trace, not the capture, since a static graph replays fixed allocations and launch parameters. That is why we insist on equal-count, and why holding it under MD dynamics is part of the contribution. Those codes also exchange a handful of scalars per ghost atom, where ours carries a projected equivariant SKV payload (Section~\ref{sec:reduce_volume}).

The difference is worth stating in the currency of the two costs. A ghost-atom halo is sized by the cutoff and the domain surface, and its per-step volume is small because the imported object is a position and a species tag; a per-layer feature exchange is sized by the receptive field of one layer and the feature width, and it recurs once per layer in each pass. The former can be recomputed from scratch every step at negligible cost, so classical codes can rebuild their halo every step and need no separate topology/geometry split, even when they do reuse topology. The latter cannot be rebuilt cheaply, which is why the split of Section~\ref{sec:multilevel_search} is load-bearing rather than an optimization---and why its displacement bound must make staleness impossible rather than small, where parameter training can absorb a stale neighborhood as noise but a force evaluation integrates it into the trajectory.

\textbf{Distributed GNN training.} DistDGL~\cite{zheng2020distdgl} and ROC~\cite{jia2020roc} partition large graphs across machines---METIS-style and through an offline learned cost model, respectively---then pipeline cross-partition communication for mini-batch GNN training. Ours also splits a graph across devices with boundary message passing, but three properties differ. Atoms carry a 3D embedding, so a geometry-aware curve suffices where a generic sparse graph needs METIS-style cuts. GNN training partitions a \emph{fixed} graph once and sub-samples from it; ours is re-derived every step from current coordinates, with the entire simulation state as the ``batch.'' That is why a plan built once and reused for a partition's lifetime (Section~\ref{sec:reduce_volume}) is worth the engineering, where a mini-batch system's routing changes every batch. And both backpropagate, but over different objects: GNN training differentiates a loss with respect to \emph{parameters}, where stale neighborhoods are an accepted trade---PipeGCN~\cite{wan2022pipegcn} buys its overlap by pipelining one-epoch-stale boundary features---whereas conservative forces are the analytic gradient with respect to \emph{coordinates}, taken over the exact neighborhood every step, so the exchange recurs in the backward pass and needs the \texttt{pre}/\texttt{post} pair to stay overlapped under reverse traversal (Section~\ref{sec:overlap}).

\textbf{Communication--computation overlap in large-scale transformer training.}
Overlapping collectives with compute is standard in LLM training~\cite{narayanan2021megatron, rajbhandari2020zero}, and systems work pushes that overlap below the operator boundary, into kernel fusion, tile-level triggering, and hierarchical collective scheduling~\cite{jangda2022coconet, pati2024t3, chen2024centauri}. We follow the same principle (Section~\ref{sec:overlap}) but hide a different collective: those methods overlap \emph{dense} collectives of rank-uniform size fixed for training's duration, where ours is \emph{sparse}, variable-length, per-edge, and re-derived from the evolving neighbor graph. Mixture-of-Experts routing is the closest analogue, variable-length at runtime~\cite{he2022fastermoe,hwang2023tutel}, but it is learned and so negotiable---tokens rebalanced or dropped under a capacity factor---where ours is fixed by geometry and cannot drop an edge without changing the force. That is why a reusable plan, sender-side SKV fusion, and the asymmetric \texttt{pre}/\texttt{post} operators earn their place here, where a fixed-size all-reduce needs none of them. The bound itself is the same arithmetic in all three settings---the smaller concurrent phase is what disappears---but its value differs: here the exchange is sized by a subdomain's boundary and the local branch by its interior, so the hidden fraction improves with per-rank load rather than degrading with the machine.

\section{Discussion and Outlook}
\label{sec:discussion}

\textbf{Scaling beyond the present deployment.}
The per-step path is not tied to our 512-GPU scale: per-GPU cost follows the \emph{local} problem size---fixed atom count, constant-size scalar reductions (Sections~\ref{sec:morton_partition}, \ref{sec:md_engine})---rather than device count. Peer degree is the exception in kind, though not in magnitude: it rises from $6$ to $19$ over $64$--$512$ GPUs at $25$k atoms/GPU, near $K^{0.5}$, so it grows with the machine far more slowly than the $K$ an all-gather contacts. No weak-scaling ceiling is reached within the measured range: efficiency stays above $77\%$ through $512$ GPUs, though it declines with $K$---the $K\!\times\!K$ request matrix and the world reductions' latency still grow with $K$, the rank-$0$ sort with $N$, and the metadata all-gather remains the next target the telemetry names (Section~\ref{sec:neighbor-scaling}). Saying the design extends in principle toward billion-atom systems is a statement about the exchange, not about every term in the step.

\subsection{Threats to Validity}
\label{sec:threats}

\emph{Internal.} Every scaling point is one official run, so no error bars are drawn; where a gap is smaller than the run-to-run spread we do not treat it as separable. The neighbor pipeline's sub-stage timers are MAX-reduced across ranks independently, and different sub-stages' maxima may come from different ranks, so the pipeline is not decomposed additively---why the attribution of spatial-bin's cost to plan overhead is described as consistent with the telemetry rather than as a measured stage decomposition (Section~\ref{sec:multilevel_search}), and why the forward-pass split is not read as a stage breakdown. The composed neighbor-search figure carries a further caveat: its cached rows combine a rebuild measured on a separate interval-$1$ run with a hit measured on an interval-$10$ run, and the interval-$1$ arm searches at each layer's bare cutoff while an interval-$10$ rebuild builds the wider $r_c+\delta$ superset, so the composed $15.7\times$ is an upper estimate rather than a measured speedup (Section~\ref{sec:experiments}). The figures the argument leans on---the directly measured hit latencies and the full-step $1.87\times$ A/B---do not depend on it.

\emph{External.} Every system measured is near-uniform in density, bulk water or the KCl solution, the easiest case for a cut that equalizes atom counts; the one heterogeneous system, the Norovirus capsid, is a single device-count pair, and we probe neither sharper density contrasts nor anisotropic cells, where AABBs could inflate past equal-count's unfixed peer bound. One model is measured throughout---a UBio-MolFM variant built on E2Former~V2, at $23.5$M parameters (Section~\ref{sec:exp-setup})---so the claim is about this class of equivariant Transformer. And throughput and weak scaling come from an H100/InfiniBand cluster, the trajectory and strong-scaling results from an A100/PCIe one, so numbers from the two are not interchangeable.

\emph{Construct.} The central invariant is atom-count balance, and atom count is not work: work is per-edge, so a protein-dense region carries more edges per atom than the water beside it. The capsid pair is consistent with this account, its $13$--$33\%$ per-step margin pointing to edges per atom rather than atoms (Section~\ref{sec:md-validation}); node shapes survive untouched, fixed by counts alone, while edge shapes and the peer set follow geometry and hold only within a cache epoch, so what degrades is the work behind the invariants. The exact-search backend also clamps to FAISS's $k$-selection limit of $2048$ wherever the per-query candidate set is large enough to saturate it---a function of local density and cutoff rather than of device count---not necessarily truncation, but the range's two ends may not run an identical search, so the edge set is the complete radius graph only where that selection is unsaturated.

\emph{Limitations that remain.} \emph{Precision:} keeping only precision-sensitive contractions in higher precision could recover TF32's throughput without its drift; we have not built that split. \emph{Cadence:} two fixed clocks set the schedule---re-partition every $100$ steps, neighbor rebuild every $10$---where a displacement-driven trigger would act only when locality degrades. Faster dynamics push the rebuild clock toward every step, an A/B measured at $1.87\times$ on the full step (Section~\ref{sec:e2e_scaling}), not collapse. The re-partition clock is a separate cost: its partition-and-broadcast half runs $\approx\!87$--$88$\,ms per event, amortized over $100$ steps here. \emph{No external baseline:} we compare only our own configurations, since the closest message-passing implementation stops at $32$ GPUs (Section~\ref{sec:related}). \emph{Timescale:} the measured $\approx\!8$\,ps/day covers the checks reported here, not the nanosecond windows a binding study needs; closing it is strong scaling at fixed atom count, the axis on which returns go flat past $160$ GPUs on the $1.28$M-atom strong-scaling system. The drift and structure check runs on a $10.24$M-atom trajectory, $3\times$ under the $30.7$M scaling run, and one trajectory is not a replicate; the ps/day figure comes from the $10.06$M-atom capsid run.

\subsection{Complexity Accounting}
\label{sec:complexity}

Table~\ref{tab:complexity} lists what still grows per step. Four terms grow with the machine---peer count, the $K\!\times\!K$ request matrix, the metadata all-gather, and world-collective latency---one with the problem, the rank-$0$ sort, and one with the local density, per-GPU edges, which weak scaling holds fixed at constant atoms per GPU. None is removed by the design. Their sum grows slowly enough that measured weak-scaling efficiency stays between $77\%$ and $90\%$ through $512$ GPUs---$77.3\%$ and $84.7\%$ at the two loads that run without activation recomputation, and $89.7\%$ at the $60$k load, which carries it and so is not a like-for-like continuation---because each term is small against the local compute window at these sizes. Whether that holds at a billion atoms, about $33\times$ the $30.7$M measured here, and the correspondingly larger $K$, is an extrapolation, which is why we say ``extends in principle'' rather than predicting an efficiency.
\begin{table}[t]
\centering
\caption{What still grows per step, and in what. Weak scaling holds per-GPU atom count fixed, so the question is which terms grow while the local forward--backward work does not. The last column gives what the design does about each term; none is removed outright. The table scopes the extension argument rather than supporting it.}
\label{tab:complexity}
\small
\begin{tabular}{@{}lll@{}}
\toprule
Term & Grows in & What the design does \\
\midrule
Per-GPU edges & local density & $O(1)$ in $K$ at fixed atoms/GPU \\
Peer count & geometry, $6\!\to\!19$ over $64$--$512$ GPUs & sublinear in $K$, roughly $K^{0.5}$ \\
$K\times K$ request matrix & $K^2$ nominally & sparse rows, coarse AABBs \\
Metadata all-gather & $K$ & small fixed descriptor ($<2$\,KiB) \\
World-collective latency & $\log K$ & constant payload per step \\
Rank-0 $O(N\log N)$ sort & $N$ & amortized over $100$ steps \\
\bottomrule
\end{tabular}
\end{table}

The $K$-dependent terms are the ones the telemetry already names. The request matrix is $K^2$ only nominally: SFC locality keeps each row sparse, but the test is still a loop over all $K$ descriptors, and the metadata all-gather feeding it is the term still rising with $K$ at the top of our range (Section~\ref{sec:neighbor-scaling}). The world collectives are constant in payload but not in latency---a barrier over $K$ ranks has a floor that does not shrink with per-GPU work. The rank-$0$ sort is $O(N\log N)$ in the full coordinate set, measured at $\approx\!70.7$\,ms at $10.24$M atoms and amortized over $100$ steps; a distributed sort would remove it rather than amortize it, and we have not built one.

\subsection{What Generalizes, and What Does Not}
\label{sec:generalizes}

The placement mechanism is not specific to this model: its three guarantees follow from the placement argument, and apply to any potential that exchanges boundary features per layer over a live graph. The requirements are stated in Section~\ref{sec:bg-problem} without reference to E2Former, and nothing in the cut, the cache invariant, or the plan-reuse accounting depends on the particular tensor product or attention form.

What is bound to our stack is worth naming. The kernel work of Section~\ref{sec:single_gpu} assumes sparse equivariant attention with $l_{\max}=2$ and a scalar-per-head score, though the CSR/CSC duality argument applies to any two-pass sparse operator. The payload arithmetic of Section~\ref{sec:reduce_volume} uses the fact that only the Value must travel equivariant, which holds for attention-style aggregation but not for an architecture that ships full features between layers. And the empirical support for the generalization is itself stack-bound, and splits into measurement and derivation: the $6\!\to\!19$ peer degree over $64$--$512$ GPUs and the $1.87\times$ cache A/B are measured, while the $80\%$ local-fraction figure is a bound derived from the box covariogram, falling to $\approx\!78\%$ at the $8$k-atom/GPU strong-scaling points; all are properties of $\rho=0.1$\,atoms\,\AA$^{-3}$, this cutoff stack, and this range of device counts, and the formulas behind them are not.

\section{Conclusion}
\label{sec:conclusion}

This paper built a runtime for message-passing equivariant molecular dynamics at tens-of-millions-of-atom scale, treating the model and the distribution as one object of design. Its foundation is a placement mechanism: quantize coordinates onto a space-filling curve, sort, and cut the order into equal-count contiguous shards. That single decision supplies exact per-rank atom counts and with them invariant tensor shapes, a peer set that grows as roughly $K^{0.5}$ rather than $K$, and a deterministic ownership map, so each cache epoch's whole forward--backward pass replays as a static CUDA Graph. On it rest custom Triton kernels with dual CSR/CSC indexing and a unified MCH layout, a conservative three-level neighbor search with exact cross-step topology reuse, an All-to-All reduced to its payload and overlapped in both passes, and a decentralized on-GPU MD engine.

The system sustains $30.7$M atoms on 512 GPUs at $89.7\%$ weak-scaling efficiency ($77$--$90\%$ across three per-GPU loads), with graph reuse cutting full-step time $1.87\times$ and neighbor-search cost growth over an $8\times$ device-count increase held to $1.7\times$ rather than $3.0\times$. On a live $10.24$M-atom trajectory the same code path holds energy drift to $0.42\,\mu$eV/atom/ps and reproduces the serial reference's structure; a $10.06$M-atom Norovirus capsid locates the cut's limit in edges per atom.

What remains is scale and scope. The exchange grows only slowly with the machine, but it does grow, and the request matrix, the metadata all-gather, and world-collective latency still scale with it, while the rank-$0$ sort scales with the problem; a distributed sort and a sparser request structure are the next terms to remove. Beyond that, the timescale a binding study needs is a strong-scaling problem at fixed atom count, and the placement mechanism is stated without reference to this model, so any potential exchanging boundary features per layer over a live graph is the natural next object to carry it.

\section*{Contributors}

\begin{itemize}
  \item \textbf{Neural Network Architecture Optimization}: Lin Huang, Jia Zhang, Arthur Jiang
  \item \textbf{Core Operator Optimization}: Jiajun Cheng, Lin Huang, Zion Wang, Junping Zhao
  \item \textbf{Communication Optimization}: Hao Yin, Hao Li, Zion Wang, Ji Zhang, Junping Zhao
  \item \textbf{MD Engine Development}: Jia Zhang
  \item \textbf{Experimental Design and Analysis}: Zion Wang, Hao Yin, Hao Li, Lin Huang, Jia Zhang
  \item \textbf{Computing Cluster Support}: Ji Zhang, Lucas Li, Allex Kan
  \item \textbf{Project Management and Support}: Owen Fu, Xunxing Chen
  \item \textbf{Technical Report Writing}: All authors
\end{itemize}

\makeatletter
\renewenvironment{thebibliography}[1]{  \section*{\refname}  \@mkboth{\MakeUppercase\refname}{\MakeUppercase\refname}  \list{\@biblabel{\@arabic\c@enumiv}}       {\settowidth\labelwidth{\@biblabel{#1}}        \leftmargin\labelwidth \advance\leftmargin\labelsep
        \@openbib@code
        \usecounter{enumiv}        \let\p@enumiv\@empty
        \renewcommand\theenumiv{\@arabic\c@enumiv}        \footnotesize
        \itemsep\z@ \parsep\z@ \topsep\z@}  \sloppy
  \clubpenalty4000 \@clubpenalty\clubpenalty
  \widowpenalty4000 \sfcode`\.\@m}
  {\def\@noitemerr{\@latex@warning{Empty `thebibliography' environment}}   \endlist}
\makeatother
\bibliographystyle{plain}
\bibliography{references}

\providecommand{\apptablesize}{\small}

\clearpage

\appendix

The appendices are organized one topic per letter. Appendix~\ref{sec:app-config}
records the exact model configuration behind every reported number.
Appendices~\ref{sec:app-kernels}--\ref{sec:app-comm} then take the mechanisms of
Section~\ref{sec:engine} one at a time---single-GPU kernels, then the
implementation detail behind the partition, the neighbor search and the
communication path---each supplying what the main text defers.
Appendix~\ref{sec:app-protocol} gives the measurement protocol, the
full-inference validation, and the per-run telemetry behind
Section~\ref{sec:experiments}, and Appendix~\ref{sec:app-hardware} records the
hardware and software behind every reported timing. Nothing here is needed to
follow the main argument; each appendix is for a reader who wants one mechanism
spelled out in full or one number reproduced.

\section{Model and Hyperparameters}
\label{sec:app-config}
\label{sec:appendix}

Table~\ref{tab:model-hparams} lists the backbone hyperparameters of the E2Former~V2 model~\cite{huang2025e2former, huang2026e2formerv2} used throughout, with the exact key/value pairs of our run configuration (keys shown are under \texttt{backbone\allowbreak\_config}). The stack is a $4$-layer backbone---three short-range equivariant layers at a $5$\,\AA{} cutoff followed by one long-range layer at $8$\,\AA{}, for the $23$\,\AA{} per-atom receptive field of Section~\ref{sec:experiments}. Node features carry channels up to degree $l_{\max}=2$ (irreps \texttt{256x0e+256x1e+256x2e}, i.e.\ $C=256$ channels per degree), attention uses $128$ heads with an $8$-dimensional scalar head, and every run uses the Triton tensor-product path of Section~\ref{sec:single_gpu}, in TF32 mixed precision except where stated otherwise (the trajectory validation of Section~\ref{sec:md-validation} runs with TF32 off). The total parameter count is $23.5$M.

\begin{table*}[h!]
\centering
\apptablesize
\begin{tabular}{lll}
\toprule
\textbf{Model / Backbone setting} & \textbf{Symbol} & \textbf{Value} \\
\midrule
Backbone & --- & \texttt{e2former} (E2Former~V2) \\
\# layers & $R$ & \texttt{num\_layers=4} ($3$ short-range $+$ $1$ long-range) \\
Cutoff radius & --- & short-range $5$\,\AA{} ($\times 3$ layers), long-range $8$\,\AA{} ($\times 1$) \\
Clustering & --- & \texttt{with\_cluster=False} \\
Node embedding irreps & $C,\ l_{\max}$ & \texttt{"256x0e+256x1e+256x2e"} ($C=256$, $l_{\max}=2$) \\
Head irreps & --- & \texttt{"2x0e+2x1e+2x2e"} \\
Scalar head dim & $d_k$ & \texttt{attn\_scalar\_head=8} \\
\# attention heads & $H$ & \texttt{num\_attn\_heads=128} \\
Radial basis & --- & \texttt{number\_of\_basis=256} \\
Neighbor cap & --- & none (all in-cutoff pairs retained) \\
Attention dropout & --- & \texttt{alpha\_drop=0} \\
Basis type & --- & \texttt{gaussiansmear} \\
Normalization & --- & \texttt{rms\_norm\_sh} \\
Attention type & --- & \texttt{so2-zero-order} \\
Tensor product impl. & --- & \texttt{tp\_type=QK+triton} \\
Edge embedding & --- & \texttt{default} \\
FFN type & --- & \texttt{s3} \\
Encoder & --- & \texttt{default} \\
\bottomrule
\end{tabular}
\caption{E2Former~V2 backbone hyperparameters used for all runs in this paper. Keys are configuration fields under \texttt{backbone\_config}; the \textbf{Symbol} column gives the notation used for each in this appendix ($C$ channels, $H$ attention heads, $l_{\max}$ maximum degree, $d_k$ scalar head dimension, $R$ Transformer blocks).}
\label{tab:model-hparams}
\end{table*}

\section{Single-GPU Kernel Implementation}
\label{sec:app-kernels}

This appendix details the kernel-level design summarized in Section~\ref{sec:single_gpu}.

\paragraph{Triton Kernel Design.}
Two of the four axes of Section~\ref{sec:single_gpu} are standard practice and detailed here rather than in the main text. The Clebsch--Gordan tensor products these kernels accelerate are the primitive introduced by Tensor Field Networks~\cite{thomas2018tensorfield} and implemented generically in \texttt{e3nn}~\cite{geiger2022e3nn}; our kernels target the specific shapes and access patterns E2Former~V2's attention imposes on that primitive, not a general replacement for the library.

\textbf{Hardware-aware compute saturation.}
Autotuning explores the tile-size space (e.g., $BLOCK\_H, BLOCK\_K$) under the shared-memory constraint $BH \cdot BK + BH \cdot BD + BK \cdot BD \leq 64{,}000$; the QK kernels use 2D tiled parallelism with a 3-stage pipeline ($num\_stages=3$) to overlap FMA with irregular gathers.

\textbf{Deep operator fusion.}
Single-pass kernels fuse sparse gathers, dot products, and gated multiplications---including the SpMV-based key-gradient backward and the outer-product reduction $out = \sum sh \otimes fea$---without materializing intermediates, minimizing HBM round-trips. All loads use 2D/3D tiling with stride-based indexing for memory coalescing.

\paragraph{Allocator Policy.}
The recomputation interface takes a retention fraction in $[0,1]$: $0$ is full recomputation (classical gradient checkpointing) and $1$ retains everything, so the compiler rather than the author chooses where to trade memory for recompute---the same trade selective activation recomputation~\cite{korthikanti2023reducing} makes at transformer-training scale, applied here per-operator by the compiler's own cost model rather than per-layer by the author. Alongside it we run a two-tier allocator policy. \textbf{Threshold-triggered recovery:} above 98\% GPU occupancy, a lazy reclamation path calls \texttt{gc.\allowbreak collect()} and \texttt{torch.\allowbreak cuda.\allowbreak empty\allowbreak\_cache()}, reclaiming fragmented memory before an allocation fails while keeping CPU--GPU synchronization off the common path. \textbf{Deterministic deallocation:} transient tensors (e.g.\ \textit{batched\_data}, \textit{step\_ctx}) are explicitly \texttt{del}'d at each loop's end so the caching allocator can reuse their blocks immediately, which is what keeps \texttt{cudaMalloc} failures from appearing during continuous serving.

\begin{figure*}[t]
    \centering
    \definecolor{cblue}{HTML}{2F6DB5}
\definecolor{cteal}{HTML}{2E9E96}
\definecolor{corange}{HTML}{E08A3C}
\definecolor{cslate}{HTML}{5B6573}
\definecolor{cink}{HTML}{2B2F36}
\definecolor{cred}{HTML}{B5423A}

\providecommand{\tkkey}[2]{  \tikz[baseline=-0.55ex]{\draw[#1, line width=0.9pt] (0,0) -- (0.34,0);
    \node[#2] at (0.17,0) {};}}

\pgfplotsset{
  tksty2/.style={
    width=3.05cm, height=1.95cm, scale only axis,
    scaled ticks=false,
    tick align=outside, tick pos=left,
    grid=major, grid style={black!10, line width=0.3pt},
    axis line style={black!45, line width=0.5pt},
    label style={font=\scriptsize, text=cink},
    tick label style={font=\tiny, text=cink},
    title style={font=\scriptsize\bfseries, text=cink, yshift=-2pt},
    every axis plot/.append style={line width=0.8pt, mark size=1.2pt},
    xmode=log, log basis x=2, xmin=430, xmax=39000,
    xtick={512,8192}, xticklabels={512,8192},
  },
  tktriton/.style={cblue,   solid, mark=diamond*, mark options={solid}},
  tkvanilla/.style={cslate, solid, mark=*,        mark options={solid, fill=white, line width=0.6pt}},
  tkmask/.style={corange,   solid, mark=square*,  mark options={solid, fill=white, line width=0.6pt}},
  tkdense/.style={cteal,    solid, mark=triangle*, mark options={solid}},
  tkpeak/.style={cred, dashed, no marks, line width=0.8pt},
}

\input{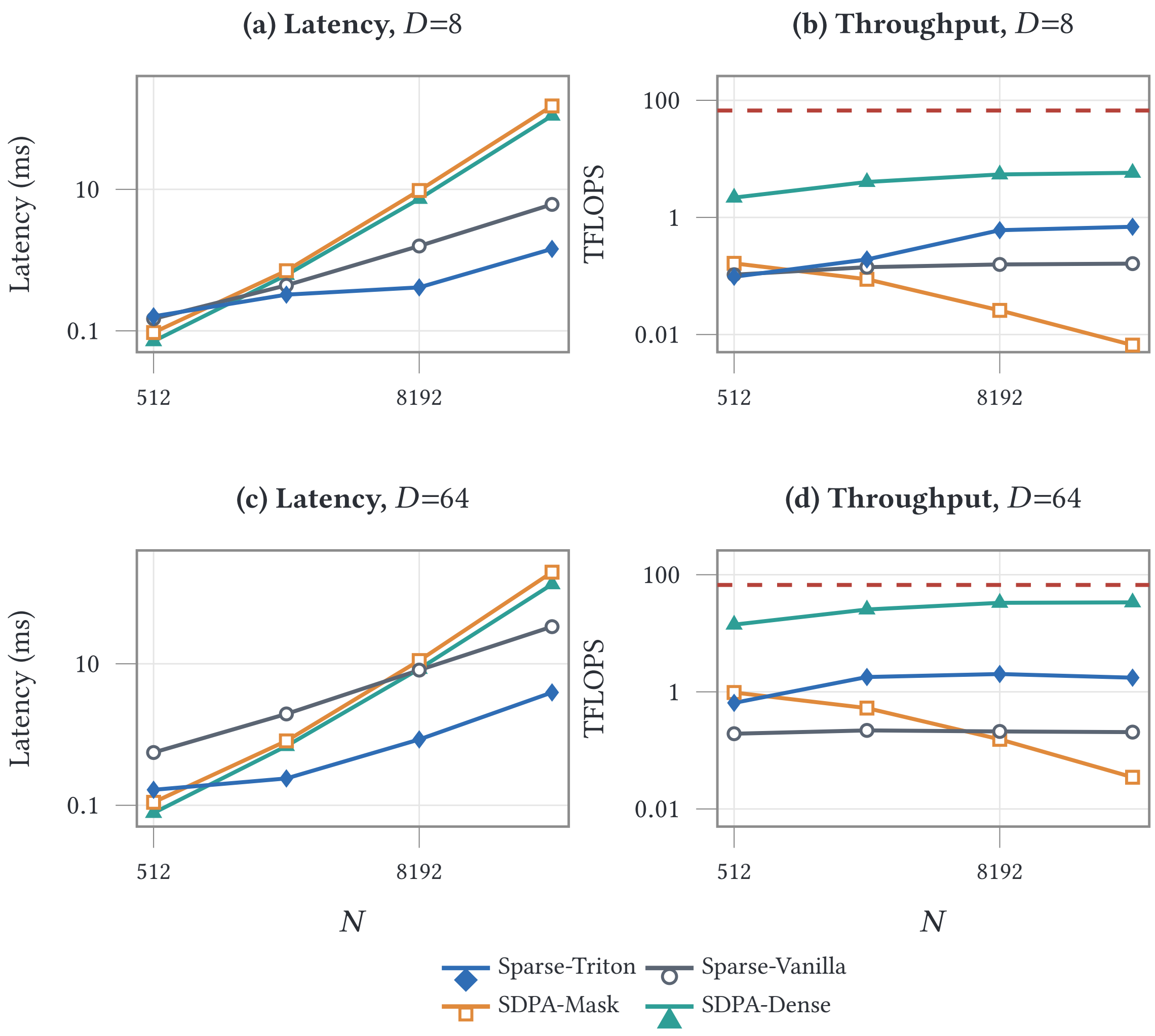}

\resizebox{\figKbScale\linewidth}{!}{\begin{tikzpicture}
\begin{groupplot}[
      group style={group size=2 by 2, horizontal sep=1.05cm, vertical sep=1.4cm},
    tksty2,
  ]

\nextgroupplot[
    title={(a) Latency, $D{=}8$},
    ylabel={Latency (ms)},
    ymode=log, log ticks with fixed point, ymin=0.05, ymax=400, ytick={0.1,10},
  ]
\addplot[tkdense]   coordinates {(512,0.0713)(2048,0.6165)(8192,7.3315)(32768,109.918)};
\addplot[tkmask]    coordinates {(512,0.0947)(2048,0.7079)(8192,9.6304)(32768,150.934)};
\addplot[tkvanilla] coordinates {(512,0.1481)(2048,0.4404)(8192,1.5826)(32768,6.1297)};
\addplot[tktriton]  coordinates {(512,0.1605)(2048,0.3250)(8192,0.4115)(32768,1.4391)};

\nextgroupplot[
    title={(b) Throughput, $D{=}8$},
    ylabel={TFLOPS},
    ymode=log, log ticks with fixed point, ymin=0.005, ymax=260, ytick={0.01,1,100},
  ]
\addplot[tkpeak] coordinates {(430,66.9)(39000,66.9)};
\addplot[tkdense]   coordinates {(512,2.1780)(2048,4.0276)(8192,5.4189)(32768,5.7830)};
\addplot[tkmask]    coordinates {(512,0.16439)(2048,0.08795)(8192,0.02586)(32768,0.00660)};
\addplot[tkvanilla] coordinates {(512,0.10511)(2048,0.14138)(8192,0.15736)(32768,0.16251)};
\addplot[tktriton]  coordinates {(512,0.09696)(2048,0.19155)(8192,0.60515)(32768,0.69220)};

\nextgroupplot[
    title={(c) Latency, $D{=}64$},
    xlabel={$N$}, ylabel={Latency (ms)},
    ymode=log, log ticks with fixed point, ymin=0.05, ymax=400, ytick={0.1,10},
  ]
\addplot[tkdense]   coordinates {(512,0.0779)(2048,0.6871)(8192,8.4551)(32768,132.952)};
\addplot[tkmask]    coordinates {(512,0.1102)(2048,0.8129)(8192,11.1053)(32768,197.232)};
\addplot[tkvanilla] coordinates {(512,0.5570)(2048,1.9553)(8192,8.1483)(32768,33.472)};
\addplot[tktriton]  coordinates {(512,0.1656)(2048,0.2396)(8192,0.8522)(32768,3.9362)};

\nextgroupplot[
    title={(d) Throughput, $D{=}64$},
    xlabel={$N$}, ylabel={TFLOPS},
    ymode=log, log ticks with fixed point, ymin=0.005, ymax=260, ytick={0.01,1,100},
  ]
\addplot[tkpeak] coordinates {(430,66.9)(39000,66.9)};
\addplot[tkdense]   coordinates {(512,14.045)(2048,25.491)(8192,33.145)(32768,33.726)};
\addplot[tkmask]    coordinates {(512,0.97417)(2048,0.52804)(8192,0.15461)(32768,0.03482)};
\addplot[tkvanilla] coordinates {(512,0.19266)(2048,0.21953)(8192,0.21072)(32768,0.20519)};
\addplot[tktriton]  coordinates {(512,0.64804)(2048,1.79170)(8192,2.01485)(32768,1.74487)};

\end{groupplot}

\node[anchor=north, yshift=-0.78cm, text=cink]
  at ($(group c1r2.south)!0.5!(group c2r2.south)$) {  \tiny
  \begin{tabular}{@{}l@{\;}l@{}}
    \tkkey{cblue}{draw=cblue, fill=cblue, diamond, inner sep=1.1pt} Sparse-Triton &
    \tkkey{cslate}{draw=cslate, fill=white, circle, inner sep=1.0pt, line width=0.6pt} Sparse-Vanilla \\[1pt]
    \tkkey{corange}{draw=corange, fill=white, rectangle, inner sep=1.2pt, line width=0.6pt} SDPA-Mask &
    \tkkey{cteal}{draw=cteal, fill=cteal, regular polygon, regular polygon sides=3, inner sep=0.9pt} SDPA-Dense \\
  \end{tabular}};
\end{tikzpicture}}
    \caption{\textbf{Single-GPU attention-kernel benchmark on H100.} Latency \textbf{(a,c)} and effective throughput \textbf{(b,d)} of Sparse-Triton against a vanilla sparse baseline and two dense SDPA variants, at $D{=}8$ (a,b) and $D{=}64$ (c,d); the dashed red line marks the device's dense FP32 peak. Dense SDPA is a hardware reference point, not a deployable alternative for this sparse workload: it computes the full $N^2$ attention matrix, which is why its throughput is high while its latency explodes with $N$. The takeaway is the comparison \emph{within} sparse implementations---Sparse-Triton dominates the vanilla baseline at practical sizes, while SDPA-Mask collapses because it still materializes dense score matrices. Exact values in Table~\ref{tab:triton_kernel_metrics}.}
    \label{fig:triton_h100}
\end{figure*}

\paragraph{Kernel benchmark.}
Fig.~\ref{fig:triton_h100} and Table~\ref{tab:triton_kernel_metrics} give the benchmark Section~\ref{sec:experiments} frames, with its exact per-configuration values. Irregular gathers and masked reductions never reach the Tensor Core path, which is why the ceiling is sparsity rather than the dense peak. What the Triton path removes is avoidable sparse-execution overhead; the intrinsic MFU penalty stays underneath it.

\begin{table*}[t]
    \centering
    \caption{Exact throughput and latency of the attention kernels of Fig.~\ref{fig:triton_h100}. Each cell reports effective throughput in TFLOPS, with kernel latency in milliseconds in parentheses. The dense SDPA result is a hardware-oriented reference point rather than a deployable alternative for the sparse molecular workload. The sweep runs at $16$ heads to isolate the $N$ and $D$ axes; production uses $128$ (Appendix~\ref{sec:app-config}), so these are kernel-level comparisons rather than production-shape measurements.}
    \apptablesize
    \begin{tabular}{ccccccc}
        \toprule
        $N$   & head & $D$ & \textbf{Sparse-Triton} & \textbf{Sparse-Vanilla} & \textbf{SDPA-Mask} & \textbf{SDPA-Dense} \\
        \midrule
        512   & 16   & 8   & 0.097 (0.161)          & 0.105 (0.148)           & 0.164 (0.095)      & 2.178 (0.071)       \\
        2048  & 16   & 8   & 0.192 (0.325)          & 0.141 (0.440)           & 0.088 (0.708)      & 4.028 (0.617)       \\
        8192  & 16   & 8   & 0.605 (0.412)          & 0.157 (1.583)           & 0.026 (9.630)      & 5.419 (7.331)       \\
        32768 & 16   & 8   & 0.692 (1.439)          & 0.163 (6.130)           & 0.007 (150.934)    & 5.783 (109.918)     \\
        \midrule
        512   & 16   & 64  & 0.648 (0.166)          & 0.193 (0.557)           & 0.974 (0.110)      & 14.045 (0.078)      \\
        2048  & 16   & 64  & 1.792 (0.240)          & 0.220 (1.955)           & 0.528 (0.813)      & 25.491 (0.687)      \\
        8192  & 16   & 64  & 2.015 (0.852)          & 0.211 (8.148)           & 0.155 (11.105)     & 33.145 (8.455)      \\
        32768 & 16   & 64  & 1.745 (3.936)          & 0.205 (33.472)          & 0.035 (197.232)    & 33.726 (132.952)    \\
        \bottomrule
    \end{tabular}
    \label{tab:triton_kernel_metrics}
\end{table*}

\section{Implementation Detail: Partition, Neighbor Search, and Communication}
\label{sec:app-partition}
\label{sec:app-neighbor}
\label{sec:app-comm}

This appendix carries what the main text defers about three mechanisms in
turn: the one caveat behind the partition-curve comparison, the metadata
descriptor the Level-1 filter publishes along with the definition of the peer
and candidate-population figures, and the buffer and autograd detail behind the
fused SKV exchange. The derivations those mechanisms rest on---the count-drift
estimate, the three-level filter comparison, the cache exactness invariant,
plan-reuse volume, sender-side projection, the overlap bound, and the
local-neighbor fraction---are in Sections~\ref{sec:morton_partition},
\ref{sec:multilevel_search} and~\ref{sec:reduce_volume}.

\paragraph{Morton/Hilbert Quantization Caveat.}
Our Morton implementation normalizes all three quantization axes by the largest extent, while Hilbert normalizes per axis. On strongly anisotropic cells the two therefore differ in effective resolution as well as in curve topology, so part of the measured Morton--Hilbert gap is attributable to quantization rather than to locality. We do not claim the two contributions are separated at every scale. The gap does persist at 64 GPUs, where the cubic benchmark cell makes the two normalizations coincide and only the topology can differ, which is the cleanest evidence that the locality effect is real.

\paragraph{Level-1 Periodic-Image Metadata.}
Each rank publishes its query AABB plus one AABB per occupied periodic image of its shard, at most $27$ including the identity image, with the corresponding counts: $8 + 27\times 8$ doubles, under $2$\,KiB per rank. Because the descriptor is this small and its size fixed, a single metadata all-gather lets every rank reconstruct the complete rank$\times$rank$\times$image request matrix locally, with no second exchange of request masks. For contrast, the spatial-bin variant publishes $4 + 8192\times 16$ doubles ($\approx 1$\,MiB) per rank and does need that second all-gather, which is the dominant reason it loses despite pruning the candidate side hardest.

\paragraph{Peer Degree and Candidate Population.}
Fig.~\ref{fig:neighbor_ablation}(c,d) and the peer/pool figures quoted throughout Section~\ref{sec:neighbor-scaling} are both per-step cross-rank maxima, read off the request matrix above for whichever rank is heaviest that step (not necessarily the same rank for both quantities). Peer degree is the number of source ranks it requests candidates from, self included; candidate population is the atom count it then holds after the exchange, before per-chunk filtering, periodic images included. All-gather has no sparse mask to bound either and is $=K$ by construction. The equal-volume comparator quoted against these figures is $27$, not the textbook $26$: a grid cell's $26$ face-, edge-, and corner-adjacent neighbors plus the cell itself, counted the same self-inclusive way.

\paragraph{Level-3 Auxiliary Costs.}
Beyond the pairwise search itself, chunking adds an $O(Q_{\mathrm{loc}}\log Q_{\mathrm{loc}})$ median bisection to build the chunk tree and one AABB test per chunk, linear in the gathered atoms. Both are small relative to the $Q_{\mathrm{loc}}C_{\mathrm{cand}}-\sum_i Q_iC_i$ pairwise work they remove, and the per-chunk searches are issued as a single batched GPU call rather than one launch per chunk.

\paragraph{Fused SKV Buffer Layout and Autograd.}
Making the SKV fusion practical requires reorganizing both data layout and operator dependencies. We redesign the communication buffer layout so subsequent attention kernels consume the received SKV data directly from the fused buffer without materializing redundant intermediate copies, and rewrite the corresponding operators' autograd dependencies so the backward pass propagates gradients through the fused path in a well-defined manner. The result preserves end-to-end differentiability with no activation-memory footprint beyond the communication buffers the forward pass already requires.

\paragraph{Pre-registered Buffer Pool.}
Send and receive buffers are carved from a pre-allocated pool whose Memory-Region registration is done in advance during plan construction, so steady-state transfers repeat neither buffer creation nor dynamic MR registration and each exchange launches with minimal bookkeeping (Fig.~\ref{fig:mempool}).

\section{Measurement Protocol and Full-Inference Validation}
\label{sec:app-protocol}

This appendix gives the measurement protocol behind Section~\ref{sec:experiments}
in full. The three tables the main text uses---the neighbor-search composition,
the full-inference cache validation, and the strong-scaling detail---are set in
Sections~\ref{sec:neighbor-scaling} and~\ref{sec:md-validation} rather than
repeated here.

\paragraph{Single-GPU ablation timing.}
Each \texttt{Latency} entry in Table~\ref{tab:toy_infer_ablation} is the mean
over the last $50$ of $100$ repetitions on the fixed $10$k-atom system, so the
first half absorbs \texttt{torch.\allowbreak compile} tracing, autotuning, and
first-touch allocation and the reported mean is the steady-state cost.
\texttt{Mem} is the peak allocation over the same run.

\paragraph{Neighbor-stage timing.}
Beyond the scope limit stated in Section~\ref{sec:neighbor-scaling}, two
details govern how the neighbor-pipeline numbers should be read. The sub-stage
timers are each reduced across ranks with MAX independently, and the maxima of
different sub-stages may originate from different ranks, so we do not decompose
the pipeline additively---this is why our attribution of spatial-bin's cost to
plan overhead is described as consistent with the telemetry rather than as a
measured stage decomposition. Step~0 is excluded from every series, since it
carries warm-up and first-touch allocation. The $25$k and $50$k configurations
run in neighbor-only mode; the cache result is separately confirmed in
full-inference runs at $20$k atoms per GPU
(Table~\ref{tab:neighbor_cache_fullinfer}, Section~\ref{sec:neighbor-scaling}).
One official run per point, so no error bars are drawn.

\paragraph{Neighbor-search composition and its full-inference check.}
Table~\ref{tab:neighbor_composition} (Section~\ref{sec:neighbor-scaling}) gives
the exact per-mechanism numbers behind Section~\ref{sec:neighbor-scaling}'s
composed $15.7\times$ reduction, including each row's own
$64$-to-$512$-GPU growth---a scaling behavior Figure~\ref{fig:neighbor_cache}(b)
does not show, since it plots only the 512-GPU endpoint.

How the cached rows are built matters for how much weight the exact figure
carries. Each is $(t_{\mathrm{rebuild}}+9\,t_{\mathrm{hit}})/10$, where
$t_{\mathrm{hit}}$ is the cache-hit cost measured on an interval-$10$ run and
$t_{\mathrm{rebuild}}$ is the rebuild cost measured on a \emph{separate}
interval-$1$ run of the same configuration. The two are not interchangeable,
for two reasons that both point the same way. The interval-$1$ arm runs with the
cache off and therefore searches at each layer's bare cutoff, while an
interval-$10$ rebuild builds the $r_c\!+\!\delta$ superset the following nine
steps consume---$1.33\times$ the candidate volume at the $5$\,\AA{} layers and
$1.20\times$ at the $8$\,\AA{} one, so $t_{\mathrm{rebuild}}$ is measured on a
strictly cheaper search than the one it stands in for. Nor can we simply
substitute the right quantity: the interval-$10$ arm's only rebuild is step~$0$,
which carries compile and first-touch cost and is excluded as warm-up
everywhere in this study, so these runs contain no steady-state interval-$10$
rebuild at all. Where both quantities do exist in one place---the $20$k
atoms/GPU full-inference runs of Table~\ref{tab:neighbor_cache_fullinfer}
(Section~\ref{sec:neighbor-scaling}), whose rebuilds fall on steps $10$ and
$20$---a steady-state interval-$10$ rebuild runs $0.98$--$1.62\times$ its
interval-$1$ counterpart, the larger radius accounting for much of that. The
composition therefore understates the amortized cost by up to $1.4\times$ at
those points, and the composed $15.7\times$ should be read as an upper estimate
rather than a measured speedup. Two things it does not affect: the cache-hit
latencies of Fig.~\ref{fig:neighbor_cache}(a), measured directly, and the
full-step $1.87\times$ of Section~\ref{sec:e2e_scaling}, an A/B on the complete
pipeline. Table~\ref{tab:neighbor_cache_fullinfer} likewise reports a directly
timed interval-$10$ window; its \emph{Amortized} column is the mean over that
window, and the algebraic identity between that mean and
$(t_{\mathrm{rebuild}}+9\,t_{\mathrm{hit}})/10$ within a single $2$:$18$ run is
arithmetic, not independent confirmation that an interval-$1$ rebuild may stand
in for an interval-$10$ one.

\paragraph{Overlap A/B: what the timelines are not.}
Section~\ref{sec:overlap-results}'s overlap result is a single-flag A/B on the
configuration of Section~\ref{sec:e2e_scaling}, so it inherits that protocol.
Nsight timelines corroborate it qualitatively, showing the layer-wise exchange
largely masked behind the local branch, but we do not count them as a second
measurement: they are an aggregate, run-level view on a different denominator
than the per-step timers the A/B compares.

\paragraph{End-to-end timing.}
Every point in Section~\ref{sec:e2e_scaling} is the mean over steps
$300$--$399$ of a $500$-step run. The window is chosen so that it excludes the
\texttt{torch.\allowbreak compile} first step and spans exactly ten rebuild
periods at interval $10$, which makes the mean the amortized per-step cost
rather than a sample biased by where rebuilds happen to land. One official run
per point, so no error bars are drawn.

\paragraph{Trajectory Settings and Achievable Timescale.}
The $10.24$M-atom KCl and $10.06$M-atom capsid trajectories of
Section~\ref{sec:md-validation} rebuild the neighbor graph every $10$ steps and
re-partition every $100$, matching Sections~\ref{sec:multilevel_search}
and~\ref{sec:md_engine}'s defaults, with the $0.5$\,\AA{} displacement skin
triggering an early rebuild past half that margin ($0.25$\,\AA{}). At the
$0.5$\,fs step used throughout, the capsid run is $200{,}000$ steps, or about
$12$ days of wall time at its measured $5.2$\,s per step---$\approx\!8$\,ps/day.
The $30.7$M-atom configuration of Section~\ref{sec:e2e_scaling} reaches
$\approx\!26$\,ps/day at $1.684$\,s per step. Tens of picoseconds per day is
enough for the equilibration, structural and energy-stability checks reported
here, and short of the nanosecond-to-microsecond windows a binding or folding
study would need; closing it is a strong-scaling problem at fixed atom count,
the axis on which Fig.~\ref{fig:strong_scaling}'s returns go flat past $160$
GPUs at $8$k atoms per GPU.

\paragraph{MD trajectory: rebuild-reason breakdown.}
Section~\ref{sec:md-validation}'s ``cache-only rebuild'' reasons condense two of
the four categories the underlying per-rank-step log actually distinguishes.
Over the NVT segment, \texttt{scheduled\_interval} (the every-$10$-step
schedule) accounts for $48.05\%$ of rank-step records and
\texttt{max\_displacement} (a rank's own skin crossing) for $1.77\%$---a
$27.09$:$1$ ratio ($35.52$:$1$ over NVE). That ratio, however, compares
\emph{initiating} ranks, and is not the share of rebuild work displacement is
responsible for; the paragraph below resolves the two. The other two categories
split along a different axis than schedule vs.\ displacement, not a third
independent cause: \texttt{scheduled\_interval} and \texttt{dispatch}
($4.81\%$) are both unconditional---the driver's global-sync logic invokes
every rank identically on the same step, with no triggering-rank/following-rank
distinction, so \texttt{dispatch} is every rank's own record of a repartition
event, not a subset that caused it. Only \texttt{max\_displacement} has that
triggering structure, since it is a genuinely local, per-rank measurement:
whichever ranks independently cross the skin record \texttt{max\_displacement},
and the all-reduce of Section~\ref{sec:multilevel_search} then forces every
other rank to rebuild too, purely to keep the decision unanimous. That induced
remainder is \texttt{passive\_follow} ($45.37\%$)---specifically
max\_displacement's, not dispatch's, since only a small minority of ranks cross
the skin on any given displacement-triggered step. Attributing cost rather than
blame, then, displacement is responsible for $1.77+45.37=47.14\%$ of rank-step
rebuild records against the schedule's $48.05\%$: the two are near parity,
which is what the main text reports and what the $90\%\!\rightarrow\!79\%$
hit-rate gap independently confirms. The $27$:$1$ figure above is the ratio of
initiating ranks only, and would badly understate displacement's cost if read
as an event frequency. The capsid trajectory of
Section~\ref{sec:density-heterogeneity} splits the same four ways, and the split
separates cleanly into system-independent and system-dependent causes. The two
globally scheduled ones match the figures above---\texttt{scheduled\allowbreak\_interval}
$48.06\%$ against $48.05\%$, \texttt{dispatch} $4.81\%$ in both---which is what
a fixed interval should do, since neither depends on what is being simulated.
The one genuinely per-rank cause does not: \texttt{max\allowbreak\_displacement}
is $1.23\%$ of the capsid's rank-step records against $1.77\%$ here, $31\%$
lower, with the induced \texttt{passive\allowbreak\_follow} share rising to
match ($45.89\%$ against $45.37\%$). Capsid atoms are covalently constrained and
tightly packed, so they displace less per step than bulk water and fewer ranks
independently cross the $\delta/2$ threshold.
\paragraph{Strong scaling: recomputation off, and the forward-pass split.}
Each point in Figure~\ref{fig:strong_scaling} and
Table~\ref{tab:strong-scaling-detail} (Section~\ref{sec:md-validation}) is the
mean step time over the last $5$ logged windows of its run, the steady-state
value used throughout for scaling comparisons. The table extends
Figure~\ref{fig:strong_scaling} two ways. Recomputation off is undefined at
$32$/$64$ GPUs (OOM, Section~\ref{sec:e2e_scaling}) but runs $5$--$23\%$ faster
where both do, flattening at the same $160$-GPU knee. Its margin narrows with
device count---$22.9\%$ at $96$ GPUs, $5.0\%$ at $256$---as recomputation's own
cost follows the shrinking local batch, so on a common $96$-GPU baseline it
reaches only $64.0\%$ at $256$ against on's $75.0\%$. The knee appears in both
modes, but they do not scale identically, and the efficiency figure quoted for
either is only comparable against a shared reference point. The forward-pass
hit/rebuild split shows no consistent ordering ($\pm 35$\,ms either way), unlike
the $5.6$--$8.4\%$ hit-slower-than-rebuild gap
Table~\ref{tab:density-heterogeneity} finds at the much larger per-step scale
of Section~\ref{sec:density-heterogeneity}.

\paragraph{TF32: What the Precision Mode Costs.}
Section~\ref{sec:single_gpu}'s TF32 mode is characterized statically in
Table~\ref{tab:toy_infer_ablation}; what it does to a trajectory we measure
directly, as a single-flag A/B on $64$ GPUs over the $1.28$M-atom system of
Fig.~\ref{fig:strong_scaling}, $10$\,ps per arm with the drift fitted over the
last $5$\,ps ($101$ samples each). Energy drift is $-6.316$\,$\mu$eV/atom/ps
with TF32 on against $-0.966$ with it off, a factor of $6.54$. Two readings say
that the extra drift is systematic rather than noise, and that it does not carry
the thermodynamics with it: the residual about each linear fit is the same in
both arms ($0.328$ vs.\ $0.320$\,eV), and temperature and its rate of change are
indistinguishable ($201.37\pm1.71$ vs.\ $201.45\pm1.72$\,K; $+1.16$ vs.\
$+1.17$\,K/ps). The mode is worth $1416.9\pm16.0 \to 912.1\pm9.4$\,ms per step,
$1.55\times$. The pair runs on the same A100 cluster and with the same
activation recomputation as the strong-scaling sweep, so it anchors to an
already-reported point: the off arm reproduces
Table~\ref{tab:strong-scaling-detail}'s $64$-GPU step time to $0.9\%$. That is
the split Section~\ref{sec:exp-setup} makes: of the two observables resolved
here TF32 moves only the drift, so Section~\ref{sec:md-validation}, where drift
\emph{is} the measurement, runs without it, while the throughput and weak-scaling
results, which the $6.54\times$ does not touch, run with it. Two scope limits:
this pair sits at $\approx\!201$\,K and is still warming at $1.16$\,K/ps, not at
Section~\ref{sec:md-validation}'s equilibrated $300$\,K, and it is a $1.28$M-atom
system---so its off-arm $-0.966$ is not the same measurement as the $0.42$
quoted there for $10.24$M atoms, and the two should not be read against each
other.

\begin{figure*}[t]
    \centering
    \input{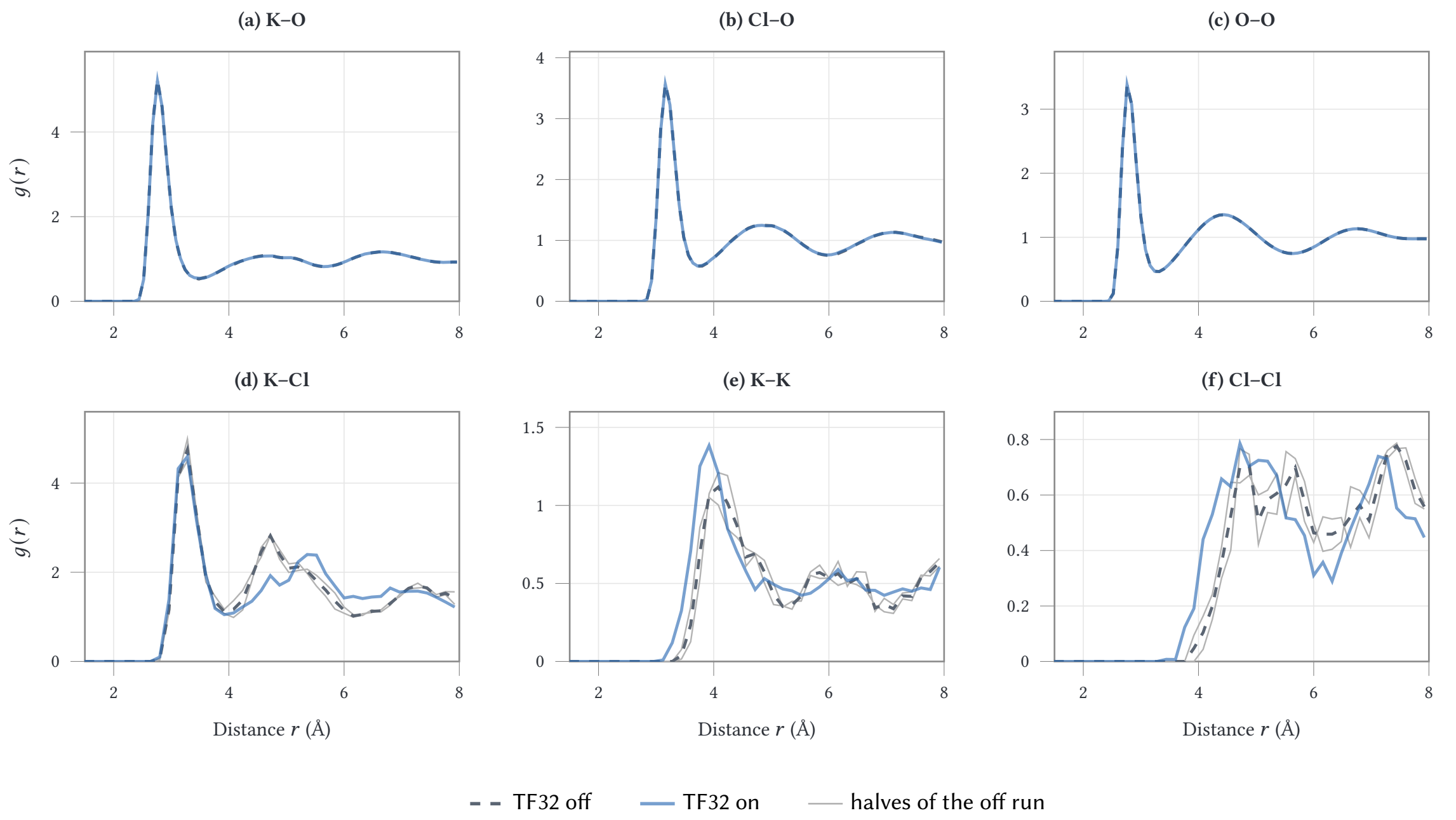}
    \caption{\textbf{TF32 on/off structure over the same $5$\,ps window.} All six radial distribution functions of the $1.28$M-atom KCl-in-water run of Fig.~\ref{fig:strong_scaling}, TF32 off (dashed) against TF32 on (solid), $101$ frames each. \textbf{(a--c)}~Ion--water and water--water structure is indistinguishable. \textbf{(d--f)}~The ion--ion pairs differ, but the two gray curves are the two halves of the TF32-off run \emph{by itself}, $51$ frames each and no precision change involved: they set a noise floor of $10.3\%$, $14.1\%$ and $12.5\%$ on these first peaks, which is the scale of the on/off gap. With $1181$ ions of each sign and ion decorrelation slower than the window, the effective sample behind a K--K peak is tens of pairs, not the $10^3$ the frame count suggests.}
    \label{fig:tf32_rdf}
\end{figure*}

\paragraph{TF32 and Structure.}
Drift is not the only thing a precision mode could move, so we also compare all
six radial distribution functions of the same pair of runs over the same $5$\,ps
window (Fig.~\ref{fig:tf32_rdf}). The solvation structure is reproduced
essentially exactly: first-peak heights agree to $0.5\%$ (K--O), $0.2\%$
(Cl--O) and $0.0\%$ (O--O), and, integrated to a common cutoff, both ion--water
coordination numbers agree to $0.2\%$. The three ion--ion pairs are where the
two runs visibly part---first-peak heights differ by $-3.2\%$ (K--Cl), $+23.8\%$
(K--K) and $+11.6\%$ (Cl--Cl)---and that gap is not evidence about TF32.
Splitting the TF32-\emph{off} trajectory alone into its two halves of $51$
frames, with no precision change anywhere, moves the same three peaks by
$10.3\%$, $14.1\%$ and $12.5\%$ (the gray curves in
Fig.~\ref{fig:tf32_rdf}d--f): K--Cl and Cl--Cl sit inside that floor and K--K
is $1.7\times$ it. The reason is sample size. The system carries $1181$ K$^+$
and $1181$ Cl$^-$, so the K--K first-peak shell holds on the order of $10$ pairs
per frame and $10^3$ over the window---but ion diffusion decorrelates far more
slowly than $5$\,ps, so those $101$ frames contain only $O(1$--$3)$ independent
ion configurations and an effective sample of order $10$--$30$ pairs. That
predicts $1/\sqrt{N_{\text{eff}}} \approx 18$--$32\%$ scatter, which brackets
the observed $23.8\%$ and sits above the half-split floor---as it should, since
two halves of one trajectory share an initial condition and so understate the
scatter between independent runs. Resolving whether TF32 touches ion--ion
structure at all needs a trajectory long enough to decorrelate the ions, not a
longer averaging window over the same one; what these runs do establish is that
it leaves water and the solvation shells alone.

\paragraph{Guard and Rebuild Stability over $100$\,ps.}
Section~\ref{sec:density-heterogeneity}'s capsid arm is long enough to ask
whether the shape invariant survives a real trajectory rather than a
steady-state window. Over its $200{,}000$ steps the run performed $200{,}000$
force evaluations---$158{,}800$ on a cached graph and $41{,}200$ on a rebuild, a
$79.40\%$ hit rate---and re-cut the partition throughout on the fixed
$100$-step schedule. \texttt{torch.\allowbreak compile}'s guards are the
mechanism that would catch a shape moving, and they held: zero graph breaks,
zero fallbacks to eager, zero backend compilation failures, and a compiled-code
cache that never evicted---a limit of $128$ against a high-water mark of three
variants. Past the $1{,}738$ one-time compilations at step~$0$, the whole
trajectory triggered about $41$ further recompilations, all in the energy/force
output head and all traced to one cause that is not shape drift: two dimensions
that happened to be equal when that function was first traced (both $34{,}921$)
were specialized into an equality by duck-shaped sizing, so later steps that
break the coincidence recompile. Turning duck shapes off removes them.
\texttt{suppress\allowbreak\_errors} was armed throughout and never fired, so
the zero-fallback counts are measured rather than an artifact of silent
recovery. The division of labor behind those counts is worth stating, since
$41{,}200$ rebuilds might otherwise be expected to cost a recompilation apiece:
the node path is count-fixed and compiled static, while the operators whose
extents follow the edge set are compiled for dynamic shapes, so a rebuild moves
their sizes without asking for new code. Three compiled variants and $41$
attributable recompilations across $200{,}000$ steps are the durability
evidence behind Section~\ref{sec:morton_partition}'s claim that the equal-count
cut pins node shapes for the life of a partition.

\paragraph{Hit-Slower-Than-Rebuild: Communication Wait, Not Compute.}
Table~\ref{tab:density-heterogeneity}'s cache-hit forward costs more than
rebuild forward in both columns, despite skipping the search a rebuild pays for.
A $256$-GPU, $2000$-step run with kernel-level profiling (steps $2$--$13$,
spanning $2$ rebuilds and $10$ hits) resolves this. The model-compute kernel is
not slower on hit steps---if anything rebuild's is $1.6\%$ higher ($1833.4$
vs.\ $1804.3$\,ms, consistent with a guard check on the changed edge shape)---
while the cross-rank NCCL exchange is $724$\,ms \emph{cheaper} on rebuild
($1235.4$ vs.\ $1959.6$\,ms). Per-step communication time on hit steps ranges
$735$--$2451$\,ms (a $3.3\times$ spread) while compute holds within $1\%$
($1797$--$1812$\,ms) over the same steps, so the gap is entirely in how long a
rank waits at the attention All-to-All, not in what it computes. Aggregated
over the full run (weighted by each window's actual hit/rebuild count, step~$0$
excluded), forward-on-hit costs $3465.8$\,ms against forward-on-rebuild's
$3243.7$\,ms, a $6.85\%$ gap consistent with the kernel-level breakdown above.

We had previously ruled out cross-rank drift because
Algorithm~\ref{alg:md-step}'s per-step collectives---the rebuild-decision
\texttt{all\allowbreak\_reduce}, the per-layer All-to-All, the closing scalar
reduction---complete every step. That reasoning conflated \emph{completing}
with \emph{arriving together}: a collective forces the exchange to finish before
the step proceeds, not that every rank reaches it at the same wall-clock
instant. A rebuild's ghost-atom exchange is a larger, different collective that
in practice re-aligns ranks; consecutive hit steps carry no equivalent, so
small per-step timing differences build up until the next attention All-to-All
waits on whichever rank is currently slowest. The effect is scale-dependent:
across $2$/$4$/$8$/$32$/$256$-GPU configurations, compute alone never favors hit
over rebuild (tied at $2$ and $8$ GPUs, rebuild $1.3$--$1.6\%$ higher at
$4$/$32$/$256$), but the full forward timer's hit-slower gap needs per-step work
to drift against as well as ranks: it shows at $256$ GPUs here and in
Table~\ref{tab:density-heterogeneity}, and not in
Table~\ref{tab:strong-scaling-detail}'s strong-scaling points, whose $256$-GPU
forward is a sixth as long at $5$k atoms/GPU and where hit is the faster path at
four of the eight points. We report it as a property of the loaded regime, not a
universal one.

\section{Hardware and Software}
\label{sec:app-hardware}

This appendix records the hardware and software behind every reported timing.

The scaling study runs on $64$ compute nodes of $8$ NVIDIA H100 GPUs each
($80$\,GB HBM3, NVLink~4.0), linked by a $400$\,Gbps InfiniBand NDR fabric;
the largest configuration uses all $512$ GPUs. The strong-scaling sweep, the
TF32 A/B pair, and the two MD trajectory studies run on a separate cluster of
$256$ A100 GPUs ($40$\,GB HBM2e, PCIe, no NVLink). Both stacks use CUDA 13.0,
PyTorch 2.11, NCCL 2.28, Triton 3.7 and FAISS 1.14.2 with the E2Former~V2
backbone of Appendix~\ref{sec:app-config}. Every timing is the per-step value
reduced across ranks with MAX, so it measures the distributed critical path
rather than a rank-$0$ view.

\end{document}